%% file: main.tex
\documentclass[9pt,twocolumn,twoside]{pnas-new}

\templatetype{pnasresearcharticle} 

\usepackage[utf8]{inputenc} 
\usepackage[T1]{fontenc}    
\usepackage{url}            
\usepackage{booktabs}       
\usepackage{amsfonts}       
\usepackage{nicefrac}       
\usepackage{microtype}      
\usepackage{xcolor}         
\usepackage{graphicx}
\usepackage{caption}
\usepackage{multirow}
\usepackage{advdate}
\usepackage{amssymb}
\usepackage{amsmath}
\usepackage{bm} 

\usepackage{mathrsfs} 
\usepackage{systeme}
\usepackage{mathtools}
\usepackage{pdfpages}
\usepackage{paralist}
\usepackage{algorithm}
\usepackage{xspace}
\usepackage{array}
\usepackage{comment}
\usepackage[noend]{algpseudocode}
\usepackage{tikz}
\usetikzlibrary{matrix, positioning, arrows.meta, calc, shapes, decorations, backgrounds, arrows, decorations.pathreplacing, automata}
\usepackage[round]{natbib}
\usepackage{hyperref}%
\hypersetup{colorlinks=true, citecolor=[rgb]{0,0,1}, urlcolor=black}
\usepackage{pgfplots}
\usepackage{soul}
\usepackage{longtable}
\usepackage{xcolor}
\usepackage{hhline}
\usepackage{pgfplots}
\usepackage{bbm}
\usepgfplotslibrary{groupplots}

\algrenewcommand\algorithmicindent{1.1em}%
\usepackage{xpatch}

\makeatletter
\xpatchcmd{\algorithmic}
  {\ALG@tlm\z@}{\leftmargin\z@\ALG@tlm\z@}
  {}{}
\makeatother

\renewcommand{\cite}[1]{\citep{{#1}}} 
\newcommand{\namecite}[1]{\citet{{#1}}} 

\input{defs}

\begin{document}

\title{Sampling-based Continuous Optimization with Coupled Variables for RNA Design}

\author[a,b]{Wei Yu Tang}
\author[a]{Ning Dai}
\author[a]{Tianshuo Zhou}
\author[d,e,f]{David H.~Mathews}
\author[a,c,1]{Liang Huang}

\affil[a]{{School of EECS}}
\affil[b]{{Dept.~of Quantitative and Computational Biology}, {University of Southern California}}
\affil[c]{{Dept.~of Biochemistry~\& Biophysics}, {Oregon State University},  {{Corvallis}, {97330}, {OR}, {USA}}}
\affil[d]{{Dept.~of Biochemistry~\& Biophysics}}
\affil[e]{{Center for RNA Biology}}
\affil[f]{{Dept.~of Biostatistics~\& Computational Biology}, {University of Rochester Medical Center}, {{Rochester}, {14642}, {NY}, {USA}}}

\leadauthor{Tang}

\significancestatement{
Since structure determines function in biology, we 
often want to design (non-coding) RNA sequences that fold into a particular  structure.
This is an important problem with wide applications in science and medicine, yet is notoriously difficult due to the vast design space and numerous competing structures.
Previous work mostly uses local search methods such as random walk, which can not keep up with the exponential growth of the design space. 
We instead cast this discrete search problem as continuous optimization, and develop a general-purpose sampling-based framework which can be used for any  objective function.
Our work substantially outperforms existing methods in almost all metrics, especially on long and hard-to-design structures.
}

\authorcontributions{L.H.~conceived and directed the project, and developed the sampling framework.
W.Y.T.~implemented the whole system, and analyzed and visualized the results.
N.D.~suggested the coupled variable for pairs,
guided the softmax parameterization,
and ran with the baseline \cite{matthies+:2023}.
T.Z.~implemented the projected gradient descent and 
contributed to data analysis, esp.~about the SAMFEO baseline.
D.H.M.~suggested the coupled variable for mismatches,
and guided the data analysis and visualizations.
All authors wrote the manuscript.
}
\authordeclaration{The authors have no competing interests.}
\correspondingauthor{\textsuperscript{1}To whom correspondence should be addressed. E-mail: \href{email:liang.huang.sh@gmail.com}{liang.huang.sh@gmail.com}}


\begin{abstract}
The task of RNA design given a target structure
aims to find a sequence that can fold into that structure.
It is a computationally hard problem where
some version(s) have been proven to be NP-hard.
As a result, heuristic methods such as local search have been popular
for this task, but by only exploring a fixed number of candidates.
They can not keep up with the exponential growth of the design space, 
and often perform poorly on longer and harder-to-design structures.
We instead formulate these discrete problems 
as continuous optimization,
which starts with a distribution over all possible candidate sequences,
and uses gradient descent to improve the expectation of an objective function.
We define novel distributions based on coupled variables to rule out invalid sequences given the target structure and 
to model the correlation between nucleotides.
To make it universally applicable to any objective function,
we use sampling to approximate the expected objective function, to estimate the gradient,
and to select the final candidate.
Compared to the state-of-the-art methods,
our work consistently outperforms them in key metrics
such as Boltzmann probability, ensemble defect, and energy gap,
especially on long and hard-to-design puzzles in the Eterna100 benchmark.
Our code is available at: \url{http://github.com/weiyutang1010/ncrna_design}.
\end{abstract}

\dates{This manuscript was compiled on \today}
\doi{\url{www.pnas.org/cgi/doi/10.1073/pnas.XXXXXXXXXX}}

\maketitle
\thispagestyle{firststyle}
\ifthenelse{\boolean{shortarticle}}{\ifthenelse{\boolean{singlecolumn}}{\abscontentformatted}{\abscontent}}{}



\firstpage[3]{4} 
\label{sec:introduction}
\input{introduction}

\section{RNA Design as Discrete Optimization}
\label{sec:discrete}
\input{discrete}

\section{RNA Design as Continuous Optimization}
\label{sec:continuous}
\input{continuous}

\section{Sampling for Objective Evaluation, Gradient Estimation, and  Design Space Exploration}
\label{sec:sampling}
\input{sampling}

\section{Parameterization and Optimization}
\label{sec:params}
\input{parameterization}

\section{Related Work}
\label{sec:related}
\input{related}

\section{Evaluation Results on Eterna100 Dataset}
\label{sec:results}
\input{results}

\section{Conclusions and Future Work}
\label{sec:conclusions}
\input{conclusions}



\acknow{This work was supported in part by 
NSF grants 2009071 (L.H.) and 2330737 (L.H.~and D.H.M.)
and NIH grant R35GM145283 (D.H.M.).}

\showacknow{} 


\bibliography{references}

\clearpage
\input{appendix}

\end{document}

%% file: defs.tex
\renewcommand{\vec}[1]{\ensuremath{\boldsymbol{{#1}}}\xspace}

\newcommand{\nts}{\ensuremath{\text{\it nt}}\xspace}

\newcommand{\pairs}{\ensuremath{\mathit{pairs}}\xspace}

\newcommand{\unpaired}{\ensuremath{\mathit{unpaired}}\xspace}

\newcommand{\mismatches}{\ensuremath{\mathit{mismatches}}\xspace}

\newcommand{\trimismatches}{\ensuremath{\mathit{trimismatches}}\xspace}

\newcommand{\thetau}{\ensuremath{\theta^{\mathrm{u}}}\xspace}
\newcommand{\thetap}{\ensuremath{\theta^{\mathrm{p}}}\xspace}
\newcommand{\thetam}{\ensuremath{\theta^{\mathrm{m}}}\xspace}
\newcommand{\thetatm}{\ensuremath{\theta^{\mathrm{tm}}}\xspace}

\newcommand{\vecthetau}{\ensuremath{\vectheta^{\mathrm{u}}}\xspace}
\newcommand{\vecthetap}{\ensuremath{\vectheta^{\mathrm{p}}}\xspace}
\newcommand{\vecthetam}{\ensuremath{\vectheta^{\mathrm{m}}}\xspace}
\newcommand{\vecthetatm}{\ensuremath{\vectheta^{\mathrm{tm}}}\xspace}

\newcommand{\vecthetahat}{\ensuremath{\widehat{\vectheta}}\xspace}
\newcommand{\vecThetahat}{\ensuremath{\widehat{\vecTheta}}\xspace}

\newcommand{\obj}{\ensuremath{\mathcal{J}}\xspace}

\newcommand{\pystar}{\ensuremath{p_{\vecystar}}\xspace}

\newcommand{\pu}{\ensuremath{p^{\mathrm{u}}}\xspace}
\newcommand{\pp}{\ensuremath{p^{\mathrm{p}}}\xspace}
\renewcommand{\pm}{\ensuremath{p^{\mathrm{m}}}\xspace}
\newcommand{\ptm}{\ensuremath{p^{\mathrm{tm}}}\xspace}

\newcommand{\notes}[1]{}

\newcommand{\ith}[1]{\ensuremath{i^{{th}}}}

\newcount\permx
\newcount\permy
\def\permdot#1#2{
\permx=#1 \advance\permx by-1
\permy=#2 \advance\permy by-1
\psframe[fillcolor=black, fillstyle=solid]
(\permx,\permy)(#1, #2)
}

\newcommand\sign{\mathrm{sign}}

\newcommand{\argmax}{\operatornamewithlimits{\mathrm{argmax}}}

\newcommand{\argmin}{\operatornamewithlimits{\mathrm{argmin}}}

\newcommand{\pluseq}{\mathrel{+}=}

\newcommand{\ystar}{\ensuremath{y^*}\xspace}

\newcommand{\vecx}{\ensuremath{\vec{x}}\xspace}
\newcommand{\vecy}{\ensuremath{\vec{y}}\xspace}

\newcommand{\vecX}{\ensuremath{\mathbf{X}}\xspace}
\newcommand{\vectheta}{\ensuremath{\boldsymbol{\theta}}\xspace}
\newcommand{\vecTheta}{\ensuremath{\bm{\Theta}}\xspace}

\newcommand{\subscript}[2]{_{{#1}, {#2}}}
\newcommand{\Qf}[2]{\ensuremath{Q\subscript{{#1}}{{#2}}}\xspace}

\newcommand{\vecystar}{\ensuremath{\vecy^{\!\!\;\text{$\star$}}}\xspace}
\newcommand{\vecytilde}{\ensuremath{\tilde{\vecy}}\xspace}
\newcommand{\vecxstar}{\ensuremath{\vecx^\star}\xspace}

\newcommand{\vecyhat}{\ensuremath{\hat{\vecy}}\xspace}

\newcommand{\smallnt}[1]{\ensuremath{_{\mbox{\tiny PP}}}\xspace}

\iffalse

\else

\fi

\newcommand{\GEN}{\ensuremath{\mathcal{Y}}\xspace}
\newcommand{\GENX}{\ensuremath{\mathcal{X}}\xspace}

\newcommand{\SAMPLES}{\ensuremath{\mathcal{S}}\xspace}

\newcommand{\defeq}{\ensuremath{\stackrel{\Delta}{=}}\xspace}

\newcommand{\smallurl}[1]{{\scriptsize \url{#1}}}

\newcommand{\leftb}{\ensuremath{\text{\tt (}}\xspace}
\newcommand{\rightb}{\ensuremath{\text{\tt )}}\xspace}

\newcommand{\mydot}{\ensuremath{\text{\tt .}}\xspace}

\newcommand{\nucA}{\ensuremath{\text{\tt A}}}
\newcommand{\nucU}{\ensuremath{\text{\tt U}}}
\newcommand{\nucC}{\ensuremath{\text{\tt C}}}
\newcommand{\nucG}{\ensuremath{\text{\tt G}}}

\newcommand{\allpairtypes}{\ensuremath{\{\nucA\nucU, \nucU\nucA, \nucC\nucG, \nucG\nucC, \nucG\nucU, \nucU\nucG\}}\xspace}

\newcommand{\freeenergy}{\ensuremath{\Delta\:\!\! G^{\scriptsize\circ\!}}\xspace}

\newcommand{\MFE}{\ensuremath{\text{MFE}}\xspace}
\newcommand{\MFEs}{\ensuremath{\text{MFEs}}\xspace}

\newcommand{\UMFE}{\ensuremath{\text{uMFE}}\xspace}

\newcommand{\NED}{\ensuremath{\text{NED}}\xspace}
\newcommand{\ED}{\ensuremath{\text{ED}}\xspace}

\definecolor{intnull}{RGB}{213,229,255}
\definecolor{inteins}{RGB}{128,179,255}
\definecolor{intvier}{RGB}{42,127,255}
\definecolor{intdrei}{RGB}{0,85,212}
\definecolor{intvier}{RGB}{0,51,128}
\definecolor{intfunf}{RGB}{0,34,85}

\newsavebox\CBox

\DeclareMathOperator{\E}{\mathbb{E}}

\newcommand{\expQ}{\ensuremath{\widetilde{Q}}\xspace}

\newcommand{\nucset}{\ensuremath{\mathcal{N}}\xspace}
\newcommand{\pairset}{\ensuremath{\mathcal{P}}\xspace}

\newcommand{\distri}{\ensuremath{\mathbb{D}}\xspace}
\newcommand{\DDG}{\ensuremath{\Delta \freeenergy}\xspace}
\newcommand{\BPD}{\ensuremath{\text{BPD}\xspace}}

\definecolor{pair_color}{rgb}{0.0,0.0,1.0}
\definecolor{mismatch_color}{rgb}{1.0,0.0,0.0}
\definecolor{trimismatch_color}{rgb}{0.71,0.4,0.11}

\definecolor{designable}{rgb}{0.0,0.0,0.0}
\definecolor{undesignable}{rgb}{0.71,0.4,0.11}
\definecolor{unknown}{rgb}{0.5,0.5,0.5}

\newcommand{\proj}{\ensuremath{\mathit{proj}}\xspace}

%% file: introduction.tex

Ribonucleic acid (RNA) is vital for fundamental cellular processes
such as transcription and
translation, catalyzing reactions, and controlling gene expression~\cite{eddy:2001,doudna+cech:2002,bachellerie+:2002}.
Its importance is also evidenced by
COVID-19, an RNA virus, 
as well as 
the last two Nobel Prizes in Physiology and Medicine
for messenger RNA vaccines (2023) and microRNAs (2024). 
The problem of 
{\em RNA design} aims to find sequences that is capable of folding into a target structure \cite{zhou+:2023samfeo,bellaousov+:2018,portela:2018nemo,garcia+:2013rnaifold,zadeh+:2010}. 
This process enables the creation of artificial RNA molecules 
with specific function, such as artificial ribozymes~\cite{dotu+:2014, yamagami+:2018}, 
artificial miRNAs~\cite{schwab+:2006}, 
artificial RNA aptamers~\cite{hamada:2018}, 
and artificial riboswitches~\cite{bauer+suess:2006, findeiss+:2017}. 

Computationally, the RNA design problem is extremely challenging
due to its exponentially large search space, $\{\nucA,\nucC,\nucG,\nucU\}^n$, which has the size of $O(4^n)$.
Indeed, it has been proved NP-hard at least for a simplified energy model~\cite{bonnet+:2020}. 
Therefore probably the most popular approach for this problem uses
heuristic methods such as local search~\cite{portela:2018nemo,zhou+:2023samfeo},
which starts with a single sequence and tries to optimize an objective function by revising one or a few nucleotides in each step.
However, such methods are only capable of exploring a fixed number of candidate sequences, thus not able to keep up with the exponential growth of the design space.
As a result, they tend to perform poorly on longer and harder-to-design structures.

We instead cast the RNA design problem as {\em continuous optimization} \cite{matthies+:2023, dai+:2024}. 
The basic idea is to start with a {distribution} over {\em all possible} candidate sequences and use gradient descent to gradually sharpen the distribution,
with 
each step changing {\em all} positions
 simultaneously in contrast to 
local search methods.
In this work, we first define novel sequence distributions for any given RNA structure using coupled variables for paired and mismatch positions, which not only rules out invalid sequences but also models the positional correlations explicitly.
We then aim to optimize the
{\em expectation} of an {\em arbitrary} objective function
 over these distributions.
However, given the diversity of various objective functions in RNA design (such as Boltzmann probability or ensemble defect), it is often computationally prohibitive to 
compute the exact values of 
the expected objective function or its gradient
over the whole distribution of exponentially many sequences.
Therefore, in contrast to previous work, we use sampling 
to approximate the expected objective function and its gradient.
At the end, 
we return the best sample in terms of the objective function among all samples collected, which yields high-quality designs.


When tested on the Eterna100 benchmark,
our work consistently outperforms the state-of-the-art RNA design methods \cite{portela:2018nemo,zhou+:2023samfeo}
in (almost) all metrics 
such as Boltzmann probability, ensemble defect, and free energy gap.
The advantage of our work is especially salient
on longer and harder-to-design structures,
demonstrating the advantage of this distributional approach that models the whole design space over local search methods that performs local changes on a single sequence.

%% file: discrete.tex
An RNA sequence \vecx of length $n$ is specified as a string of  base nucleotides $x_1 x_2\dots x_n,$  where $x_i \in \nucset$ ($\nucset \defeq \{\nucA, \nucC, \nucG, \nucU\}$ is the set of nucleotides) for $i=1, 2,...,n$. 
A pseudoknot-free secondary structure for sequence \vecx is 
a well-balanced dot-bracket string~$\vecy=y_1 y_2\dots y_n$
where $y_i = ``\mydot"$ indicates that $x_i$ is unpaired,
and $y_i=``\leftb"$ indicates that $x_i$ is paired with some downstream $x_j$ 
and $y_i=``\rightb"$ indicates that $x_i$ is paired with some upstream $x_j$. 
The set of unpaired indices is denoted $\unpaired(\vecy)$ and the set of paired indices 
$\pairs(\vecy)$. For example,
if $\vecx= \verb|CCCAAAGGG|$ and $\vecy=\verb|(((...)))|$,
we have $\unpaired(\vecy)=\{4,5,6\}$ and $\pairs(\vecy)=\{(1,9),(2,8),(3,7)\}$.
We assume each base-pair is a Watson-Crick-Franklin or wobble 
pair, i.e., $\forall (i,j)\in \pairs(\vecy), x_i x_j\in \mathcal{P}$ where 
$\mathcal{P} \defeq 
\{\verb|CG|, \verb|GC|, \verb|AU|, \verb|UA|, \verb|GU|, \verb|UG|\}$.
%
%

\subsection{RNA Folding}
The \emph{ensemble} of an RNA sequence $\vecx$  is the set of all possible secondary structures of  $\vecx$, denoted as $\mathcal{Y}(x)$. In thermodynamic RNA folding models, \emph{Gibbs free energy change} $\freeenergy(\vecx, \vecy)$ is used to characterize the stability of $\vecy \in \GEN(\vecx)$. The lower the free energy~$\freeenergy(\vecx, \vecy)$, the more stable the secondary structure $\vecy$ for $\vecx$. The structure with the \emph{minimum free energy} is the most stable structure in the ensemble, i.e., 
{\em \MFE structure},
\begin{equation}
\MFE(\vecx) \defeq \argmin_{\vecy \in \GEN (\vecx)} \freeenergy(\vecx, \vecy).
\end{equation}
%
Note that for most methods for secondary structure prediction, ties for $\argmin$ are broken arbitrarily when there are multiple lowest free energy structures.
This issue was often neglected in the literature, but it deserves clarification here. To be precise, we define 
\begin{equation}
\MFEs(\vecx) \defeq \{\vecy \mid 
\freeenergy(\vecx, \vecy) = \min_{y' \in \GEN(\vecx)} \freeenergy(\vecx, \vecy')\}
\end{equation}
to be the {\em set of MFE structures} for \vecx. When it is a singleton set,
we say \vecx has a {\em unique MFE} (uMFE) structure. 

The \emph{partition function} sums the contribution of all structures in an ensemble:

\begin{equation}
Q(\vecx) \defeq  \sum_{\vecy \in \GEN(\vecx)} e^{-\freeenergy(\vecx, \vecy)/RT},
\end{equation}
where $R$ is the molar gas constant and $T$ is the absolute temperature. Accordingly, the equilibrium probability of a sequence $\vecx$ folding into a structure $\vecy$ is defined as

\begin{equation}
p(\vecy \mid \vecx) = \frac{e^{-\freeenergy(\vecx, \vecy)/RT}}{Q(\vecx)}.
\end{equation}

\begin{figure}
\centering
\includegraphics[width=\linewidth]{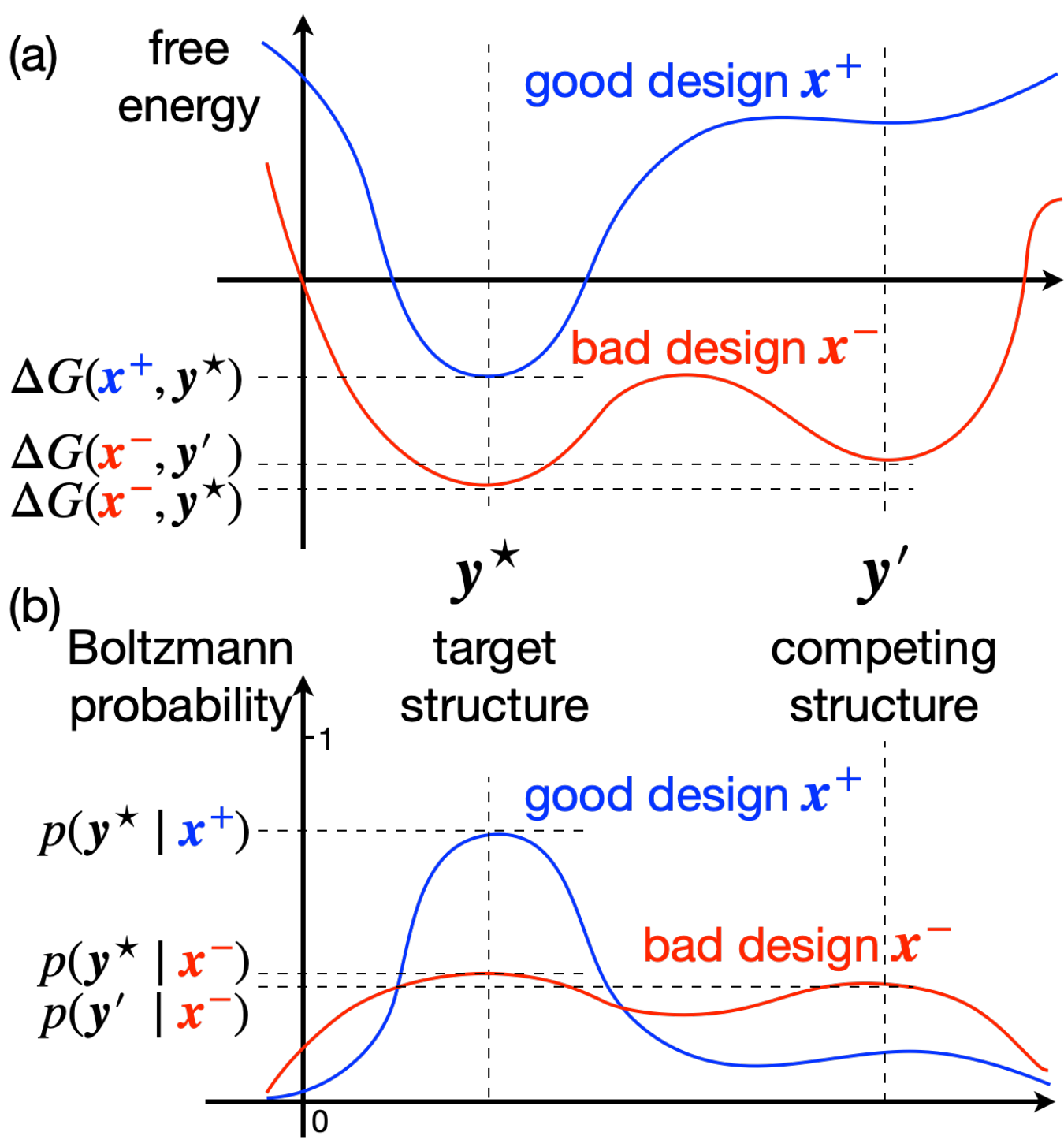}
\caption{RNA design criteria: (a) MFE vs.~(b) Boltzmann probability. In (a), both designs $\vecx^+$ and $\vecx^-$ are MFE solutions 
for the target structure \vecystar, but $\vecx^-$
is  not a good design
due to the competing structure $\vecy'$
having similar free energy, which results in low 
probability of \vecystar in the ensemble (b).
By contrast, $\vecx^+$ is a better design
with sharper energy landscape (less competition),
thus higher probability of \vecystar
(i.e., it is more likely to fold into \vecystar).}
\label{fig:landscape}
\end{figure}

\subsection{RNA Design as Inverse Folding}
Given a target structure~$\vecystar$, RNA design aims to find a suitable RNA sequence~$\vecx$ 
that can naturally and easily fold into \vecystar,
within the design space $\GENX(\vecystar)$ 
of all valid sequences for \vecystar:
\begin{equation}
\GENX(\vecystar) \defeq \{ \vecx \in \nucset^{|\vecystar|} \mid \forall (i,j)\in \pairs(\vecystar), x_i x_j \in \mathcal{P}\}
\end{equation}
But there are different ways to quantify how ``naturally'' or ``easily'' 
\vecx folds into \vecystar, which 
we categorize into two broad groups:
(a) MFE-based  and (b) ensemble-based criteria.

\paragraph{MFE criteria and uMFE criteria} 
Sequence \vecx is said to be an {\em MFE solution} of \vecystar if
\vecystar is one of the MFE structures of \vecx:
%
\begin{equation}
\vecystar \in \MFEs(\vecx)
\end{equation}
Or equivalently,
$\forall \vecy \neq \vecystar, \freeenergy(\vecx, \vecy) \leq \freeenergy(\vecx, \vecystar).$ 
%
As a stricter criteria, 
sequence \vecx is 
said to be a {\em uMFE solution} of \vecystar
if \vecystar is the unique MFE structure of \vecx,
or equivalently:
\begin{equation}
\forall \vecy \neq \vecystar, \freeenergy(\vecx, \vecy) < \freeenergy(\vecx, \vecystar) 
\end{equation}

To search for an MFE or uMFE solution,
we can start from a random sequence \vecx and gradually
update it to minimize one of the following metrics:
\begin{itemize}
\item \emph{structural distance} $d(\vecystar, \MFE(\vecx))$,
where 
$d(\cdot, \cdot)$ is a standard distance metric 
between 
two secondary structures, 
 returning the number of differently folded nucleotides:
\begin{equation}
\begin{aligned}
d(\vecy, \vecy') \defeq |\vecy| - 2\cdot |\pairs(\vecy)\cap \pairs(\vecy')| \\ - |\unpaired(\vecy)\cap \unpaired(\vecy')|.\label{eq:dist}
\end{aligned}
\end{equation}

\item \emph{free energy gap} $\DDG(\vecx, \vecystar)$, which is the difference between 
the free energies of \vecystar and $\MFE(\vecx)$: 
\begin{equation}
\begin{aligned}
\DDG(\vecx, \vecystar) 
& \defeq 
\freeenergy(\vecx, \vecystar) - \freeenergy(\vecx, \MFE(\vecx))\\
& = \freeenergy(\vecx, \vecystar) - \min_{\vecy} \freeenergy(\vecx, \vecy) \geq 0
\end{aligned}
\end{equation}

\end{itemize}
Clearly, when $d(\vecystar, \MFE(\vecx))$
or $\DDG(\vecx, \vecystar)$ reaches $0$, we have an MFE solution.

\paragraph{Ensemble-based criteria: Boltzmann probability and ensemble defect}

However, the above two criteria only consider MFE structures,
and neglect the other competing structures.
Even if \vecystar is the unique MFE structure, there could still be many highly competitive structures that are very close in energy to \vecystar;
see design $\vecx^-$ in Fig.~\ref{fig:landscape}(a) for an example. 
As a result,
the Boltzmann probability $p(\vecystar \mid \vecx)$ could still be arbitrarily small due to competition,
which means \vecx is highly unlikely to fold into \vecystar in equilibrium (see Fig.~\ref{fig:landscape}(b)).
So a better criteria is to look at the whole Boltzmann ensemble to minimize the competition from alternative structures. We consider two such metrics:

\begin{itemize}
\item \emph{conditional (i.e., Boltzmann) probability}
$p(\vecystar \mid \vecx)$. 
Since each \vecx has exponentially many possible structures in the ensemble, this probability can be arbitrarily small. So for numerical stability, we {\em minimize} the negative log probability $-\log p(\vecystar \mid \vecx)$ instead.

\item \emph{ensemble defect} \ED(\vecx, \vecystar),
which is 
the expected structural distance between \vecystar and all structures in the ensemble \cite{zadeh+:2011}. 
This metric not only considers competition, but also how (dis)similar the competing structures are from \vecystar; we want to penalize highly competitive structures that are very different from \vecystar. 
The value of ensemble defect can be normalized to between 0 and 1, known as \emph{normalized ensemble defect} (\NED):
\begin{equation}
\begin{aligned}
\NED(\vecx, \vecystar)  &\defeq \frac{1}{|\vecx|}~\mathbb{E}_{\vecy \sim p(\cdot \mid \vecx)}~ d(\vecystar, \vecy) \\
                     &= \frac{1}{|\vecx|} \sum_{\vecy\in \mathcal{Y}(\vecx)} p(\vecy \mid \vecx) \cdot d(\vecystar, \vecy). \label{eq:ned} 
\end{aligned}
\end{equation}
By plugging in Eq.~\ref{eq:dist} and some simplifications \cite{zadeh+:2011,zhang+:2020}, we get
\begin{equation}
\begin{aligned}
    \NED(\vecx, \vecystar) \! =\! 1 \! -\! \frac{2}{|\vecx|} \!\!\sum_{(i, j) \in \pairs(\vecystar)}\!\!\!\! p_{ij} \!-\! \frac{1}{|\vecx|}\!\! \sum_{j \in \unpaired(\vecystar)} \!\!\!\!q_{j},
\end{aligned}
\end{equation}
where $p_ij$ is the base-pairing probability of nucleotides $i$ and $j$, while $q_j = 1 - \sum_i p_{ij}$ is the probability of $j$ being unpaired. $\NED(\vecx, \vecystar)$ can also be further decomposed into the sum of \textit{positional defect} ($\epsilon_i$):
\begin{equation}
\begin{aligned}
    \NED(\vecx, \vecystar) &= \frac{1}{|\vecx|}\sum_{1 \leq i \leq |\vecx|} \epsilon_i(\vecx, \vecystar) 
\end{aligned}
\end{equation}
where
\begin{equation}
\begin{aligned}
    \epsilon_i &\!=\! \begin{cases}
        1 \!-\! q_i & \text{if } i \in \unpaired(\vecystar);\\
        1 \!-\! p_{ij} & \text{if } (i, j) \in \pairs(\vecystar) \text{ for some } j > i;\\
        1 \!-\! p_{ji} & \text{if } (j, i) \in \pairs(\vecystar) \text{ for some } j < i.
    \end{cases}
\end{aligned}
\end{equation}

\end{itemize}

Now we can formulate the RNA design problem as optimizing some 
objective function
$f(\vecx, \vecystar)$ 
over the design space $\GENX(\vecystar)$:
\begin{equation}
\vecxstar = \argmin_{\vecx \in \GENX(\vecystar)} f(\vecx, \vecystar)
\end{equation}
where
the objective function can be one of these four:
\begin{equation}
f(\vecx, \vecystar) = 
\begin{cases}
d(\MFE(\vecx), \vecystar) & \text{structural distance} \\
\DDG(\vecx, \vecystar) & \text{free energy gap} \\
-\log p(\vecystar \mid \vecx) & \text{conditional probability}\\
\NED(\vecx, \vecystar)    & \text{ensemble defect}
\end{cases}
\end{equation}

%% file: continuous.tex

%
However, it is well known that the above discrete optimization formulation is hard to optimize. For any target structure, the RNA design space is exponentially large:
\begin{equation}
|\GENX(\vecystar)| = 4^{|\unpaired(\vecystar)|} \cdot 6^{|\pairs(\vecystar)|}
\end{equation}
But most commonly used local search methods \cite{hofacker+:1994,andronescu+:2004, busch+backofen:2006, zadeh+:2011nupack} 
considers only one (or a few) candidate sequence in each step and 
only modifies one (or a few) nucleotides,
which seems highly inefficient in exploring the exponentially large design space.

Can we instead modify {\em all} positions of the candidate sequence in each step, or consider {\em all} candidate sequences simultaneously and promote the better ones?
Here we replace the discrete representation of a single candidate sequence 
by a probability distribution $p_{\vecystar}(\vecx)$ over all possible sequences \vecx in $\GENX(\vecystar)$.
Essentially, we 
propose the following continuous relaxation of the optimization problem minimizing the following objective function:
\begin{equation}
\mathcal{J} 
\defeq \mathbb{E}_{\vecx \sim \pystar(\cdot)} \left[ f(\vecx, \vecystar) \right]
= \!\!\! \sum_{\vecx \in \GENX(\vecystar)} \!\!\! p_{\vecystar} (\vecx)  f(\vecx, \vecystar) 
\end{equation}

This new objective function is to find a {\em distribution $\pystar(\cdot)$ of RNA candidates} whose expectation of the objective function $f(\vecx, \vecystar)$ is minimized. 
If the probability mass concentrates on only one sequence, then this new relaxed objective degenerates to the original discrete objective.


The way of modeling the probability distribution $p_{\vecystar}(\cdot)$ over the design space $\mathcal{X}(\vecystar)$ could potentially affect the complexity of the optimization and the convergence of the final solution. We aim to find a method that can represent the design space efficiently while being easy to manage. 

\subsection{Independent distributions (v0)}
\label{sec:v0}

The most obvious modeling of $p_{\vecy}(\vecx)$
is to use an independent distribution over $\nucset=\{\nucA,\nucC,\nucG,\nucU\}$ for each position, so that 
the distribution over sequences is simply the product 
of individual distributions:
\begin{equation}
p^0_{\vecy}(\vecx) \defeq \prod_i p_i (x_i)
\end{equation}
This is the same distribution in previous work \cite{matthies+:2023}.
However, this distribution is  simplistic and 
overlooks the fact that 
each base-pair $(i,j)\in \vecy$ 
requires
$x_i x_j$ to be one of the 6 possible pairs in $\mathcal{P}$, which is impossible with independent variables ($4\times 4=16$ choices for $x_i x_j$).
As a result, the domain of this $p^0_{\vecy}(\cdot)$ 
distribution is {\em all possible sequences}
$\nucset^{|\vecy|}$ (of size $4^{|\vecy|}$) rather than the  set $\GENX(\vecy)$ of {\em valid sequences} 
(of size $4^{|\unpaired(\vecy)|}\cdot 6^{|\pairs(\vecy)|}$). In other words, for any input structure \vecystar (except for the trivial case of fully-unpaired), this naive distribution includes (exponentially many) invalid sequences.

\begin{figure}[!t]
    \centering
    \includegraphics[width=.8\linewidth]{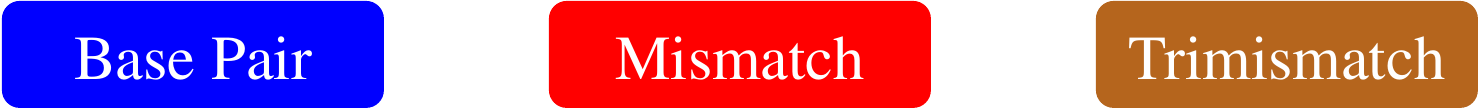}\\[6pt]
    \begin{tabular}{cc}
        (a) running example
        &
        \hspace{-0.52cm}
        (b) larger example
        \\
        \begin{tikzpicture}
            \node[anchor=south west,inner sep=0] (image) at (0,0) {\includegraphics[width=.44\linewidth]{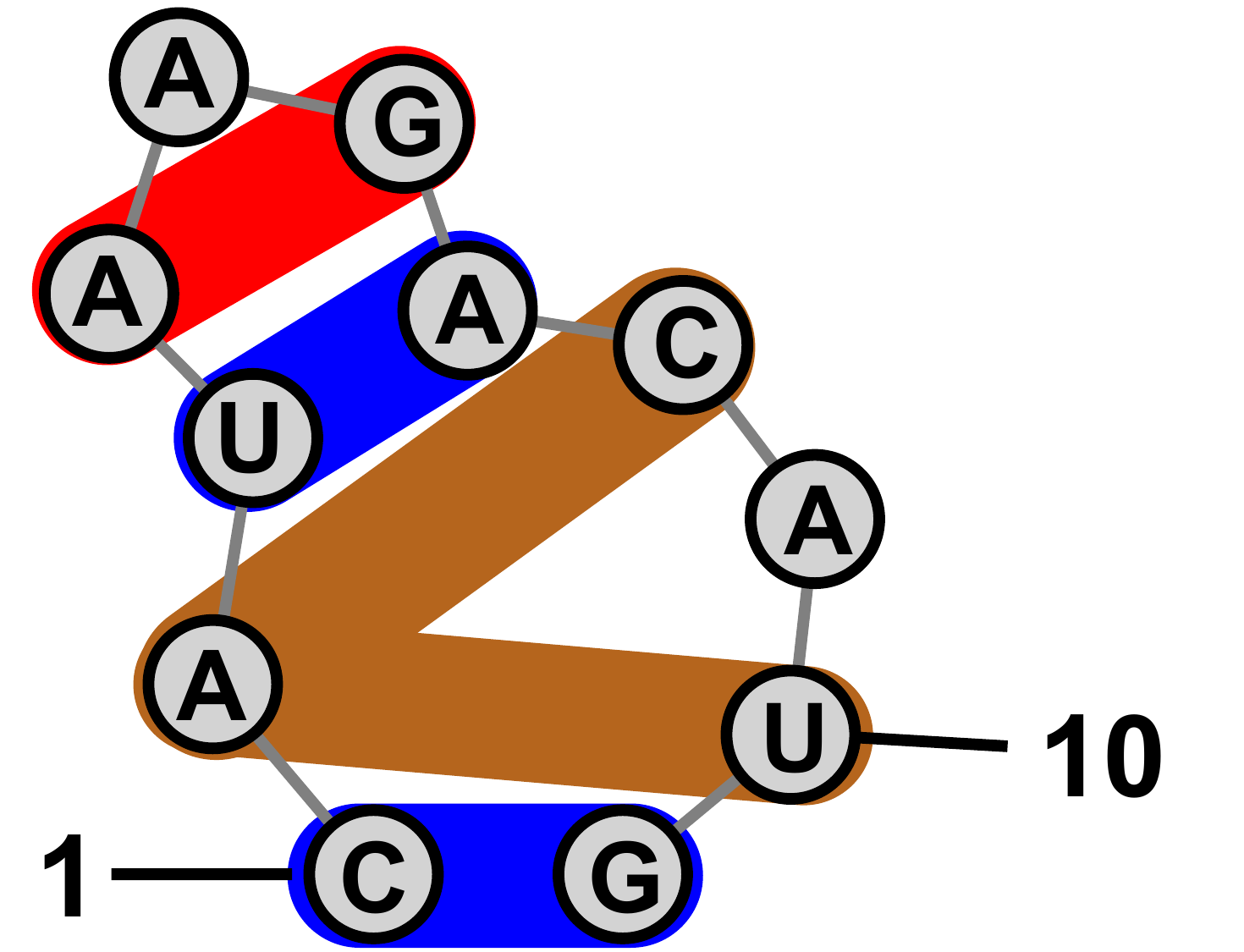}};
            \node at (2.9,0.2) {\color{pair_color}{$\pp_{1, 11}(\nucC \nucG)$}};
            \node at (3.2,1.35) {$\pu_{9}(\nucA)$};
            \node at (3.2,1.9) {\color{trimismatch_color}{$\ptm_{2, 8, 10}(\nucA\nucC\nucU)$}};
            \node at (2.25,2.25) {\color{pair_color}{$\pp_{3,7}(\nucU\nucA)$}};
            \node at (2.1,2.7) {\color{mismatch_color}{$\pm_{4,6}(\nucA\nucG)$}};
            \node at (1.1,3.0) {$\pu_5(\nucA)$};
        \end{tikzpicture}
        &
        \hspace{-0.4cm}\includegraphics[width=.5\linewidth]{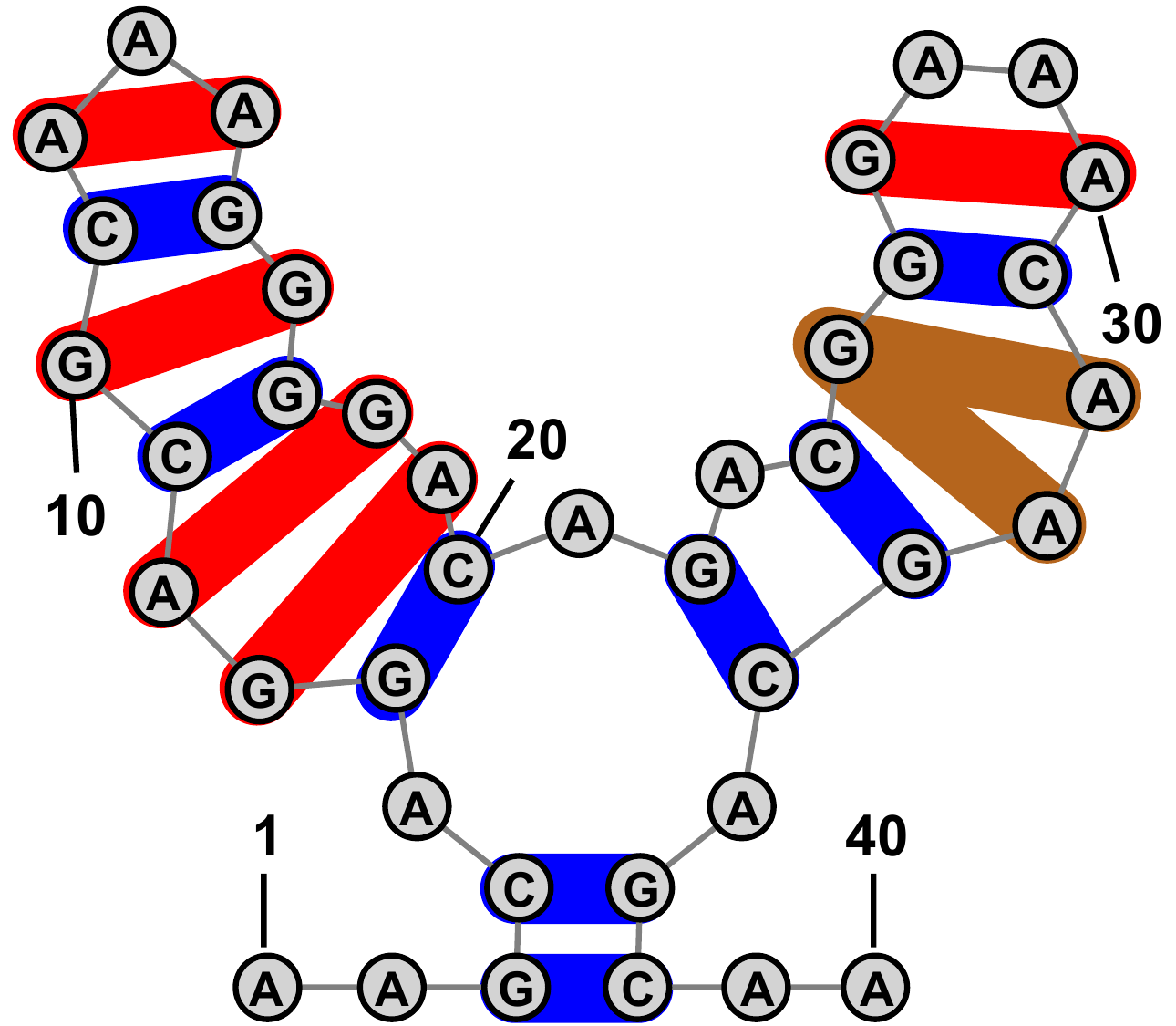}
    \end{tabular}
    \caption{Positions of mismatches and trimismatches in different types of loops.}
    \label{fig:mismatch}
\end{figure}
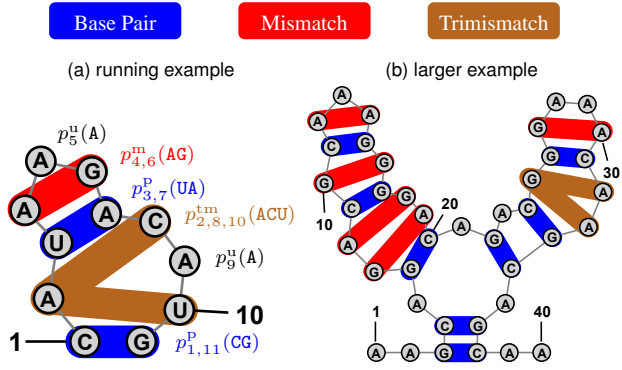

\subsection{Coupled variables for pairs (v1)}
\label{sec:v1}

In order to model the dependencies between paired positions, we separate the positions into 
two groups: the set of unpaired indices, denoted $\unpaired(\vecy)$, and the set of paired indices, $\pairs(\vecy)$. 
We now factorize the joint distribution of the entire sequence as
\begin{align*}
p^1_{\vecy}(\vecx) \defeq
\prod_{i \in \unpaired(\vecy)} \pu_i(x_i) \cdot
{\color{pair_color}{\prod_{(i, j) \in \pairs(\vecy)} \pp_{i,j}(x_i x_j)}}
\end{align*}
where $\pu_i(\cdot)$ is the local distribution (over $\nucset$) for unpaired position $i$, 
and $\pp_{i,j}(\cdot)$ is the local distribution 
(over 6 choices in \pairset) for paired positions $(i, j)$. 
This is the first distribution 
over the set of valid sequences $\GENX(\vecy)$ for
a given \vecy.


As an example, consider Fig.~\ref{fig:mismatch}(a). Here $\vecystar = \texttt{(.(...)...)}$, so $\unpaired(\vecystar) = \{2, 4, 5, 6, 8, 9, 10\}$ and $\pairs(\vecystar) = \{(1, 11), (3, 7)\}$.
The probability distribution of the design space is factorized as: 
\begin{align*}
p^1_{\vecystar}(\vecx) 
= \;
& \pu_2(x_2)\cdot 
 \pu_4(x_4) \cdot
 \pu_5(x_5) \cdot
 \pu_6(x_6) \cdot
 \pu_8(x_8) \\
& \cdot \pu_9(x_9) \cdot
 \pu_{10}(x_{10})\cdot
 \color{pair_color}{\pp_{1,11}(x_1 x_{11})}\cdot
 \color{pair_color}{\pp_{3,7}(x_3 x_7)}
\end{align*}
Therefore, for the particular design in 
Fig.~\ref{fig:mismatch}(a),
\begin{align*}
p^1_{\vecystar}(\texttt{CAUAAGACAUG})= \;
& \pu_2(\nucA) \cdot
 \pu_4(\nucA) \cdot
 \pu_5(\nucA) \cdot
 \pu_6(\nucG) \cdot
 \pu_8(\nucC) 
 \\
&
 \pu_9(\nucA) \cdot
 \pu_{10}(\nucU) \cdot
 \color{pair_color}{\pp_{1, 11}(\nucC\nucG)}\cdot
 \color{pair_color}{\pp_{3, 7}(\nucU\nucA)}
\end{align*}



\subsection{Coupled variables for terminal mismatches (v2 and v3)}
\label{sec:v2v3}

The next two versions (v2 and v3) are refinements
of the above v1. 
First, we note that in the standard energy models (Turner rules~\cite{mittal+:2024}), there are terminal mismatches lookup tables. For example, for a hairpin loop defined by the pair $(i,j)$, the first and last nucleotide of the loop, $x_{i+1}$ and $x_{j-1}$, are the terminal mismatch, and will be looked up together in the energy table (such as $x_4$ and $x_6$ in Fig.~\ref{fig:mismatch}(a)). Therefore, it is better to make a coupled variable over $\nucset^2$ ($4\times 4=16$ choices) for each terminal mismatch position-pair:
\begin{align*}
p^2_{\vecy}(\vecx) \defeq
& 
\!\!\!\!\!\prod_{i \in \unpaired(\vecy)} \!\!\!\!\!\!\!\pu_i(x_i) \cdot \!\!\!\!\!
	\color{pair_color}{\prod_{(i, j) \in \pairs(\vecy)} \!\!\!\!\!\!\! \pp_{i,j}(x_i x_j)} 
 \cdot \!\!\!\!\!\!\!\!\!\color{mismatch_color}{\prod_{(i, j) \in \mismatches(\vecy)} \!\!\!\!\!\!\!\!\!\!\pm_{i,j}(x_i x_j)}
\end{align*}

Moreover, there is a special case that 
deserves our attention. 
Let us consider the 1-by-3 internal loop in Fig.~\ref{fig:mismatch}.
For such 1-by-$x$ ($x>1)$ internal loops,
there is exactly one unpaired nucleotide
on one of the two branches,
and that single unpaired nucleotide ($x_2$ in our example)
is included in two mismatches ($x_2$ and $x_8$ on one side and $x_2$ and $x_{10}$ on the other).
Therefore, it is better to model all these three nucleotides together in a coupled variable 
over $\nucset^3$ ($4^3=64$ choices), which we call a ``trimismatch'':
\begin{align*}
p^3_{\vecy}(\vecx) \defeq
& \prod_{i \in \unpaired(\vecy)} \pu_i(x_i) \cdot
	\color{pair_color}{\prod_{(i, j) \in \pairs(\vecy)} \pp_{i,j}(x_i x_j)} \\
& \cdot \!\!\!\!\!\!\!\!  \color{mismatch_color}{\prod_{(i, j) \in \mismatches(\vecy)} \!\!\!\!\!\!\!\!  \pm_{i,j}(x_i x_j)} 
 \ \ \ \cdot \!\!\!\!\!\!\!\!  \color{trimismatch_color}{\prod_{(i, j, k) \in \trimismatches(\vecy)} \!\!\!\! \!\!\!\! \ptm_{i,j,k}(x_i x_j x_k)} \\
\end{align*}
Note that, a 1-by-1 internal loop is a special case of mismatch (see Fig.~\ref{fig:mismatch}(b), nucleotides 10 and 16).

Now for the example structure \vecystar in Fig.~\ref{fig:mismatch}(a),
our final 
joint distribution is:
\begin{align*}
p^3_{\vecystar}(\vecx)
= \;
& \pu_5(x_5) \cdot
\pu_9(x_9) \cdot 
\color{pair_color}{\pp_{1, 11}(x_1 x_{11})} \cdot
\color{pair_color}{\pp_{3, 7}(x_3 x_7)} \\
& \cdot \color{mismatch_color}{\pm_{4, 6}(x_4 x_6)}\cdot 
\color{trimismatch_color}{\ptm_{2, 8, 10}(x_2 x_8 x_{10})}
\end{align*}
And for the particular design in Fig.~\ref{fig:mismatch}(a),
\begin{align*}
p^3_{\vecystar}(\texttt{CAUAAGACAUG})= \;
& \pu_5(\nucA) \cdot \pu_9(\nucA) \cdot
\color{pair_color}{\pp_{1, 11}(\nucC\nucG)}\cdot
\color{pair_color}{\pp_{3, 7}(\nucU\nucA)} \\
& \cdot \color{mismatch_color}{\pm_{4, 6}(\nucA\nucG)}\cdot
\color{trimismatch_color}{\ptm_{2, 8, 10}(\nucA\nucC\nucU)}
\end{align*}

%% file: sampling.tex


Given the complexity and variety of RNA design problem settings, a method that can seamlessly switch between various objective functions $f(\vecx, \vecystar)$ is desirable. Even though we factorize $\pystar(\cdot)$ into many tractable local distributions, without making any assumptions or requirements about the structure of $f(\vecx, \vecystar)$, the exact calculation of the expectation $\obj=\mathbb{E}_{\vecx \sim \pystar(\cdot)} \left[ f(\vecx, \vecystar) \right]$ is generally intractable.
Therefore, we employ gradient descent for optimization and adopt random sampling for estimating the objective and its gradients. 

\subsection{Sampling for Objective Evaluation}
We approximate the expectation by  averaging over a set of samples \SAMPLES  from the distribution:
\begin{align*}
\obj & = \E_{\vecx \sim \pystar(\cdot)} \left[ f(\vecx, \vecystar) \right] 
\approx 
\frac{1}{|\SAMPLES|} \sum_{\vecx \in \SAMPLES} f(\vecx, \vecystar)\\
\text{where\quad} \SAMPLES & = \{\vecx^{(l)} \sim \pystar(\cdot)\}_{l=1}^{|\SAMPLES|}
\end{align*}
Theoretically, as $|\SAMPLES| \rightarrow \infty$, this approximation will converge to the true expectation. 

\input{fig-sampling}

\subsection{Sampling for Gradient Estimation}
\label{sec:gradient}
Next, we derive the true gradient of \obj with respect to $\pystar(\cdot)$ as:
\begin{align*}
&\nabla_{\pystar} \obj \\
= & \nabla_{\pystar}  \E_{\vecx \sim \pystar(\cdot)} \left[ f(\vecx, \vecystar) \right] \\
=&
 \nabla_{\pystar} \!\!\! \sum_{\vecx \in \GENX(\vecystar)} \!\!\! p_{\vecystar} (\vecx)  f(\vecx, \vecystar)  & \text{(def.~of expectation)}\\
=&
\!\!\! \sum_{\vecx \in \GENX(\vecystar)} \!\!\!\nabla_{\pystar} p_{\vecystar} (\vecx) f(\vecx, \vecystar)  & \text{(linearity of $\nabla$)}\\
=&
 \!\!\!\sum_{\vecx \in \GENX(\vecystar)}\!\!\! p_{\vecystar} (\vecx) \frac{\nabla_{\pystar} p_{\vecystar} (\vecx)}{p_{\vecystar} (\vecx)} f(\vecx, \vecystar) \\
 =&
\!\!\! \sum_{\vecx \in \GENX(\vecystar)}\!\!\! p_{\vecystar} (\vecx) \nabla_{\pystar} \log p_{\vecystar} (\vecx) f(\vecx, \vecystar) 
      & \!\!\!\left(\!(\log f(x))' \!=\! \frac{f'(x)}{f(x)}\! \right) \\
=&
 \E_{\vecx \sim \pystar(\cdot)} \left[ \nabla_{\pystar} \log p_{\vecystar} (\vecx) f(\vecx, \vecystar) \right] & \text{(def.~of expectation)}
\end{align*}

We again use sampling to 
estimate
the above expectation and derive the approximate gradient for step $t$:
\begin{align}
\label{eq:gradient}
\nabla_{\pystar} \obj^{(t)} & \approx \frac{1}{|\SAMPLES|} \sum_{\vecx \in \SAMPLES} \nabla_{\pystar} \log p_{\vecystar} (\vecx) f(\vecx, \vecystar)\\
\SAMPLES^{(t)} & \gets \{\vecx^{(l)} \sim \pystar^{(t)}(\cdot)\}_{l=1}^{N}
\end{align}
We then use the approximate gradient to update the distribution (see Sec.~\ref{sec:params} for details):
\begin{align}
\pystar^{(t+1)} & \gets \text{update}\left(\pystar^{(t)}, \nabla_{\pystar}\obj^{(t)}\right)
\end{align}

\subsection{Sampling-based Design Space Exploration}

At the end of this continuous optimization,
we still need to return a single sequence from the distribution, i.e., an ``integral solution''.
This can be done by ``rounding'' if the distribution is close to one-hot, 
or more generally
by taking 
the sequence with the highest probability in the final distribution 

\begin{align}
\vecxstar = \argmax_{\vecx} p_{\vecystar} (\vecx)
\end{align}


For example,
for independent distributions (v0), since each position is isolated, we simply take the best nucleotide for each position:
$x_i^\star = \argmax_{a\in \nucset} p_i(a)$.
But for the coupled variable distribution (v1),
for each unpaired position $i \in \unpaired(\ystar)$, we take $x_i^\star = \argmax_{a\in \nucset} \pu_i(a)$ same as in v0,
and for each paired position-pair $(i,j)\in \pairs(\ystar)$, we take the best pair out of the six pair types:
\(
x_i^\star x_j^\star = \argmax_{ab\in \pairset} \pp_{i,j}(ab).
\)

However, this max-probability sequence 
is not necessarily the best sequence in terms of the objective function, since the distribution is often not perfectly aligned with objective function.
Here we use an alternative approach that simply takes the best sample in terms of the objective function out of all samples collected in the optimization process:
\begin{align}
\vecx^{(t)} & = \argmin_{\vecx \in \SAMPLES^{(t)}} f(\vecx, \vecystar)\\
\vecxstar & = \argmin_t \vecx^{(t)}
\end{align}
This method, which we call ``sampling-based candidate exploration'',
outperforms the max-probability solution, because the samples offer much more diversity in the exploration of the distribution than a single sequence.

%% file: fig-sampling.tex
\definecolor{Gray1}{gray}{0.85}
\definecolor{Gray2}{gray}{0.70}
\newcolumntype{a}{>{\columncolor{Gray1}}c}
\newcolumntype{b}{>{\columncolor{Gray2}}c}

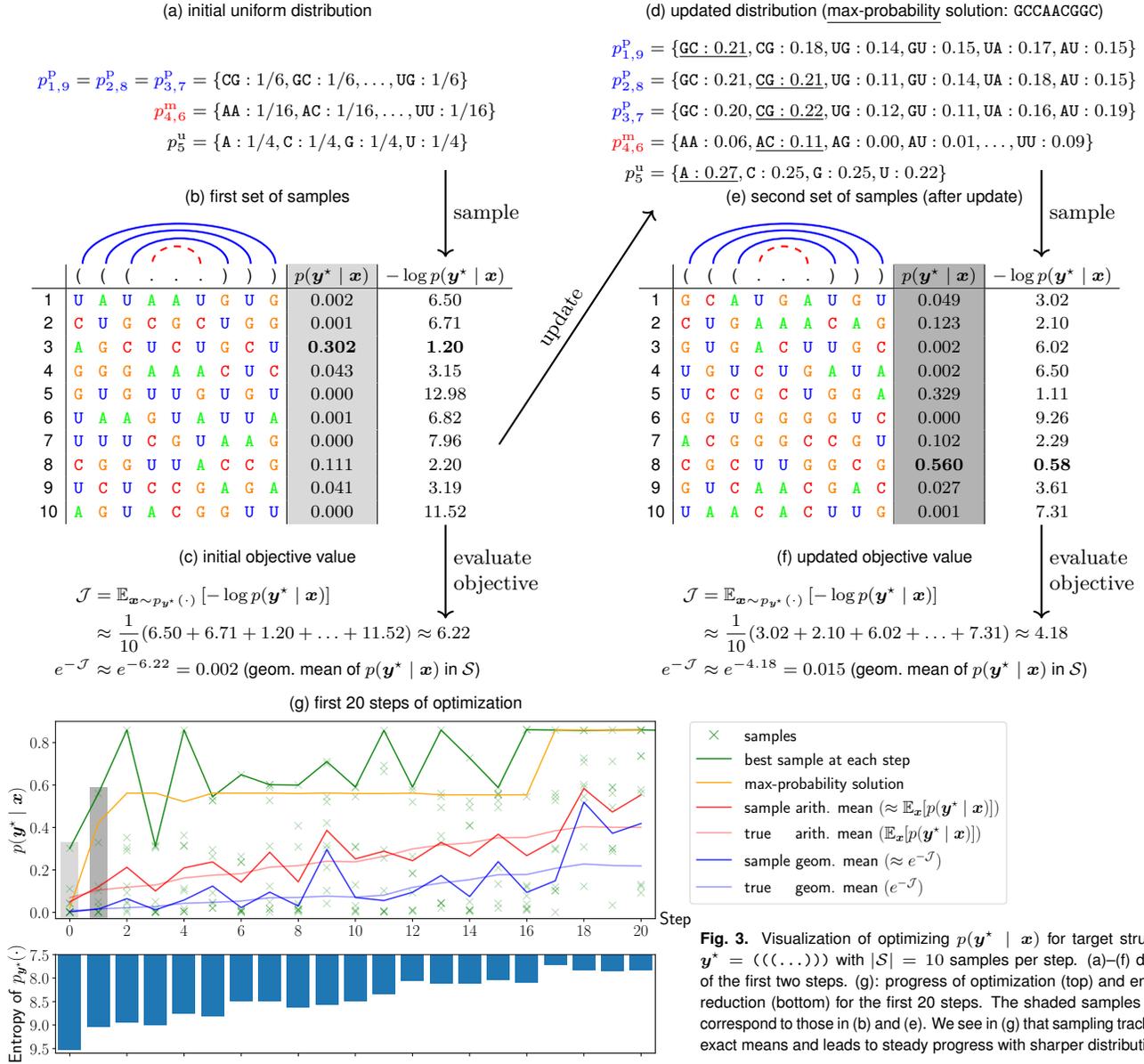
\begin{figure*}[!htb]
    \centering
    \begin{tikzpicture}
    \node[] {
    \hspace{-.8cm}
    \setlength{\tabcolsep}{20pt}
    \begin{tabular}{cc}
      (a) initial uniform distribution
      &
      (d) updated distribution (\underline{max-probability} solution: \nucG\nucC\nucC\nucA\nucA\nucC\nucG\nucG\nucC)
      \\[6pt]
          {$\begin{aligned}
                {\color{pair_color}{\pp_{1,9}}} = {\color{pair_color}{\pp_{2,8}}} = {\color{pair_color}{\pp_{3, 7}}}
                    &= \{\nucC\nucG: 1/6, \nucG\nucC: 1/6, \ldots, \nucU\nucG: 1/6\}
                \\
                \color{mismatch_color}{\pm_{4,6}} &= \{\nucA\nucA: 1/16, \nucA\nucC: 1/16, \ldots, \nucU\nucU: 1/16\}
                \\
                \pu_5 &= \{\nucA: 1/4, \nucC: 1/4, \nucG: 1/4, \nucU: 1/4\}
          \end{aligned}$}
          &
          {$\begin{aligned}
                \color{pair_color}{\pp_{1,9}} &=\{\underline{\nucG\nucC: 0.21}, \nucC\nucG: 0.18, \nucU\nucG: 0.14, \nucG\nucU: 0.15, \nucU\nucA: 0.17, \nucA\nucU: 0.15\}
                \\
                \color{pair_color}{\pp_{2,8}} &= \{\nucG\nucC: 0.21, \underline{\nucC\nucG: 0.21}, \nucU\nucG: 0.11, \nucG\nucU: 0.14, \nucU\nucA: 0.18, \nucA\nucU: 0.15\}
                \\
                \color{pair_color}{\pp_{3, 7}} &= \{\nucG\nucC: 0.20, \underline{\nucC\nucG: 0.22}, \nucU\nucG: 0.12, \nucG\nucU: 0.11, \nucU\nucA: 0.16, \nucA\nucU: 0.19\}
                \\
                \color{mismatch_color}{\pm_{4,6}} &= \{\nucA\nucA: 0.06, \underline{\nucA\nucC: 0.11}, \nucA\nucG: 0.00, \nucA\nucU: 0.01, \ldots, \nucU\nucU: 0.09\}
                \\
                \pu_5 &= \{\underline{\nucA: 0.27}, \nucC: 0.25, \nucG: 0.25, \nucU: 0.22\}
          \end{aligned}$}
          \\[8pt]
          (b) first set of samples
          &
          (e) second set of samples (after update)
          \\[6pt]
          \begin{tikzpicture}
                \node [] {
                      \setlength{\tabcolsep}{3.2pt}
                      \begin{tabular}{c|ccccccccc|a|c}
                            & ( & ( & ( & . & . & . & ) & ) & ) & $p(\vecystar \mid \vecx)$ & $- \log p(\vecystar \mid \vecx)$ \\
                           \hline
                           1 & \color{blue}{\nucU} & \color{green}{\nucA} & \color{blue}{\nucU} & \color{green}{\nucA} & \color{green}{\nucA} & \color{blue}{\nucU} & \color{orange}{\nucG} & \color{blue}{\nucU} & \color{orange}{\nucG} & $0.002$ & $6.50$\\
                           2 & \color{red}{\nucC} & \color{blue}{\nucU} & \color{orange}{\nucG} & \color{red}{\nucC} & \color{orange}{\nucG} & \color{red}{\nucC} & \color{blue}{\nucU} & \color{orange}{\nucG} & \color{orange}{\nucG} & $0.001$ & $6.71$\\
                           3 & \color{green}{\nucA} & \color{orange}{\nucG} & \color{red}{\nucC} & \color{blue}{\nucU} & \color{red}{\nucC} & \color{blue}{\nucU} & \color{orange}{\nucG} & \color{red}{\nucC} & \color{blue}{\nucU} & $\mathbf{0.302}$ & $\mathbf{1.20}$\\
                           4 & \color{orange}{\nucG} & \color{orange}{\nucG} & \color{orange}{\nucG} & \color{green}{\nucA} & \color{green}{\nucA} & \color{green}{\nucA} & \color{red}{\nucC} & \color{blue}{\nucU} & \color{red}{\nucC} & $0.043$ & $3.15$\\
                           5 & \color{orange}{\nucG} & \color{blue}{\nucU} & \color{orange}{\nucG} & \color{blue}{\nucU} & \color{blue}{\nucU} & \color{orange}{\nucG} & \color{blue}{\nucU} & \color{orange}{\nucG} & \color{blue}{\nucU} & $0.000$ & $12.98$\\
                           6 & \color{blue}{\nucU} & \color{green}{\nucA} & \color{green}{\nucA} & \color{orange}{\nucG} & \color{blue}{\nucU} & \color{green}{\nucA} & \color{blue}{\nucU} & \color{blue}{\nucU} & \color{green}{\nucA} & $0.001$ & $6.82$\\
                           7 & \color{blue}{\nucU} & \color{blue}{\nucU} & \color{blue}{\nucU} & \color{red}{\nucC} & \color{orange}{\nucG} & \color{blue}{\nucU} & \color{green}{\nucA} & \color{green}{\nucA} & \color{orange}{\nucG} & $0.000$ & $7.96$\\
                           8 & \color{red}{\nucC} & \color{orange}{\nucG} & \color{orange}{\nucG} & \color{blue}{\nucU} & \color{blue}{\nucU} & \color{green}{\nucA} & \color{red}{\nucC} & \color{red}{\nucC} & \color{orange}{\nucG} & $0.111$ & $2.20$\\
                           9 & \color{blue}{\nucU} & \color{red}{\nucC} & \color{blue}{\nucU} & \color{red}{\nucC} & \color{red}{\nucC} & \color{orange}{\nucG} & \color{green}{\nucA} & \color{orange}{\nucG} & \color{green}{\nucA} & $0.041$ & $3.19$\\
                           10 & \color{green}{\nucA} & \color{orange}{\nucG} & \color{blue}{\nucU} & \color{green}{\nucA} & \color{red}{\nucC} & \color{orange}{\nucG} & \color{orange}{\nucG} & \color{blue}{\nucU} & \color{blue}{\nucU} & $0.000$ & $11.52$\\
                   \end{tabular}
                };
                \draw [thick, dashed, color=mismatch_color] (-1.0,1.95) arc(0:180:.36cm and .26cm);
                \draw [thick, color=pair_color] (-0.59,1.95) arc(0:180:.72cm and .38cm);
                \draw [thick, color=pair_color] (-0.25, 1.95) arc(0:180:1.08cm and .5cm);
                \draw [thick, color=pair_color] (0.12,1.95) arc(0:180:1.44cm and .6cm);
          \end{tikzpicture}
          &
          \begin{tikzpicture}
                \node [] {
                      \setlength{\tabcolsep}{3.2pt}
                      \begin{tabular}{c|ccccccccc|b|c}
                            & ( & ( & ( & . & . & . & ) & ) & ) & $p(\vecystar \mid \vecx)$ & $- \log p(\vecystar \mid \vecx)$ \\
                           \hline
                           1 & \color{orange}{\nucG} & \color{red}{\nucC} & \color{green}{\nucA} & \color{blue}{\nucU} & \color{orange}{\nucG} & \color{green}{\nucA} & \color{blue}{\nucU} & \color{orange}{\nucG} & \color{blue}{\nucU} & $0.049$ & $3.02$\\
                           2 & \color{red}{\nucC} & \color{blue}{\nucU} & \color{orange}{\nucG} & \color{green}{\nucA} & \color{green}{\nucA} & \color{green}{\nucA} & \color{red}{\nucC} & \color{green}{\nucA} & \color{orange}{\nucG} & $0.123$ & $2.10$\\
                           3 & \color{orange}{\nucG} & \color{blue}{\nucU} & \color{orange}{\nucG} & \color{green}{\nucA} & \color{red}{\nucC} & \color{blue}{\nucU} & \color{blue}{\nucU} & \color{orange}{\nucG} & \color{red}{\nucC} & $0.002$ & $6.02$\\
                           4 & \color{blue}{\nucU} & \color{orange}{\nucG} & \color{blue}{\nucU} & \color{red}{\nucC} & \color{blue}{\nucU} & \color{orange}{\nucG} & \color{green}{\nucA} & \color{blue}{\nucU} & \color{green}{\nucA} & $0.002$ & $6.50$\\
                           5 & \color{blue}{\nucU} & \color{red}{\nucC} & \color{red}{\nucC} & \color{orange}{\nucG} & \color{red}{\nucC} & \color{blue}{\nucU} & \color{orange}{\nucG} & \color{orange}{\nucG} & \color{green}{\nucA} & $0.329$ & $1.11$\\
                           6 & \color{orange}{\nucG} & \color{orange}{\nucG} & \color{blue}{\nucU} & \color{orange}{\nucG} & \color{orange}{\nucG} & \color{orange}{\nucG} & \color{orange}{\nucG} & \color{blue}{\nucU} & \color{red}{\nucC} & $0.000$ & $9.26$\\
                           7 & \color{green}{\nucA} & \color{red}{\nucC} & \color{orange}{\nucG} & \color{orange}{\nucG} & \color{orange}{\nucG} & \color{red}{\nucC} & \color{red}{\nucC} & \color{orange}{\nucG} & \color{blue}{\nucU} & $0.102$ & $2.29$\\
                           8 & \color{red}{\nucC} & \color{orange}{\nucG} & \color{red}{\nucC} & \color{blue}{\nucU} & \color{blue}{\nucU} & \color{orange}{\nucG} & \color{orange}{\nucG} & \color{red}{\nucC} & \color{orange}{\nucG} & $\mathbf{0.560}$ & $\mathbf{0.58}$\\
                           9 & \color{orange}{\nucG} & \color{blue}{\nucU} & \color{red}{\nucC} & \color{green}{\nucA} & \color{green}{\nucA} & \color{red}{\nucC} & \color{orange}{\nucG} & \color{green}{\nucA} & \color{red}{\nucC} & $0.027$ & $3.61$\\
                           10 & \color{blue}{\nucU} & \color{green}{\nucA} & \color{green}{\nucA} & \color{red}{\nucC} & \color{green}{\nucA} & \color{red}{\nucC} & \color{blue}{\nucU} & \color{blue}{\nucU} & \color{orange}{\nucG} & $0.001$ & $7.31$\\
                   \end{tabular}
                };
                \draw [thick, dashed, color=mismatch_color] (-1.0,1.95) arc(0:180:.36cm and .26cm);
                \draw [thick, color=pair_color] (-0.59,1.95) arc(0:180:.72cm and .38cm);
                \draw [thick, color=pair_color] (-0.25, 1.95) arc(0:180:1.08cm and .5cm);
                \draw [thick, color=pair_color] (0.12,1.95) arc(0:180:1.44cm and .6cm);
          \end{tikzpicture}
          \\[4pt]
          (c) initial objective value
          &
          (f) updated objective value
          \\[6pt]

          {$\begin{aligned}
                \obj &= \E_{\vecx\sim p_{\vecystar}(\cdot)}\left[- \log p(\vecystar \mid \vecx)\right]\\
                &\approx \frac{1}{10} (6.50 + 6.71 + 1.20 + \ldots + 11.52)\approx 6.22\\
                e^{-\obj} & \approx e^{-6.22}=0.002 \text{\ (geom.~mean of $p(\vecystar \mid \vecx)$ in \SAMPLES)}
          \end{aligned}$}
          &
          {$\begin{aligned}
                \obj &= \E_{\vecx\sim p_{\vecystar}(\cdot)}\left[- \log p(\vecystar \mid \vecx)\right]\\
                     &\approx \frac{1}{10} (3.02 + 2.10 + 6.02 + \ldots + 7.31) \approx 4.18\\
           e^{-\obj} & \approx e^{-4.18}=0.015 \text{\ (geom.~mean of $p(\vecystar \mid \vecx)$ in \SAMPLES)}
          \end{aligned}$}
          \\[20pt]
          \multicolumn{2}{l}{
            \hspace{4cm} (g) first 20 steps of optimization
          }
          \\
        \multicolumn{2}{l}{
            \hspace{-0.3cm}
            \includegraphics[width=.85\linewidth]{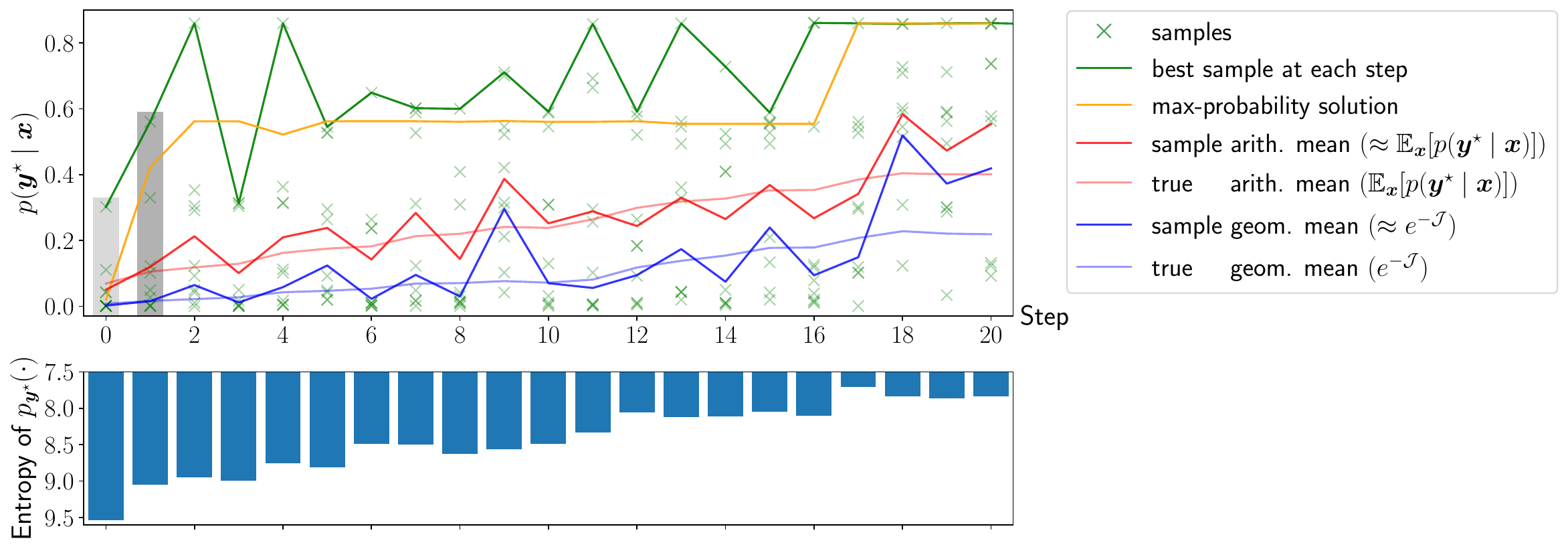}
          }

    \end{tabular}
    };

    \draw[thick,->] (-2.2, 5.5) -- (-2.2, 4.2) node[midway, right] {sample}; 
    \draw[thick,->] (-2.2, 0.2) -- (-2.2, -1.2) node[align=left,midway, right] {evaluate\\objective}; 
    \draw[thick,->] (-1.4, 1.4) -- (0.9, 4.9) node[midway, above, rotate=57] {update}; 
    \draw[thick,->] (6.7, 5.5) -- (6.7, 4.2) node[midway, right] {sample}; 
    \draw[thick,->] (6.7, 0.2) -- (6.7, -1.2) node[align=left,midway, right] {evaluate\\objective}; 
    \end{tikzpicture}
    \vspace{-1cm}
    \hspace{10.4cm} 
    \begin{minipage}{7.2cm}\vspace{-3cm}
    \caption{Visualization of optimizing  $p({\vecystar}\mid \vecx)$
    for target structure $\vecystar=\texttt{(((...)))}$
    with $|\SAMPLES|=10$ samples per step. (a)--(f) details of the first two steps. (g): progress of optimization (top) and entropy reduction (bottom) for the first 20 steps. The shaded samples in (g) correspond to those in (b) and (e). We see in (g) that sampling tracks the exact means and leads to steady progress with sharper distributions.}
    \end{minipage}
    \label{fig:samples}
\end{figure*}

%% file: parameterization.tex



Now we turn to the question of how to parameterize the factorized distribution $\pystar(\cdot)$
as $\pystar(\cdot; \vecTheta)$.
The first method (Sec.~\ref{sec:opt_PGD}) simply uses $\vecTheta$ as raw probabilities, but the update of \vecTheta needs to result in probabilities, leading to a harder \textit{constrained} optimization problem. 
The second method (Sec.~\ref{sec:opt_softmax}) models the distribution implicitly by applying softmax on $\vecTheta$, resulting in a simpler \textit{unconstrained} optimization problem.
%
%
%

\subsection{Method 1: Direct Parameterization and Constrained Optimization}
\label{sec:opt_PGD}
The obvious way of parameterization is to 
use {\em explicit} probabilities. 
For each unpaired position~$i$, 
we use a non-negative parameter vector 
\(
\vecthetau_i \defeq (\thetau_{i,\nucA}, \thetau_{i,\nucC}, \thetau_{i,\nucG}, \thetau_{i,\nucU})
\) which sums to 1
as the probability distribution over nucleotides:
\[
\forall a\in \nucset, \pu_i(a; \vecthetau_i) \defeq \thetau_{i,a}.
\]
Similarly for each paired position $(i,j)$, 
we use a non-negative parameter vector \(
\vecthetap_{i,j} \defeq (\thetap_{i,j,\nucC\nucG}, \thetap_{i,j,\nucG\nucC}, \ldots, 
\thetap_{i,j,\nucU\nucG})
\) which sums to 1,
and we have 
\[
\forall ab\in \pairset, \pp_{ij}(ab; \vecthetap_{i,j}) \defeq \thetap_{i,j,ab}.
\]
The cases for mismatches and trimismatches are also similar.
The whole parameter set \vecTheta includes all parameter vectors:
\begin{equation}
\begin{aligned}
\vecTheta = &\{\vecthetau_i \mid i \in \unpaired(\vecystar)\}\\
            &\cup \{ \vecthetap_{i,j} \mid (i,j) \in \pairs(\vecystar)\}\\
            &\cup \{ \vecthetam_{i,j} \mid (i,j) \in \mismatches(\vecystar)\}\\
            &\cup \{ \vecthetatm_{i,j,k} \mid (i,j,k) \in \trimismatches(\vecystar)\}
\end{aligned}
\end{equation}
where each $\vectheta \in \vecTheta$ is a distribution, i.e.,
\[
\forall \theta_a \in \vectheta,   
\theta_a\in [0, 1],
\text{ and } 
\textstyle\sum_{\theta_a \in \vectheta} \theta_a = 1.
\]
For example, for the structure \ystar in Fig.~\ref{fig:mismatch}(a), we have its parameters as
\(
\vecTheta=\{\vecthetau_5, \vecthetau_9, \vecthetap_{1,11}, \vecthetap_{3,7},
\vecthetam_{4,6}, \vecthetatm_{2,8,10}\}.
\)

Now we can parameterize the whole distribution as 
\begin{align*}
p^3_{\vecy}(\vecx; \vecTheta) = 
& \prod_{i \in \unpaired(\vecy)} \pu_i(x_i; \vecthetau_i) \cdot
	\prod_{(i, j) \in \pairs(\vecy)} \pp_{i,j}(x_i x_j; {\color{pair_color}{\vecthetap_{i,j}}}) \\
& \cdot \prod_{(i, j) \in \mismatches(\vecy)} \pm_{i,j}(x_i x_j; {\color{mismatch_color}{\vecthetam_{i,j}}})\\
& \cdot \prod_{(i, j, k) \in \trimismatches(\vecy)} \ptm_{i,j,k}(x_i x_j x_k; {\color{trimismatch_color}{\vecthetatm_{i,j,k}}}) \\
\defeq & 
\prod_{i \in \unpaired(\vecy)} \thetau_{i,x_i} \cdot
\prod_{(i, j) \in \pairs(\vecy)} {\color{pair_color}{\thetap_{i,j,x_i x_j}}} \cdot\\
& \prod_{(i, j) \in \mismatches(\vecy)} \hspace{-.5cm} {\color{mismatch_color}{\thetam_{i,j,x_i x_j}}} \cdot \hspace{-.3cm}
\prod_{(i, j, k) \in \trimismatches(\vecy)} \hspace{-.7cm} {\color{trimismatch_color}{\thetatm_{i,j,k,x_i x_j x_k} }}
\end{align*}


Now we adopt a parameterized version for our objective function:
\[
\obj(\vecTheta) = \E_{\vecx \sim \pystar(\cdot ; \vecTheta)} f(\vecx, \vecystar)
\]

The optimization problem can then be formulated as a
constrained optimization
\begin{equation}
\begin{aligned}\label{pro:cons}
    \min_{\vecTheta} \quad &  \obj(\vecTheta) \\
    \textrm{s.t.} \quad & \text{each $\vectheta \in \vecTheta$ is a distribution.}
 \end{aligned}
\end{equation}

To solve this constrained optimization problem, we use the Projected Gradient Descent (PGD) method~\cite{madry:2019}. 
At each step $t$,
we first perform a gradient descent (with learning rate $\alpha$):
\begin{equation}
\vecThetahat \gets \vecTheta
- \alpha \nabla_{\vecTheta}\obj(\vecTheta)
\label{eq:update}
\end{equation}
where the gradient components are computed individually for each parameter vector:
\[
\nabla_{\vecTheta} \obj(\vecTheta)
= \{ \frac{\partial \obj(\vecTheta)}{\partial \vectheta} \mid \vectheta \in \vecTheta \}
\]
For example, for an unpaired position $i$, we have:
\[
\frac{\partial \obj(\vecTheta)}{\partial \vecthetau_i}
=(
\frac{\partial \obj(\vecTheta)}{\partial \thetau_{i,\nucA}}, 
\frac{\partial \obj(\vecTheta)}{\partial \thetau_{i,\nucC}}, 
\frac{\partial \obj(\vecTheta)}{\partial \thetau_{i,\nucG}}, 
\frac{\partial \obj(\vecTheta)}{\partial \thetau_{i,\nucU}}
)
\]
The first component can be estimated using Eq.~\ref{eq:gradient} as follows:
\begin{equation}\label{eq:grad-component}
    \frac{\partial \obj(\vecTheta)}{\partial \thetau_{i,\nucA}} 
    \approx
    \frac{1}{|\SAMPLES|} \sum_{\vecx \in \SAMPLES} 
    \frac{\partial \log p_{\vecystar} (\vecx; \vecTheta)}
    {\partial \thetau_{i,\nucA}} f(\vecx, \vecystar)
\end{equation}
Expanding the term $\log p_{\vecystar} (\vecx; \vecTheta)$, the gradient can be simplified (details provided in Sec.~\ref{sec:direct-gradient}) as:
\begin{equation}\label{eq:grad-partial}
\frac{\partial \obj(\vecTheta)}{\partial \thetau_{i,\nucA}}
\approx
\frac{1}{|\SAMPLES|} \sum_{\vecx \in \SAMPLES} 
\mathbbm{1}[x_i = A]
\frac{f(\vecx, \vecystar)}
{\thetau_{i,\nucA}}
= 
\frac{1}{|\SAMPLES|} \sum_{\substack{\vecx \in \SAMPLES\\x_i = A}} 
\frac{f(\vecx, \vecystar)}
{\thetau_{i,\nucA}}
\end{equation}

After the gradient update (Eq.~\ref{eq:update}), we then project \vecThetahat back onto the set of valid distributions.
For each \vecthetahat in \vecThetahat,
we project it back
to the probability simplex by finding the vector in the simplex that is closest (in $\ell_2$ norm) to \vecthetahat:
\begin{align*}
\vecTheta' 
 \gets & \{ \proj(\vecthetahat) \mid \vecthetahat \in \vecThetahat\}\\
\proj(\vecthetahat)
   \defeq &\argmin_{\vectheta} \quad  \|\vecthetahat -\vectheta\|^2_2 \\
   & \;\; \textrm{s.t.} \quad \text{\vectheta is a distribution}.
\end{align*}



\subsection{Method 2: Softmax Parameterization and Unconstrained Optimization}
\label{sec:opt_softmax}
An alternative approach to the optimization problem is to introduce a parametrization that naturally enforces the required normalization for a valid distribution, thus converting the problem into an unconstrained optimization problem. This approach eliminates the need for performing gradient projection at each step. A common choice for achieving this normalization is the \textit{softmax} function, which inherently converts a set of real numbers into a valid probability distribution.

Instead of using a parameter vector as a distribution explicitly, now we model a distribution implicitly using softmax and our new parameter vector 
\(
\vecthetau_i \defeq (\thetau_{i,\nucA}, \thetau_{i,\nucC}, \thetau_{i,\nucG}, \thetau_{i,\nucU})
\) no longer sums to 1; instead we have:
\begin{equation}\label{eq:softmax-unpaired}
\forall a\in \nucset, \pu_i(a; \vecthetau_i) \defeq \frac{\exp(\thetau_{i,a})}{\sum_{a'} \exp(\thetau_{i,a'})}
\end{equation}
where $\thetau_{i,a}$ can be any real number. The softmax function ensures that each $\pu_i(\cdot; \vecthetau_i)$ forms a valid distribution without explicitly imposing this as a constraint. This definition can be extended for other parameter vectors $\thetap_{i,j}$, $\thetam_{i,j}$, and $\thetatm_{i,j,k}$ as follows:
\begin{align*}
    \forall a b\in \pairset, \pp_{i, j}(ab; \vecthetap_{i,j}) &\defeq \frac{\exp(\thetap_{i,j,ab})}{\sum_{a' b'\in\pairset} \exp(\thetap_{i,j,a'b'})}\\
    \forall a b\in \nucset^2, \pm_{i, j}(ab; \vecthetam_{i,j}) &\defeq \frac{\exp(\thetam_{i,j,ab})}{\sum_{a' b'\in\nucset^2} \exp(\thetam_{i,j,a'b'})}\\
    \forall a b c\in \nucset^3, \ptm_{i, j, k}(abc; \vecthetatm_{i,j,k}) &\defeq \frac{\exp(\thetatm_{i,j,k,abc})}{\sum_{a' b' c'\in\nucset^3} \exp(\thetatm_{i,j,k,a'b'c'})}
\end{align*}



With this new parametrization, the optimization problem becomes an unconstrained problem:

\begin{equation}
\begin{aligned} \label{pro:uncons}
    \min_{\vecTheta} \quad &  \obj(\vecTheta)
\end{aligned}
\end{equation}

Since the constraints have been naturally embedded into the problem formulation through the softmax function, we can directly apply the vanilla gradient descent algorithm to solve this optimization problem. The gradient can be updated by:
\[
\vecTheta'  \gets \vecTheta - \alpha \nabla_{\vecTheta}\obj(\vecTheta)
\]
where $\alpha$ is the learning rate.

Due to the softmax parametrization, the specific form of the gradient $\nabla_{\vectheta}\obj(\vecTheta)$ differs from that in the constrained optimization problem. Using the chain rule, we express it as:
\begin{equation}\label{eq:unconstraint-gradient}
    \frac{\partial \log p_{\vecystar} (\vecx; \vecTheta)}{\partial \thetau _{i,\nucA}} 
    = \sum_{a \in \nucset} 
    \underbrace{\frac{\partial \log p_{\vecystar} (\vecx; \vecTheta)}{\partial \pu_i(a; \vecthetau_i)}}_{\text{same as Eq.~\ref{eq:grad-partial}}}
    \cdot
    \underbrace{\frac{\partial \pu_i(a; \vecthetau_i)}{\partial \thetau_{i,\nucA}}}_{\text{softmax; Eq.~\ref{eq:softmax-grad}}}
\end{equation}
where the first partial derivative on the right hand side
is identical to the case of direct parameterization above (Sec.~\ref{sec:opt_PGD}; Eq.~\ref{eq:grad-partial}),
but the second partial derivative, which used to be 1 in direct parameterization, is now the 
gradient of the softmax function
(see Fig.~\ref{fig:computational-graph}
 and Sec.~\ref{sec:softmax-gradient}  
for details):
\begin{equation}
\label{eq:softmax-grad}
    \frac{\partial \pu_i(a; \vecthetau_i)}{\partial \thetau_{i,\nucA}}
    = \pu_i(a; \vecthetau_i) \cdot (\mathbbm{1}[a=\nucA] - \pu_i(\nucA; \vecthetau_i))
\end{equation}
So the gradient for the softmax parameterization is:
\begin{align*}
    &\frac{\partial \log p_{\vecystar} (\vecx; \vecTheta)}{\partial \thetau _{i,\nucA}} \\
    &\approx \! \sum_{a \in \nucset} \!\! \left( \!\!\!
    \frac{1}{|\SAMPLES|} \sum_{\substack{\vecx \in \SAMPLES\\x_i = a}} 
    \frac{f(\vecx, \vecystar)}{\thetau_{i,a}} 
    \!\! \right)
    \!\! \cdot \!
    \left[
     \pu_i(a; \vecthetau_i) \! \cdot \! (\mathbbm{1}[a=\nucA] - \pu_i(\nucA; \vecthetau_i))
     \right]
\end{align*}

We run the gradient decent step until the changes in the value of the objective function $\obj(\vecTheta)$ become sufficiently small (see Sec.~\ref{sec:results}), indicating that the solution has converged. Algorithm~\ref{alg:GD} outlines the procedure of both constraint and unconstraint optimization approach.


 \begin{algorithm}
    \caption{Sampling-based RNA Design}
    \label{alg:GD}
        \begin{algorithmic}
        \Function{Design}{$\vecystar$, $f$, projection=False} \Comment{$f$: objective} 
  
        \State $\vecxstar \gets \text{random\_init}(\vecystar)$ \Comment{current best design}
        \State $\vecTheta \gets \text{init\_params}(\vecystar)$
        \While{not converged}
            \State sample sequences, $\SAMPLES$, from distribution $p_{\vecystar}(\cdot; \vecTheta)$
            \State \vecx $\gets$ $\argmin_{\vecx'\in \SAMPLES} f(\vecx', \vecystar)$ \Comment{best sample in \SAMPLES}
            \If{$f(\vecx, \vecystar) < f(\vecxstar, \vecystar)$} \Comment{smaller means better}
               \State $\vecxstar \gets \vecx$
            \EndIf

            \State estimate objective $\obj(\vecTheta)$ using $\SAMPLES$

            \State estimate gradient $\nabla_{\vecTheta} \obj(\vecTheta)$ using $\SAMPLES$
            \State $\vecTheta \gets \vecTheta - \alpha \nabla_{\vecTheta} \obj(\vecTheta)$
            \If{projection} \Comment{projected gradient descent}
                \State $\vecTheta \gets \{\proj(\vectheta) \mid \vectheta \in \vecTheta\}$ \Comment{project onto simplex}
            \EndIf
        \EndWhile
        \State \Return \vecxstar
        \EndFunction
    \end{algorithmic}
\end{algorithm}

\input{fig-parameterization}

%

%% file: fig-parameterization.tex

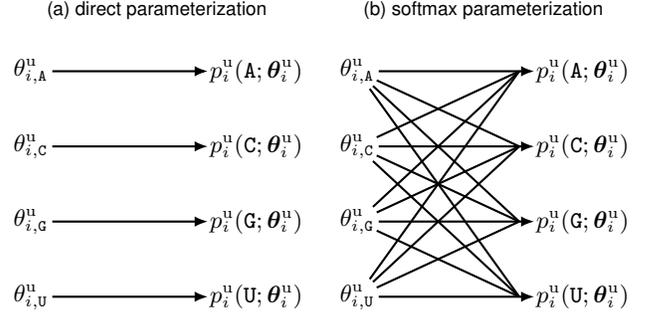
\begin{figure}[!t]
    \centering
    \begin{tabular}{cc}
        (a) direct parameterization
        &
        (b) softmax parameterization
        \\
        \begin{tikzpicture}[
            node style/.style={circle, minimum size=.5cm, inner sep=0pt, font=\normalsize},
            edge style/.style={-{Latex[length=5pt, width=4pt]}, thick}
            ]
        
            \node[node style] (T1) at (0, 4) {$\thetau_{i, \nucA}$};
            \node[node style] (T2) at (0, 3) {$\thetau_{i, \nucC}$};
            \node[node style] (T3) at (0, 2) {$\thetau_{i, \nucG}$};
            \node[node style] (T4) at (0, 1) {$\thetau_{i, \nucU}$};

            \node[node style] (P1) at (3, 4) {$\pu_{i}(\nucA;\vecthetau_{i})$};
            \node[node style] (P2) at (3, 3) {$\pu_{i}(\nucC;\vecthetau_{i})$};
            \node[node style] (P3) at (3, 2) {$\pu_{i}(\nucG;\vecthetau_{i})$};
            \node[node style] (P4) at (3, 1) {$\pu_{i}(\nucU;\vecthetau_{i})$};

        
            \draw[edge style] (T1) -- (P1);
            \draw[edge style] (T2) -- (P2);
            \draw[edge style] (T3) -- (P3);
            \draw[edge style] (T4) -- (P4);

        \end{tikzpicture}
        &
        \begin{tikzpicture}[
            node style/.style={circle, minimum size=.5cm, inner sep=0pt, font=\normalsize},
            edge style/.style={-, thick},
            arrow/.style={-{Latex[length=5pt, width=4pt]}, thick}
            ]
        
            \node[node style] (T1) at (0, 4) {$\thetau_{i, \nucA}$};
            \node[node style] (T2) at (0, 3) {$\thetau_{i, \nucC}$};
            \node[node style] (T3) at (0, 2) {$\thetau_{i, \nucG}$};
            \node[node style] (T4) at (0, 1) {$\thetau_{i, \nucU}$};

            \node[node style] (P1) at (3, 4) {$\pu_{i}(\nucA;\vecthetau_{i})$};
            \node[node style] (P2) at (3, 3) {$\pu_{i}(\nucC;\vecthetau_{i})$};
            \node[node style] (P3) at (3, 2) {$\pu_{i}(\nucG;\vecthetau_{i})$};
            \node[node style] (P4) at (3, 1) {$\pu_{i}(\nucU;\vecthetau_{i})$};

        
            \draw[arrow] (T1) -- (P1.west);
            \draw[edge style] (T1) -- ([xshift=-0.2cm]P2.west);
            \draw[edge style] (T1) -- ([xshift=-0.2cm]P3.west);
            \draw[edge style] (T1) -- ([xshift=-0.2cm]P4.west);

            \draw[edge style] (T2) -- ([xshift=-0.2cm]P1.west);
            \draw[arrow] (T2) -- (P2.west);
            \draw[edge style] (T2) -- ([xshift=-0.2cm]P3.west);
            \draw[edge style] (T2) -- ([xshift=-0.2cm]P4.west);

            \draw[edge style] (T3) -- ([xshift=-0.2cm]P1.west);
            \draw[edge style] (T3) -- ([xshift=-0.2cm]P2.west);
            \draw[arrow] (T3) -- (P3.west);
            \draw[edge style] (T3) -- ([xshift=-0.2cm]P4.west);

            \draw[edge style] (T4) -- ([xshift=-0.2cm]P1.west);
            \draw[edge style] (T4) -- ([xshift=-0.2cm]P2.west);
            \draw[edge style] (T4) -- ([xshift=-0.2cm]P3.west);
            \draw[arrow] (T4) -- (P4.west);


        \end{tikzpicture}
    \end{tabular}
    \caption{Computational graph for direct vs.~softmax parameterization.}
    \label{fig:computational-graph}
\end{figure}

%% file: related.tex

Although Matthies et al.~\cite{matthies+:2023} also used continuous optimization for RNA design, our approach is vastly different from and substantially outperforms theirs
in both scalability and design quality (by all metrics).
\begin{itemize}
\item 
First, their sequence distribution is a simple product of independent distributions for each position 
(same as our distribution v0 in Sec.~\ref{sec:v0}) 
which is ill-suited for the RNA design problem
for two reasons:
(a) that distribution
includes exponentially many 
illegal sequences for any input structure
due to pair violations and
(b) that distribution does not explicitly model the covariance between paired positions.
Instead, 
we use coupled variables  for paired and mismatch positions (our distributions v1, v2, and v3 in Secs.~\ref{sec:v1}--\ref{sec:v2v3}), which rules out invalid sequences and 
explicitly models the dependencies between
correlated positions.
\item 
Second, our sampling framework can work with arbitrary objective functions while their work is specifically designed for one such function, the Boltzmann probability. 
\item
Third, our unbiased sampling yields an {\em unbiased} approximation to the expectation of an arbitrary objective function over the distribution of sequences.
For example, for the case of Boltzmann probability,
our sampling results in an unbiased approximation
of the expected Boltzmann probability,
which converges to the true expectation as the sample size increases:
\begin{equation}
    \frac{1}{|\SAMPLES|}{\displaystyle\sum_{\vecx\in\SAMPLES} p(\vecystar \mid \vecx)}
    \approx
    \E_{\vecx}[p(\vecystar \mid \vecx)] 
\end{equation}
By contrast, they optimize a different objective (in red below) that
deviates from the true expectation
of Boltzmann probability with a bias ($\E[X/Y] \neq \E[X]/\E[Y]$):
\begin{equation}
    {
{\color{red}
    \frac{\E_{\vecx}[e^{-\freeenergy(\vecx, \vecystar) / RT}]}{\E_{\vecx}[Q(\vecx)]} 
    \!\neq
    }
    \E_{\vecx}\!\!{\Big[\frac{e^{-\freeenergy(\vecx, \vecystar) / RT}}{Q(\vecx)}\Big]} 
    \!\defeq 
    \E_{\vecx}[p(\vecystar \mid \vecx)] 
    }
\end{equation}
\item 
Fourth, our sampling-based approach is much more efficient: it scales to the longest structures in the Eterna100 benchmark (400 \nts) while their work only scaled to structures up to 50 \nts long.
\item
Last, our results substantially outperform theirs in all metrics (see Table~\ref{tab:104nucs}).
\end{itemize}

%% file: results.tex
\input{fig-mainresults}


The Eterna100 dataset~\cite{anderson+:2016eterna} is a widely used benchmark for evaluating RNA design programs. It contains 100 secondary structures (i.e., ``puzzles'') of up to 400 nucleotides, varying in design difficulty from simple hairpins to intricate multiloop structures. We evaluated this work against three baselines using this dataset:
SAMFEO~\cite{zhou+:2023samfeo}, NEMO~\cite{portela:2018nemo}, and Matthies et al.~\cite{matthies+:2023}. To compare their performance, we used the following metrics:
 \begin{enumerate}
   \item Average $p(\vecystar \mid \vecx)$ across all puzzles;
   \item Geometric mean of $p(\vecystar \mid \vecx)$ across all puzzles except those 18 that are proven to be undesignable 
   (in the sense that there is no uMFE solution)
   by our previous work \cite{zhou+:2024, zhou+:2024scalable}; these puzzles have extremely low 
   $p(\vecystar \mid \vecx)$ which bias the geometric mean towards 0;
   \item Average $\NED(\vecx, \vecystar)$ across all puzzles;
   \item Average $d(\MFE(\vecx), \vecystar)$ across all puzzles;
   \item Average $\DDG(\vecx, \vecystar)$ across all puzzles;
   \item Number of puzzles in which an MFE solution is found;
   \item Number of puzzles in which a uMFE solution is found.
 \end{enumerate}


\paragraph{Sampling-based Continuous Optimization (This Work)}
By default, our method samples 2500 sequences at each step. The number of steps is adaptive to each puzzle, specifically the run stops after 50 steps in which the objective function does not improved and the total number of steps is limited to 2000. Our main program is implemented in C++ and utilizes OpenMP for parallelization.

The default initial learning rate is set to 0.01, which works well for all puzzles under the softmax parameterization. However, we observe that the direct parameterization (\textit{projection}) requires smaller learning rates as puzzle lengths increases; otherwise, the objective value does not improve. Therefore, we apply adaptive learning rate decay for the projection method. Additionally, we implement momentum-based optimizers: Adam \cite{kingma+ba:2014} for softmax and Nesterov accelerate gradient \cite{nesterov:1983} for projection.

We adopt three types of initialization:
 \begin{itemize}
   \item \textbf{Uniform}: Each parameter is set to uniform distribution.
   \item \textbf{Targeted}: Assigns 100\% A for unpaired, 50\% CG and GC for base pairs, and uniform distribution for mismatches and trimismatches.
   \item $\epsilon$-\textbf{Targeted}: A combination of targeted and uniform distribution defined by $\epsilon \cdot \text{targeted} + (1 - \epsilon) \cdot \text{uniform}$.
 \end{itemize}
For the projection method, we use both uniform and targeted initializations. For the softmax method, we use uniform and $\epsilon$-targeted initializations with $\epsilon = 0.75$. The final solutions are selected from the best out of both initializations.

We use our previous work, LinearPartition \cite{zhang+:2020}, to compute the partition function and base-pairing probability in linear time with beam pruning. We use beam size $b = 250$ for optimizing $p(\vecystar \mid \vecx)$, and $b=100$ for optimizing $\NED(\vecx, \vecystar)$. It is only when optimizing for $p(\vecystar \mid \vecx)$, we use a larger beam size because LinearPartition tends to underapproximate the partition function, resulting in $p(\vecystar \mid \vecx) > 1$. For optimizing $d(\MFE(\vecx), \vecystar)$ and $\DDG(\vecx, \vecystar)$, we fold the sequences using LinearFold \cite{huang+:2019} with $b=100$.

At each step, we record the best sample among the 2,500 samples. After completing all steps, we reevaluate the recorded samples using ViennaRNA 2.0 \cite{lorenz+:2011}, selecting the best sequence for each metric. The values reported in this paper are thus, unaffected by the approximation error from beam search.

\paragraph{Baseline 1: SAMFEO}
SAMFEO is an iterative approach that selects a few nucleotides to mutate by (a) sampling positions based on positional defects and (b) utilizing structural information~\cite{zhou+:2023samfeo}. Similar to our work, SAMFEO is a general approach that can work with any objective function $f(\vecx, \vecystar)$. In their paper, they optimize for two ensemble objectives: $1 - p(\vecystar \mid \vecx)$ and $\NED(\vecx, \vecystar)$. SAMFEO is run five times on each puzzle under their default settings with 5000 steps, and we report the best solution obtained from these five runs.

\paragraph{Baseline 2: NEMO}
NEMO uses Nested Monte Carlo Search with domain-specific knowledge to solve puzzles~\cite{portela:2018nemo}. It maximizes a scoring function defined by:
\begin{equation}
  \text{score}(\vecx, \vecystar) = K(1 + \DDG(\vecx, \vecystar))^{-\sign(K)}
  \label{eq:nemo}
\end{equation}
where
\[K = 1 - \frac{\BPD(\MFE(\vecx), \vecystar)}{2\lvert \pairs(\vecystar) \rvert}\]
and $\BPD(\vecy, \vecy')$ is the base pairing distances,
\[\BPD(\vecy, \vecy') = \lvert \pairs(\vecy) \cup \pairs(\vecy') \rvert - \lvert \pairs(\vecy) \cap \pairs(\vecy') \rvert.\]
We ran NEMO five times with the parameters of ViennaRNA 2.5.1 and take the best solutions out of the five runs.

\paragraph{Baseline 3: Matthies et al.~\cite{matthies+:2023}}


Matthies et al.~\cite{matthies+:2023} reported the $p(\vecystar \mid \vecx)$ of puzzles up to a length of 50 (18 puzzles) using an 80 GB NVIDIA A100 GPU in their paper. We 
were able to run their code up to a length of 104 nucleotides (the shortest 51 puzzles) using 80GB NVIDIA H100 GPU under their default settings. However, extending to longer puzzles is not feasible due to GPU memory limit. 
We compare with their system in Table~\ref{tab:104nucs}
and Fig.~\ref{fig:max},
where our results substantially outperform theirs in all metrics.

\input{fig-73}

\paragraph{Main Results}
The main results of RNA Design methods on the Eterna100 dataset are shown in Figure \ref{fig:main_result}(a). Our best method uses the softmax parametererization while optimizing for $p(\vecystar \mid \vecx)$, which performed the best on the whole dataset in categories including the arithmetic mean of $p(\vecystar \mid \vecx)$, geometric mean of $p(\vecystar \mid \vecx)$ without undesignable puzzles~\cite{zhou+:2024, zhou+:2024scalable}, $\NED(\vecx, \vecystar)$, $\DDG(\vecx, \vecystar)$, and the number of \MFE solved. It also performs the second best on the other two metrics, $d(\MFE(\vecx), \vecystar)$ and the number of \UMFE solved. 

This work (\textit{softmax}) achieves arithmetic mean $p(\vecystar \mid \vecx)$ of 0.594, outperforming SAMFEO by 0.013. In terms of the geometric mean of $p(\vecystar \mid \vecx)$ without undesignable puzzles, our method performs better than other baselines by a wider margin. We obtained geometric mean of 0.512, surpassing the second-place method, SAMFEO, by 0.345. This indicates that our method is much more effective at designing solutions for longer and harder-to-design puzzles with lower $p(\vecystar \mid \vecx)$ values.

To our surprise, optimizing for $p(\vecystar \mid \vecx)$ yields excellent solutions for other metrics indirectly. Many of these metric are better optimized by optimizing p(y* | x) than when optimizing the metric directly. For example, this work (\textit{softmax}) optimizing for $p(\vecystar \mid \vecx)$ achieves average $\NED(\vecx, \vecystar)$ of 0.035, beating 
both of the approaches (this work and SAMFEO) optimizing for $\NED(\vecx, \vecystar)$ by 0.005 and 0.001 respectively.

Additionally, our method solved 79 puzzles under the \MFE criterion and 76 under the \UMFE criterion, matching NEMO's performance under the \MFE criterion and solving just one fewer puzzle under the \UMFE criterion. However, some of NEMO's advantage may stem from heuristic rules specifically tailored for the ViennaRNA energy model. Previous work \cite{zhou+:2023samfeo} showed that the vanilla version of NEMO solves only 76 puzzles under the \MFE criterion and 75 under the \UMFE criterion.

Our method uses both uniform and $\epsilon$-targeted initializations with $\epsilon = 0.75$, and take the best solution from both runs. For $p(\vecystar \mid \vecx)$, there are 14 puzzles for which the best solution comes from using the uniform initialization. For examples, the $p(\vecystar \mid \vecx)$ for puzzle \texttt{\#74} (380 \nts) improved from 0.202 to 0.457 and puzzle \texttt{\#77} (105 \nts) improved from 0.334 to 0.400.

Although this work (\textit{projection}) achieves a higher arithmetic mean of $p(\vecystar \mid \vecx)$ than SAMFEO, its geometric mean of $p(\vecystar \mid \vecx)$ without undesignable puzzles is lower than SAMFEO (0.029 vs.~.167). This suggests that the projection method performs poorly on hard to design puzzles, which is due to the difficulty in setting the appropriate learning rate for each puzzle.



\paragraph{This work vs.~SAMFEO (grouped by puzzle lengths)}
Figure \ref{fig:main_result}(b) and (c) compare the performance of this work (\textit{softmax}) against SAMFEO across different puzzle length groups, using 
the geometric mean of $p(\vecystar \mid \vecx)$ (without undesignable puzzles) and the average $\NED(\vecx, \vecystar)$.
While both methods perform similarly for shorter puzzles across both metrics, our method demonstrates advantages in the four longest length groups (116--192, 200--316, 337--387 and 389--400). Our geometric mean against SAMFEO in the longest four groups are 0.70 vs.~0.60, 0.34 vs.~0.27, 0.15 vs.~0.00 and 0.22 vs.~0.02, differing from 0.07 to 0.20. Similarly, our method outperform SAMFEO in terms of $\NED(\vecx, \vecystar)$, with margins between 0.006 to 0.033 (0.019 vs.~0.025, 0.02 vs.~0.05, 0.05 vs.~0.08, and 0.02 vs.~0.05).

Figure \ref{fig:si-main-results}(a), (b), and (c) further examine the arithmetic mean of $p(\vecystar \mid \vecx)$, the average $\DDG(\vecx, \vecystar)$, and the average $d(\MFE(\vecx), \vecystar)$. Our method outperforms SAMFEO in all four longest groups for the arithmetic mean of $p(\vecystar \mid \vecx)$  (0.76 vs.~0.74, 0.53 vs.~0.5, 0.39 vs.~0.38, 0.35 vs.~0.32). In terms of $\DDG(\vecx, \vecystar)$ and the average of $d(\MFE(\vecx), \vecystar)$, the major improvements occur in the three or four longest groups. For $\DDG(\vecx, \vecystar)$, our method surpasses SAMFEO in the longest three groups (2.23 vs.~0.47, 13.07 vs.~2.92, and 7.40 vs.~2.46). For $d(\MFE(\vecx), \vecystar)$, our method also outperforms SAMFEO in the longest four groups (0.00 vs.~1.20, 2.20 vs.~7.50, 6.70 vs.~22.70, and 4.30 vs.~11.40)

\input{fig-78}

\paragraph{\UMFE analysis}
The Venn digram in Figure~\ref{fig:main_result}(d) illustrates puzzles that are solved under the \UMFE criterion by this work, SAMFEO, and NEMO. Out of the 100 puzzles, there are 79 solved puzzles and 18 puzzles proven to be undesignable~\cite{zhou+:2024, zhou+:2024scalable}. The remaining 3 puzzles (\texttt{\#68}, \texttt{\#97}, \texttt{\#100}) have yet to be solved or proven undesignable.

Our method solved 76 puzzles under the \UMFE criterion, one fewer than NEMO. However, we solved an additional puzzle (\texttt{\#89}) that was not solved by either SAMFEO or NEMO. Although NEMO solved the most puzzles in the \UMFE sense, their solution quality is often substantially worse than ours in terms of $p(\vecystar \mid \vecx)$ and $\NED(\vecx, \vecystar)$. See Figure~\ref{fig:nemo_examples} for detailed visualizations of puzzles \texttt{\#71} and \texttt{\#79}. The examples show that NEMO's \UMFE solutions have worse positional defects overall and a lower $p(\vecystar \mid \vecx)$ due to competition from alternative structures.

\paragraph{This work vs.~SAMFEO (individual puzzle)}
The scatterplot in Figure~\ref{fig:main_result}(e) compares this work with SAMFEO across individual puzzles in terms of $p(\vecystar \mid \vecx)$. The two methods perform similarly on most puzzles, with our method showing substantial improvements in a few cases. For example, our method improved puzzle \texttt{\#74} from 0.325 to 0.457, and puzzle \texttt{\#38} from 0.569 to 0.736.

The scatterplot in Figure~\ref{fig:main_result} (f) uses a log scale for $p(\vecystar \mid \vecx)$, focusing on puzzles longer than 280\nts. In this plot, there are nine puzzles (annotated in the figure) which our solution outperforms SAMFEO's by a factor larger than 10 fold. Most of these puzzles are undesignable and have much lower values of $p(\vecystar \mid \vecx)$. Figure~\ref{fig:si-main-results}(d) compares the solutions by $\NED(\vecx, \vecystar)$, showing similar improvements for puzzles longer than 280\nts.


\paragraph{Visualized Examples}
Figures~\ref{fig:73}--\ref{fig:78} and \ref{fig:76}--\ref{fig:99} provide detailed comparisons between this work and SAMFEO for puzzles \texttt{\#73},  \texttt{\#78}, \texttt{\#76}, \texttt{\#91}, and \texttt{\#99}, respectively. In all these examples, our method consistently outperforms SAMFEO across all metrics and is able to find the \UMFE (or very close to \UMFE) solutions.

For puzzles \texttt{\#73} and \texttt{\#76} (Figs.~\ref{fig:73} and \ref{fig:76}), our method achieves better $p(\vecystar \mid \vecx)$ than SAMFEO by a substantial factor ($0.005$ vs.~$3 \times 10^{-27}$ and $0.046$ vs.~$7\times 10^{-8}$). Furthermore, our solutions meet the \UMFE criterion, whereas theirs do not satisfy either the \MFE or \UMFE criteria. In the base-pairing probabilities plots, we observe many incorrect and missing pairs from SAMFEO's \MFE structure, along with many positions with high positional defects.

Puzzles \texttt{\#78} (Fig.~\ref{fig:78}) and \texttt{\#91} (Fig.~\ref{fig:91}) are undesignable. Our solutions have structural distance of $4$ for puzzle \texttt{\#78} and $8$ for puzzle \texttt{\#91}. In both cases, the structural distances are due to missing pairs in our MFE structures, all of which belong to undesignable motifs. Therefore, we believe our MFE structures are the closest possible solutions to the target structures. In contrast, SAMFEO's \MFE structures are very different from the target structure, and the base-pairing probabilities plots show that their solution has very weak base-pairing probabilities becayse of competition from alternative structures.

Puzzle \texttt{\#99} (Fig.~\ref{fig:99}) is also undesignable. Our \MFE solution has a structural distance of 20, compared to SAMFEO's \MFE solution, which has a distance of 80. In our solution, one pair is missing from the undesignable motif, along with a few pairs from other bulge loops. Most of the missing pairs in the bulges form a 2 $\times$ 2 internal loop instead. Although these bulges are not undesignable motifs, they may still be undesignable in the context of the entire structure.

\paragraph{Time Analysis}
Figure~\ref{fig:steps} (b) shows the total number of steps to solve each puzzle based on the stopping criteria: 50 steps since the the objective function improved and a maximum of 2000 steps. Generally, as the puzzle length increases, it takes more steps to find the best solution. However, for puzzles between 100\nts to 192\nts, most puzzles terminated earlier than others, suggesting that these puzzles are easier to design.

Figure~\ref{fig:steps} (c) shows the time taken to solve each puzzle when ran on a server with 28 physical cores. The whole evaluation on the Eterna100 dataset takes about 10 days, with the longest puzzle taking up to 20 hours.

The primary time bottleneck in our method is computing the objective function $f(\vecx, \vecystar)$ for all 2500 samples. For example, computing $p(\vecystar \mid \vecx)$ takes up to more than 99\% of total time due to cubic time complexity. However, this issue is mitigated to some extent by using beam search from LinearPartition. Additionally, computing the objective function for each sample is parallelizable. With 28 physical cores, we observed a $27.4\times$ speedup for the longest puzzles with 400 nucleotides, reducing the time from 1233 seconds to 45 seconds per step.

Other speedup options include using a smaller sample size, but it tends to result in longer convergence times and greater difficulty in finding a better solution. Speed improvements can also be achieved by caching the $f(\vecx, \vecystar)$ of each sample, which avoids redundant computations for repeated samples. This approach is effective when the puzzle is short and targeted initialization is used, which tends to produce more repeated samples. For example, puzzle \texttt{\#36} with 151 nucleotides obtained a 1.95$\times$ speedup from caching, reducing the time from 4.2 seconds to 2.15 seconds per step.

\paragraph{Learning Curves}
Figure~\ref{fig:learning-curves} presents the learning curves for puzzles \texttt{\#97} and \texttt{\#98}. At each step, the plots illustrate $p(\vecystar \mid \vecx)$ for the best sample and the integral solution, along with the arithmetic and geometric means of $p(\vecystar \mid \vecx)$ computed from 2500 samples. Overall, the learning curves show that the values for the best sample, as well as the arithmetic and geometric means of $p(\vecystar \mid \vecx)$, consistently increase and converge toward the end of the process. The entropy decreases over time, while the boxplot narrows, indicating that the distribution becomes more concentrated as the optimization progresses. Notably, the best sample converges more quickly and reaches a higher $p(\vecystar \mid \vecx)$ than the integral solution. In our findings, the highest quality solutions are consistently derived from the best sample rather than the integral solution.

\paragraph{Ablation Studies}
We run two additional experiments with targeted initializations to assess the effectiveness of coupling variables on mismatches and trimismatches. First, we ablated the coupling variables for trimismatches. Then, we ablated the coupling variables for both mismatches and trimismatches. The results of these ablation studies are presented in Table~\ref{tab:ablation}.

As more coupling variables are removed, the results deteriorate across all metrics. The arithmetic mean of $p(\vecystar \mid \vecx)$ decreased from $0.589$ to $0.578$ after ablating trimismatches and further declines to $0.566$ when mismatches are also ablated. Similarly, the average $\NED(\vecx, \vecystar)$ increases from 0.035 to 0.039 and then to 0.042. This pattern continues across other metrics, suggesting that the coupled variables approach is useful.

%% file: fig-mainresults.tex

\begin{figure*}[!htb]
  \centering
\begin{tabular}{c}
(a) Main results\\
\resizebox{.99\textwidth}{!}{
\setlength{\tabcolsep}{3pt}
\begin{tabular}{l|l|@{\hskip .2cm}cc c cccc}
   Methods                        & Objective & \multicolumn{2}{c}{${p}(\vecystar \mid \vecx)(\uparrow)$}  
 &  $\NED(\vecx, \vecystar)$ & $d(\MFE(\vecx), \vecystar)$ & $\DDG(\vecx, \vecystar)$ & \# of & \# of \\
    & & arith. & geom.$\dagger$  & $(\downarrow)$  & $(\downarrow)$ & $(\downarrow)$ &  MFE $(\uparrow)$ & uMFE $(\uparrow)$ \\ 
  
\hline 
\hline

   NEMO & composite (Eq.~(\ref{eq:nemo})) & \textit{0.271} & \textit{0.083} & \textit{0.098} & \textbf{\textit{1.93}} & \textit{1.31} & \textbf{79} & \textbf{77} \\[2pt]  
  
  \hline

   SAMFEO                      & $1 - p(\vecystar \mid \vecx)$ & 0.581 & 0.167 & 0.043 & 4.57 & 2.48 & 77 & 74 \\
                               & $\NED(\vecx, \vecystar)$ & 0.512 & 0.033 & \underline{0.037} & 3.92 & 2.95 & 72 & 66 \\[2pt]  
  
  \hline

   This Work ({\em projection}) & $\E_{\vecx} [-\log p(\vecystar \mid \vecx)]$ & \underline{0.589} & 0.104 & 0.048 & 6.98 & 2.74 & 75 & 71              \\[2pt]  
  
  \hline

   This Work ({\em softmax})    & $\E_{\vecx}[-\log p(\vecystar \mid \vecx)]$ & \textbf{0.593} & \textbf{0.502} & \textbf{0.036} & \underline{2.10} & \textbf{0.78} & \textbf{79}  & \underline{76}  \\ 
                               & $\E_{\vecx}[\NED(\vecx, \vecystar)]$ & 0.541 & 0.015 & 0.041 & 5.48 & 5.07 & 63  & 62  \\ 
                               & $\E_{\vecx}[d(\MFE(\vecx), \vecystar)]$ & 0.504  & 0.022  & 0.045  & 4.88  & 4.50  & 67  & 67  \\ 
                               & $\E_{\vecx}[\DDG(\vecx, \vecystar)]$ & 0.538 & \underline{0.377} & 0.047 & 2.88  & \underline{1.14} & \underline{78}  & 72 \\[2pt]  
  
\hline 
\hline
\end{tabular}}
\end{tabular}
\\[10pt]

  \setlength{\tabcolsep}{0pt}
  \begin{tabular}{ccc}
    (b) $\text{Geometric mean}^\dagger$ of $p(\vecystar \mid \vecx)(\uparrow)$, grouped by length
    &
    (c) Average $\NED(\vecx, \vecystar)(\downarrow)$, grouped by length
    &
    (d) uMFE Solved
    \\
    \hspace{-.4cm}
    \includegraphics[width=0.39\linewidth]{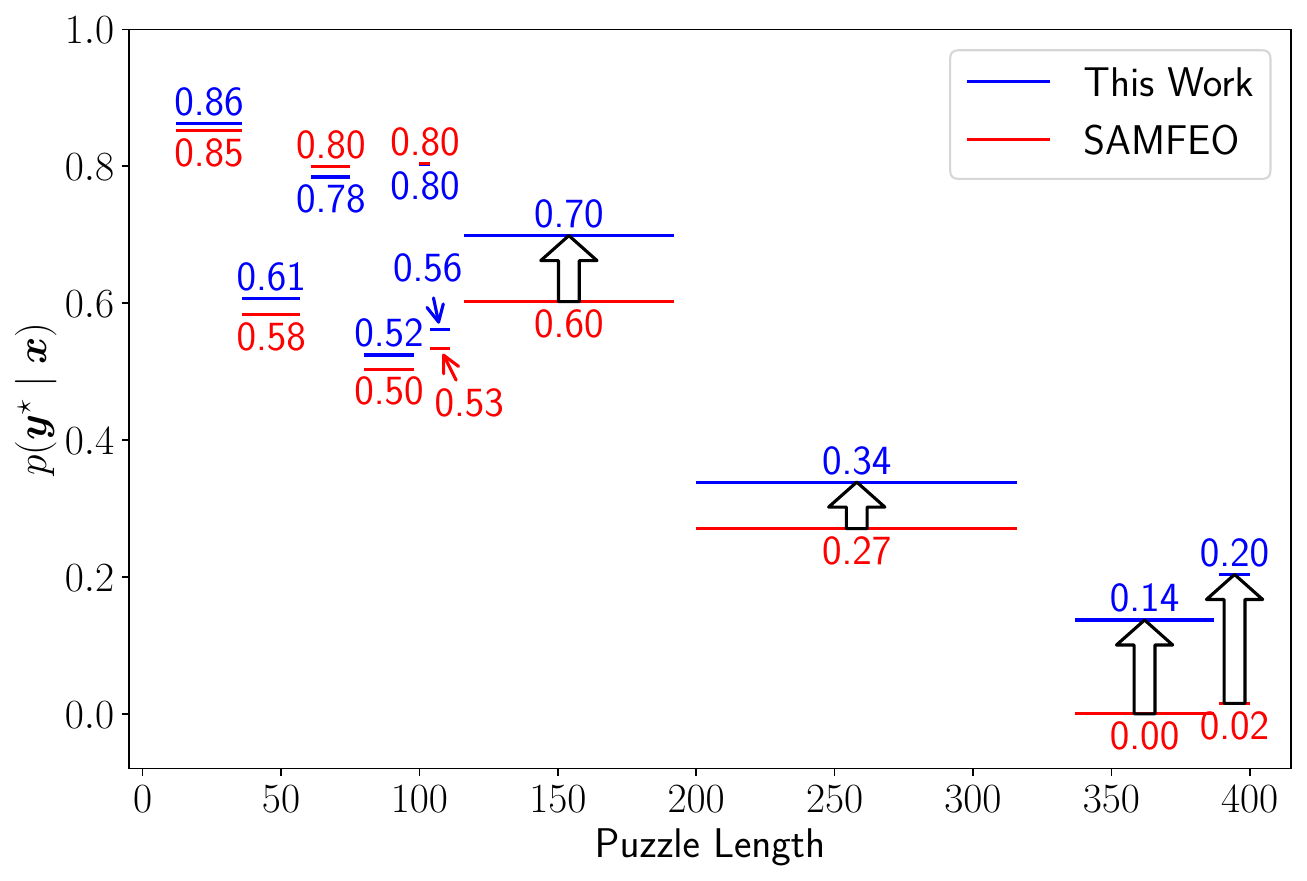}
    &
    \includegraphics[width=0.395\linewidth]{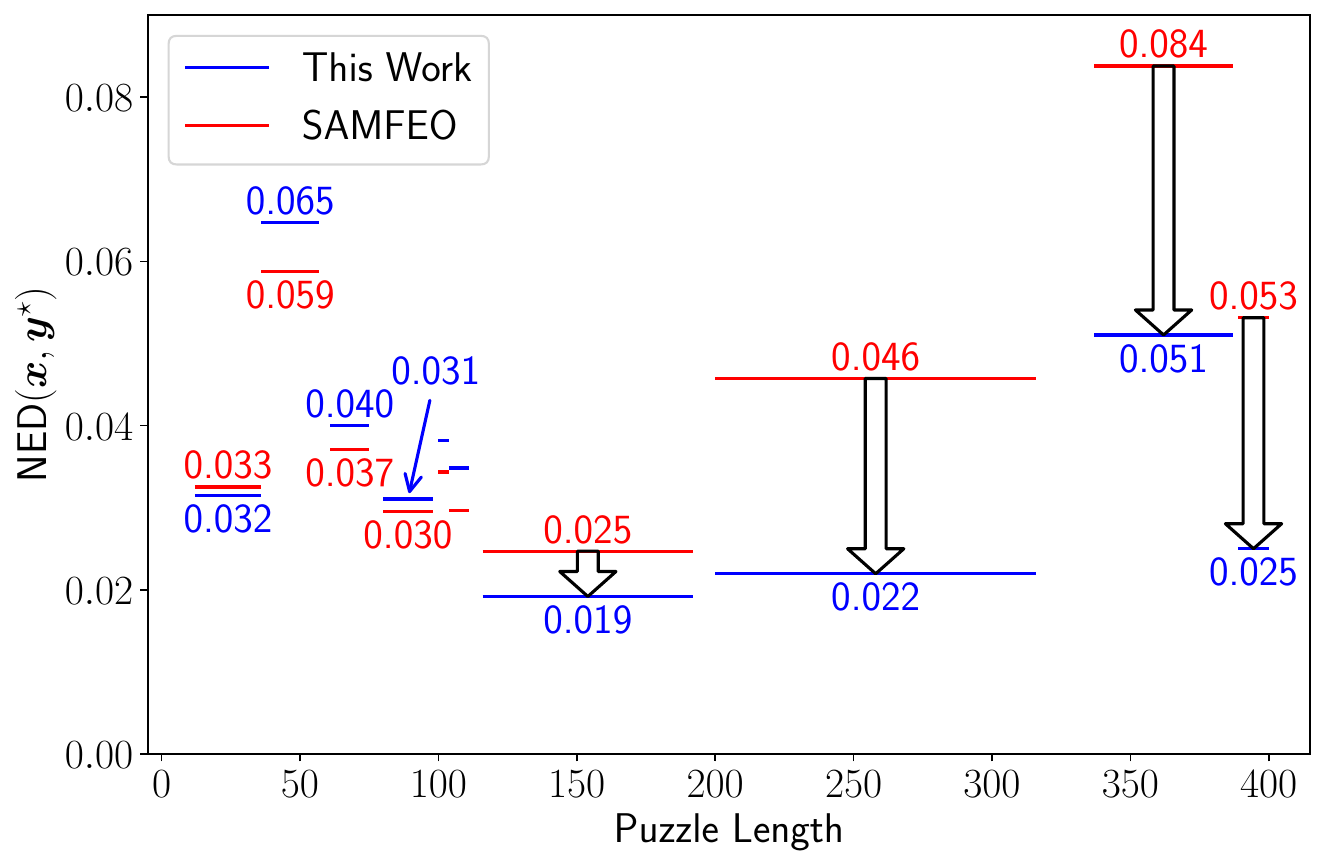}
    &
    \hspace{-.48cm}
    \resizebox{.28\textwidth}{!}{%
          \definecolor{color1}{RGB}{217, 234, 211}
          \definecolor{color2}{RGB}{244, 204, 204}
          \definecolor{color3}{RGB}{201, 218, 248}

          \def\mfenemo{(0,0) circle (1.5cm)}
          \def\mfesamfeo{(0.6, 0.5) ellipse (1.9cm and 0.88cm)}
          \def\mfesampling{(1.2,0) circle (1.5cm)}

          \def\umfenemo{(4.3,0) circle (1.5cm)}
          \def\umfesamfeo{(4.9, 0.5) ellipse (1.9cm and 0.88cm)}
          \def\umfesampling{(5.5,0) circle (1.5cm)}

        \begin{tikzpicture}
          \begin{scope}[]
            \fill[color1, opacity=0.8] \umfenemo;
            \fill[color2, opacity=0.8] \umfesamfeo;
            \fill[color3, opacity=0.6] \umfesampling;
            \draw \umfenemo node[below] {};
            \draw \umfesamfeo node [above] {};
            \draw \umfesampling node [below] {};
      
            \node at (6.1,-0.65) {\small \texttt{\#89}}; 
            \node at (6.1,0.4) {\small \texttt{\#19}}; 
            \node at (4.6,-0.6) {\small \texttt{\hyperref[fig:73]{\color{designable}{{\#73$^\star$}}}}}; 
            \node at (5.1,-0.9) {\small \texttt{\hyperref[fig:76]{\color{designable}{{\#76$^\star$}}}}}; 
            \node at (3.4,-0.5) {\small \texttt{\hyperref[fig:nemo_examples]{\color{designable}{{\#71$^\star$}}}}}; 
            \node at (3.7,-0.9) {\small \texttt{\hyperref[fig:nemo_examples]{\color{designable}{{\#79$^\star$}}}}}; 
            \node at (3.55,0.4) {\small \texttt{\#65}}; 
            \node at (4.9,0.45) {\small {72}}; 
            \node at (4.9,0.15) {\small {Puzzles}}; 
      
            \node at (4.3,0.85) {\small SAMFEO (74)};
            \node at (3.3,1.25) {\small NEMO (77)};
            \node at (6.55,1.35) {\small This Work (76)}; 2
            \node at (3.6,1.8) {\normalsize Designable (79)}; 
      
            \draw (2.5,2.0) rectangle (7.6,-1.6) node[] {};
            \draw (7.6,-1.6) rectangle (2.5,-3.3) node[] at (5.7, -1.9) {\small Undesignable (18)};
            \draw (3.8,-1.6) rectangle (2.5,-3.3) node[] at (3.15, -1.9) {\small Unknown};
            \draw node[] at (3.15, -2.2) {\small (3)};
      
            \node [text width=0.2cm] at (4.0,-2.3) {\small \color{undesignable}{\texttt{\#50}}};
            \node [text width=0.2cm] at (4.0,-2.65) {\small \color{undesignable}{\texttt{\#52}}};
            \node [text width=0.2cm] at (4.0,-3.0) {\small \color{undesignable}{\texttt{\#57}}};
            \node [text width=0.2cm] at (4.6,-2.3) {\small \color{undesignable}{\texttt{\#60}}};
            \node [text width=0.2cm] at (4.6,-2.65) {\small \color{undesignable}{\texttt{\#61}}};
            \node [text width=0.2cm] at (4.6,-3.0) {\small \color{undesignable}{\texttt{\#67}}};
            \node [text width=0.2cm] at (5.2,-2.3) {\small \color{undesignable}{\texttt{\#72}}};
            \node [text width=0.2cm] at (5.2,-2.65) {\small \hyperref[fig:78]{\color{undesignable}{\texttt{{\#78$^\star$}}}}};
            \node [text width=0.2cm] at (5.2,-3.0) {\small \color{undesignable}{\texttt{\#80}}};
            \node [text width=0.2cm] at (5.85,-2.3) {\small \color{undesignable}{\texttt{\#81}}};
            \node [text width=0.2cm] at (5.85,-2.65) {\small \color{undesignable}{\texttt{\#86}}};
            \node [text width=0.2cm] at (5.85,-3.0) {\small \color{undesignable}{\texttt{\#87}}};
            \node [text width=0.2cm] at (6.45,-2.3) {\small \color{undesignable}{\texttt{\#88}}};
            \node [text width=0.2cm] at (6.45,-2.65) {\small \color{undesignable}{\texttt{\#90}}};
            \node [text width=0.2cm] at (6.45,-3.0) {\small \hyperref[fig:91]{\color{undesignable}{\texttt{{\#91$^\star$}}}}};
            \node [text width=0.2cm] at (7.1,-2.3) {\small \color{undesignable}{\texttt{\#92}}};
            \node [text width=0.2cm] at (7.1,-2.65) {\small \color{undesignable}{\texttt{\#96}}};
            \node [text width=0.2cm] at (7.1,-3.0) {\small \hyperref[fig:99]{\color{undesignable}{\texttt{{\#99$^\star$}}}}};
      
            \node [text width=0.2cm] at (3.0,-2.5) {\small \color{unknown}{\texttt{\#68}}};
            \node [text width=0.2cm] at (3.0,-2.8) {\small \hyperref[fig:learning-curves]{\color{unknown}{{\texttt{\#97$^\star$}}}}};
            \node [text width=0.2cm] at (3.0,-3.1) {\small \color{unknown}{\texttt{\#100}}};
      
        \end{scope}
      \end{tikzpicture}
      }
  \end{tabular}

  \setlength{\tabcolsep}{4pt}
  \begin{tabular}{ccc}
      \hspace{-.4cm}
      (e) $p(\vecystar \mid \vecx)(\uparrow)$ of individual puzzles
      &
      (f) $p(\vecystar \mid \vecx)(\uparrow)$ (in log scale) of puzzles longer than 280\nts
      \\
  \hspace{-0.4cm}
  \begin{tikzpicture}
      \node [] {\includegraphics[width=.48\linewidth]{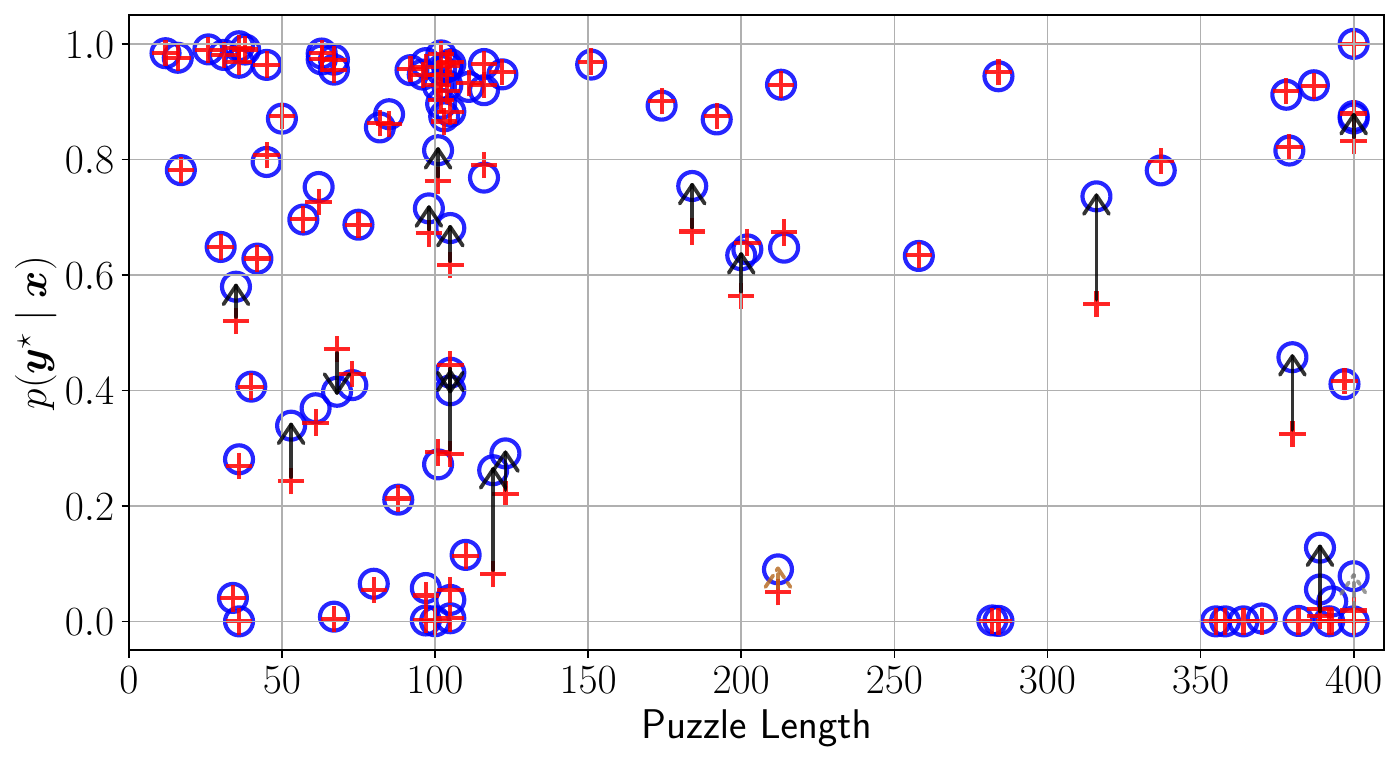}};
      \node at (-0.95, -1.25) {\small {\texttt{\color{designable}{\#83}}}};
      \node at (2.17, 0.2) {\small {\texttt{\color{designable}{\#38}}}};
      \node at (3.25, -0.48) {\small {\texttt{\color{designable}{\#74}}}};
  \end{tikzpicture}
  &
  \begin{tikzpicture}
    \node [] {\includegraphics[width=.49\linewidth]{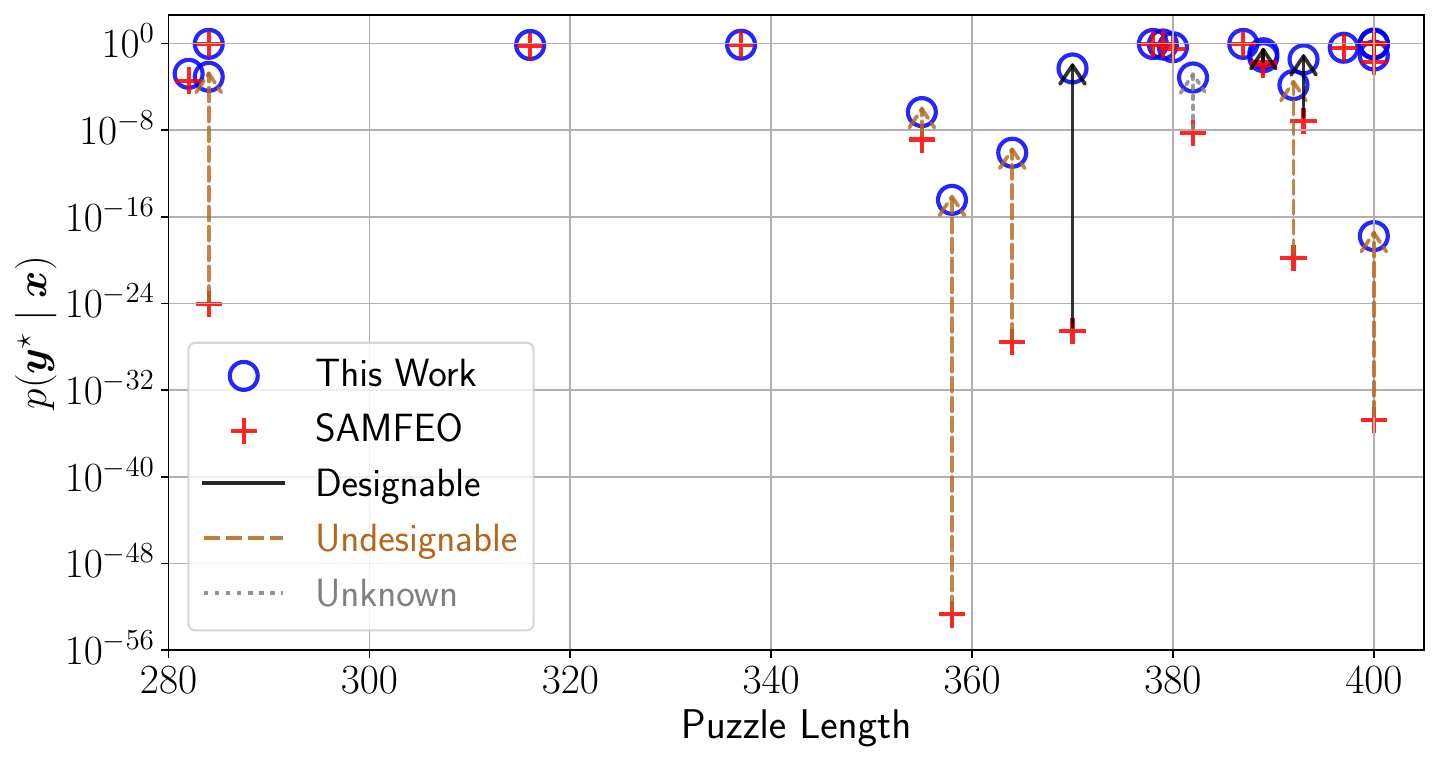}};

    \node at (-2.65,0.45) {\small {\texttt{\hyperref[fig:78]{\color{undesignable}{{\#78$^\star$}}}}}};
    \node at (1.03,1.25) {\small \color{undesignable}{\texttt{\#86}}};
    \node at (1.1,-1.52) {\small \color{undesignable}{\texttt{\#96}}};
    \node at (1.75,0.0) {\small {\texttt{\hyperref[fig:99]{\color{undesignable}{{\#99$^\star$}}}}}};
    \node at (2.4,0.1) {\small \texttt{\hyperref[fig:73]{\color{designable}{{\#73$^\star$}}}}};
    \node at (2.9,1.2) {\small \color{unknown}{\texttt{\#100}}};
    \node at (3.4,0.55) {\small {\texttt{\hyperref[fig:91]{\color{undesignable}{{\#91$^\star$}}}}}};
    \node at (3.95,1.4) {\small \texttt{\hyperref[fig:76]{\color{designable}{{\#76$^\star$}}}}};
    \node at (4.0,-0.45) {\small \color{undesignable}{\texttt{\#90}}};
  \end{tikzpicture}
  \end{tabular}

  \caption{(a) Results of various RNA Design methods on the Eterna100 dataset. \textbf{Bold}: best value. \underline{Underline}: second best value. \textit{Italic}: the byproduct is obtained by evaluating the final solution. This work and SAMFEO take the best solution across the entire optimization trajectory. $\dagger$: geometric mean without 18 undesignable puzzles. (b) -- (c) Comparison between this work and SAMFEO for $p(\vecystar \mid \vecx)$, grouped by puzzle lengths. Each group contains 10 puzzles, except for the geometric means, which exclude undesignable puzzles. (d) Puzzles solved by this work, SAMFEO, and NEMO under the \UMFE criterion. (e) -- (f) $p(\vecystar \mid \vecx)$ of solutions designed by this work vs.~SAMFEO in both original and log scale. Figure \ref{fig:si-main-results} provides similar grouped-by-length and individual plots for other metrics. Starred puzzles: 
  \texttt{\hyperref[fig:nemo_examples]{\color{designable}{\#71$^\star$}}},
  \texttt{\hyperref[fig:73]{\color{designable}{{\#73$^\star$}}}},
  \texttt{\hyperref[fig:76]{\color{designable}{{\#76$^\star$}}}},
  \texttt{\hyperref[fig:78]{\color{undesignable}{{\#78$^\star$}}}},
  \texttt{\hyperref[fig:nemo_examples]{\color{designable}{{\#79$^\star$}}}},
  \texttt{\hyperref[fig:91]{\color{undesignable}{{\#91$^\star$}}}},
  \texttt{\hyperref[fig:learning-curves]{\color{unknown}{{\#97$^\star$}}}}, and 
  \texttt{\hyperref[fig:99]{\color{undesignable}{{\#99$^\star$}}}}
  are hyperlinked to their visualizations.}
  \label{fig:main_result}
\end{figure*}

%% file: fig-73.tex

\begin{figure*}[!h]
  \centering   
    (a) \begin{tabular}{c|c|c|c|c|c|c}
      \texttt{\#73} (370 \nts) & $p(\vecystar \mid \vecx)$ & NED$(\vecx, \vecystar)$ & $d(\MFE(\vecx), \vecystar)$ & $\DDG(\vecx, \vecystar)$ & is MFE & is uMFE\\
      & $(\uparrow)$ & $(\downarrow)$ & $(\downarrow)$ & $(\downarrow)$ & & \\
    
      \hline

          This Work & \textbf{0.005} & \textbf{0.055} & \;\,\, \textbf{0} &  \; \textbf{0.0 kcal/mol} & \textbf{Yes} & \textbf{Yes}\\
          SAMFEO & 3 $\times 10^{-27}$ & 0.417 & 138 & 33.2 kcal/mol & No & No\\
      \end{tabular}\\[6pt]
  
    \includegraphics[width=.7\linewidth]{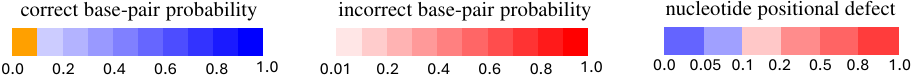}\\[6pt]
    \begin{tabular}{c|c}
          (b) Target Structure \vecystar and MFE Structure (This Work)
        &
          (c) MFE Structure (SAMFEO)
        \\
        \includegraphics[width=0.49\linewidth]{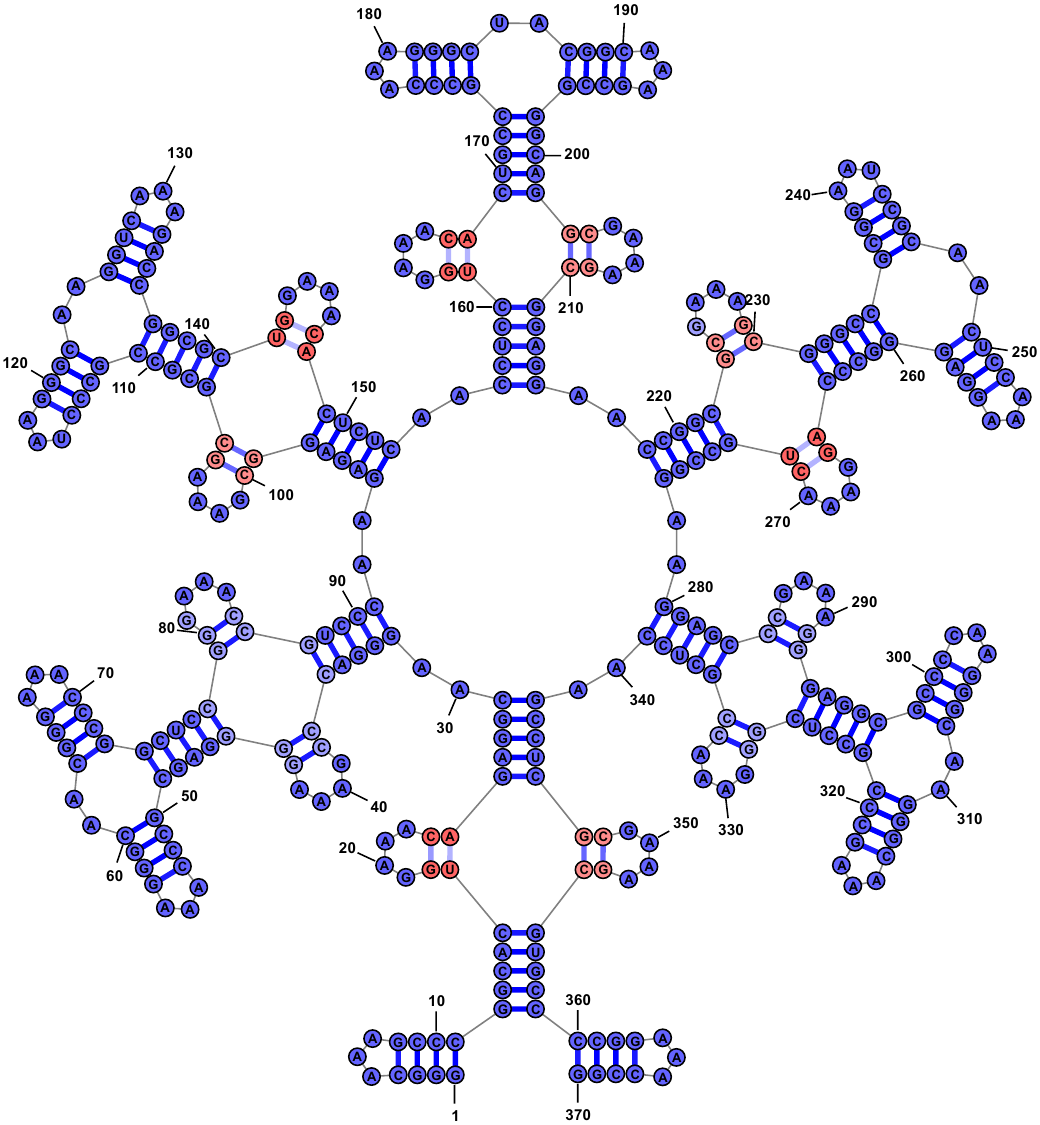}
        &
        \!\!\!\hspace{-.155cm}
        \includegraphics[width=0.49\linewidth]{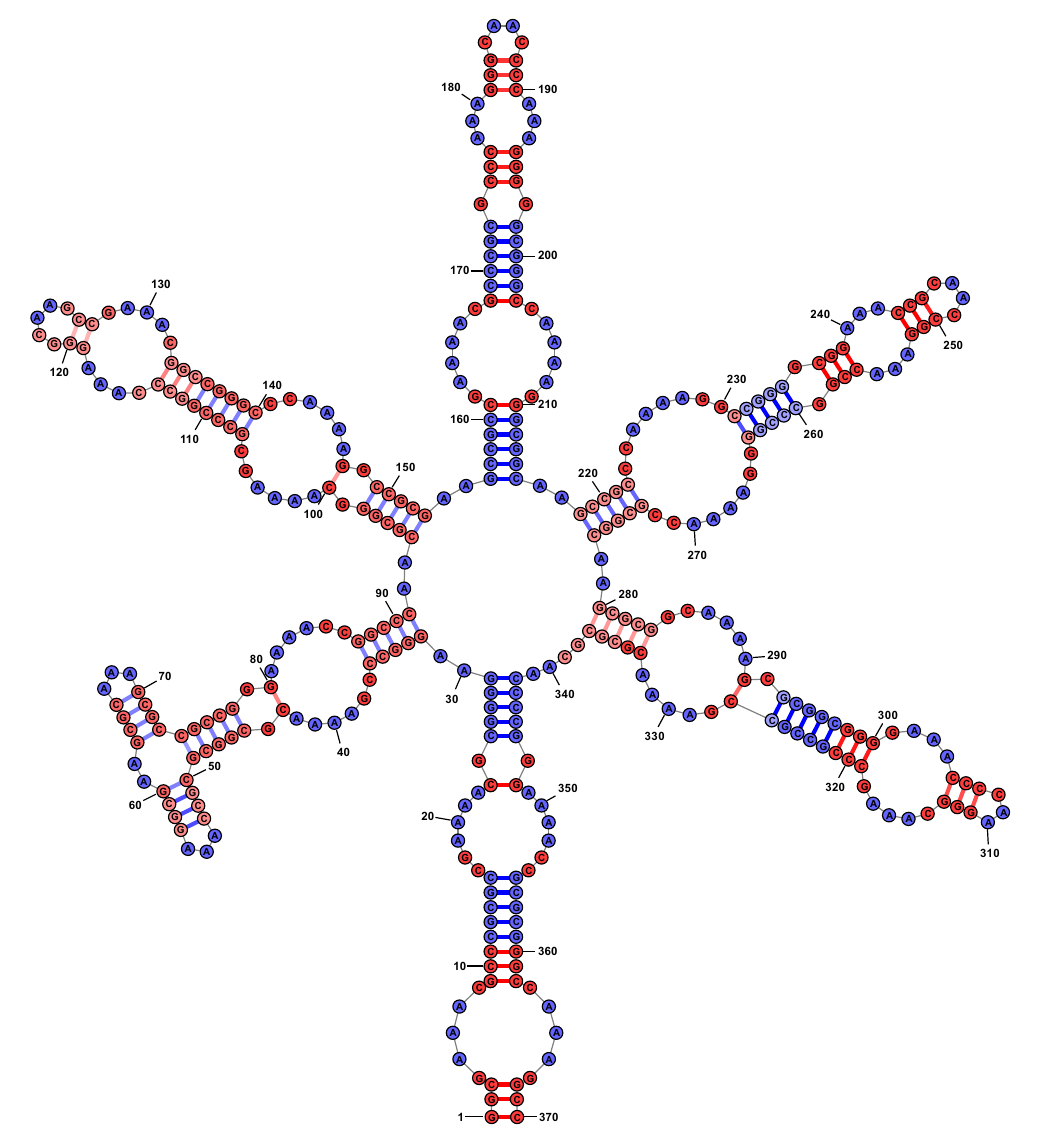}
        \\
        (d) Base-Pairing Probabilities (This Work)
        &
        (e) Base-Pairing Probabilities (SAMFEO)
        \\
          \hspace{-0.5cm}\raisebox{0.08cm}{\includegraphics[width=0.52\linewidth]{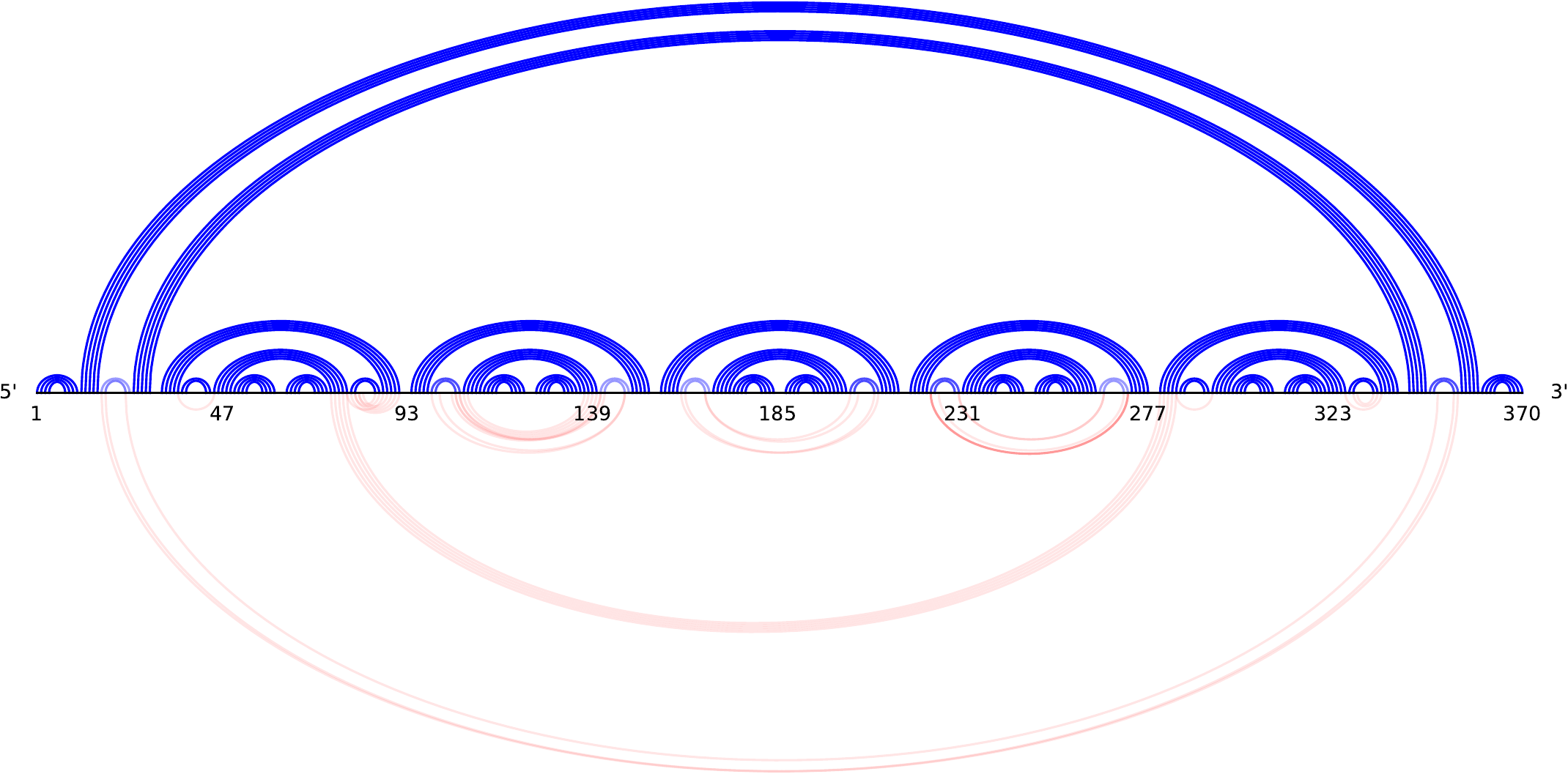}}\!\!\! 
        &
          \!\!\!\includegraphics[width=0.49\linewidth]{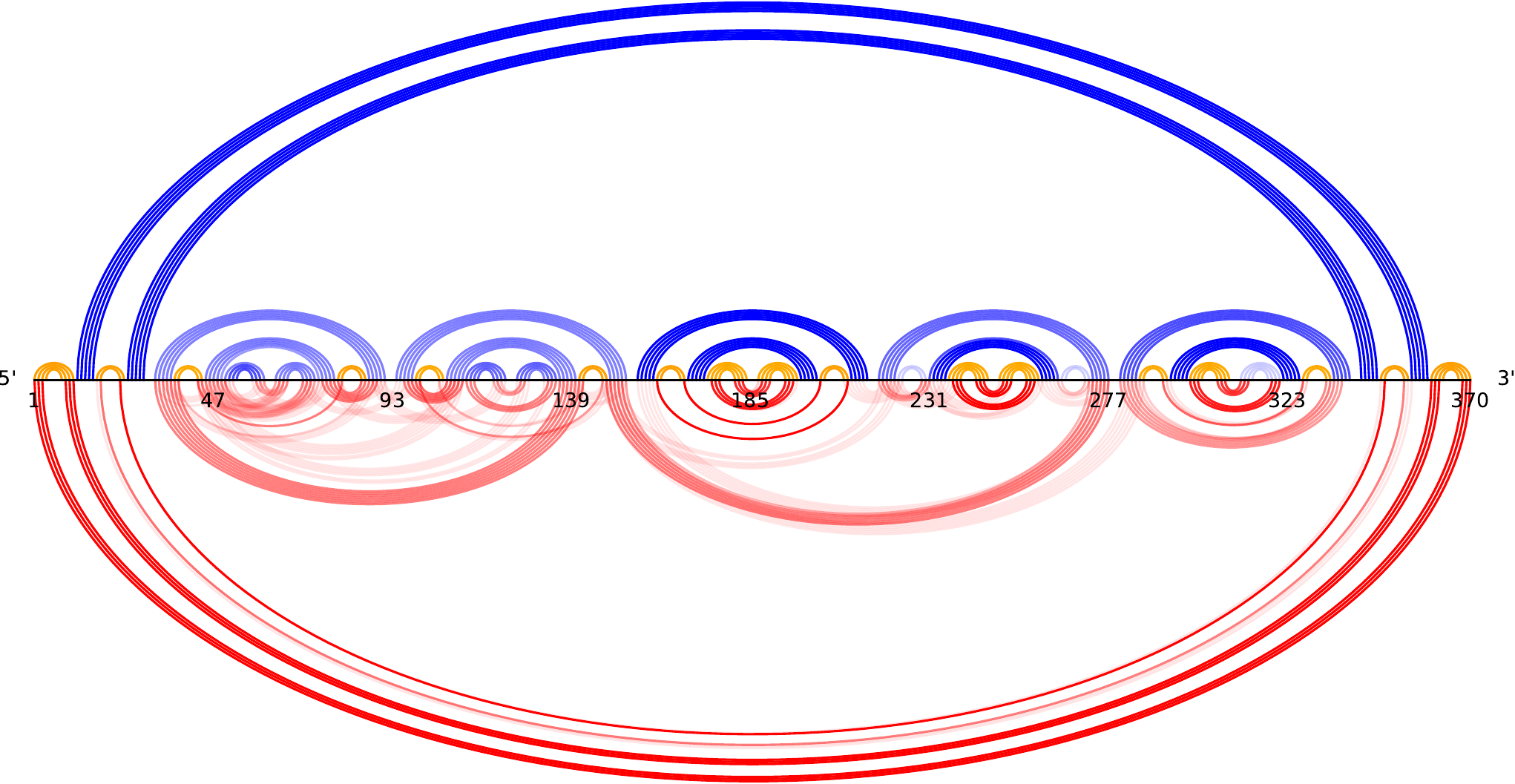}
    \end{tabular}
    \caption{Comparison of the best $p(\vecystar \mid \vecx)$ solutions designed by this work vs.~SAMFEO for Puzzle 73 (``Snowflake 4''). (b) -- (c) \MFE structures of the solutions from this work and SAMFEO. Base-pairs are colored as follows: blue for correct pairs, red for incorrect pairs, with the intensity indicating pairing probability. Nucleotide colors range from blue to red, indicating positional defect. (d) -- (e) Base-pairing probabilities of this work and SAMFEO. Orange represents missing correct pairs (i.e.~correct pairs with a pairing probability below 0.1).}
    \label{fig:73}
\end{figure*}

%% file: fig-78.tex

\begin{figure*}[!h]
  \centering   
    (a)  \begin{tabular}{c|c|c|c|c|c|c||c}
      \texttt{\#78} (284 \nts)& $p(\vecystar \mid \vecx)$ & NED$(\vecx, \vecystar)$ & $d(\MFE(\vecx), \vecystar)$ & $\DDG(\vecx, \vecystar)$ & is MFE & is uMFE & $p(\vecytilde \mid \vecx)$\\
      & $(\uparrow)$ & $(\downarrow)$ & $(\downarrow)$ & $(\downarrow)$ & & & $(\uparrow)$ \\
      \hline
      This Work & \textbf{0.001} & \textbf{0.123} & \;\,\, \textbf{4} & \; \textbf{2.6 kcal/mol} & No & No & \textbf{0.058}\\
      SAMFEO & 9 $\times 10^{-25}$ & 0.452 & 140 & 29.2 kcal/mol & No & No & 2 $\times 10^{-20}$\\
  \end{tabular}
    \includegraphics[width=.7\linewidth]{./figs/color_legend-crop}\\[6pt]
    \setlength{\tabcolsep}{3.2pt}
    \begin{tabular}{ccc}
      (b) Target Structure \vecystar
      &
      (c) MFE Structure (This Work)
      &
      (d) MFE Structure (SAMFEO)
      \\
      \hspace{-.4cm}
      \includegraphics[width=0.33\linewidth]{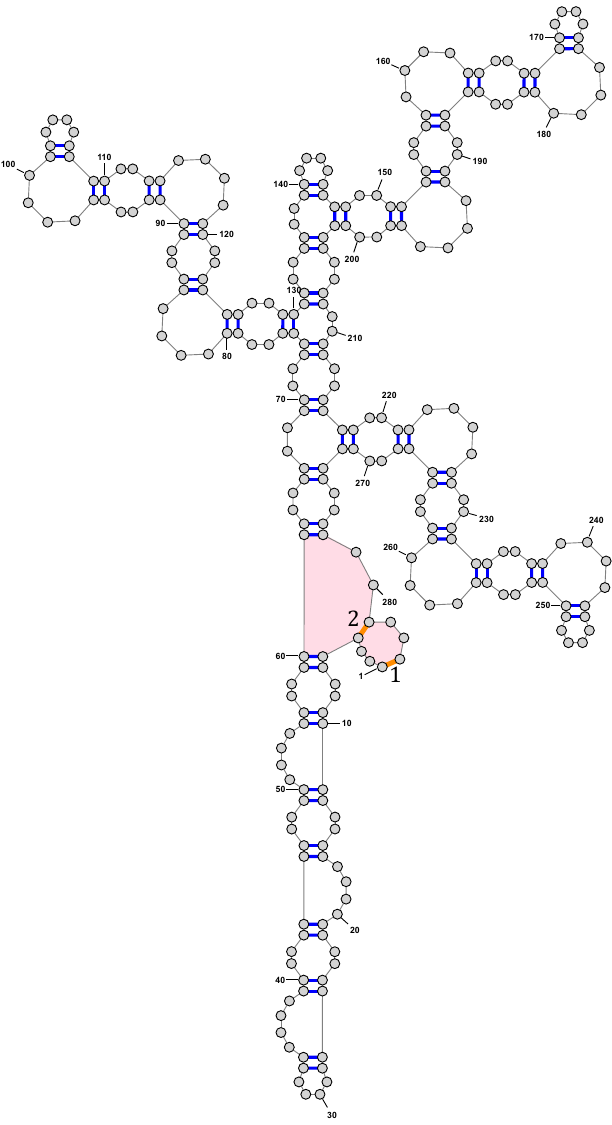}
      &
      \includegraphics[width=0.33\linewidth]{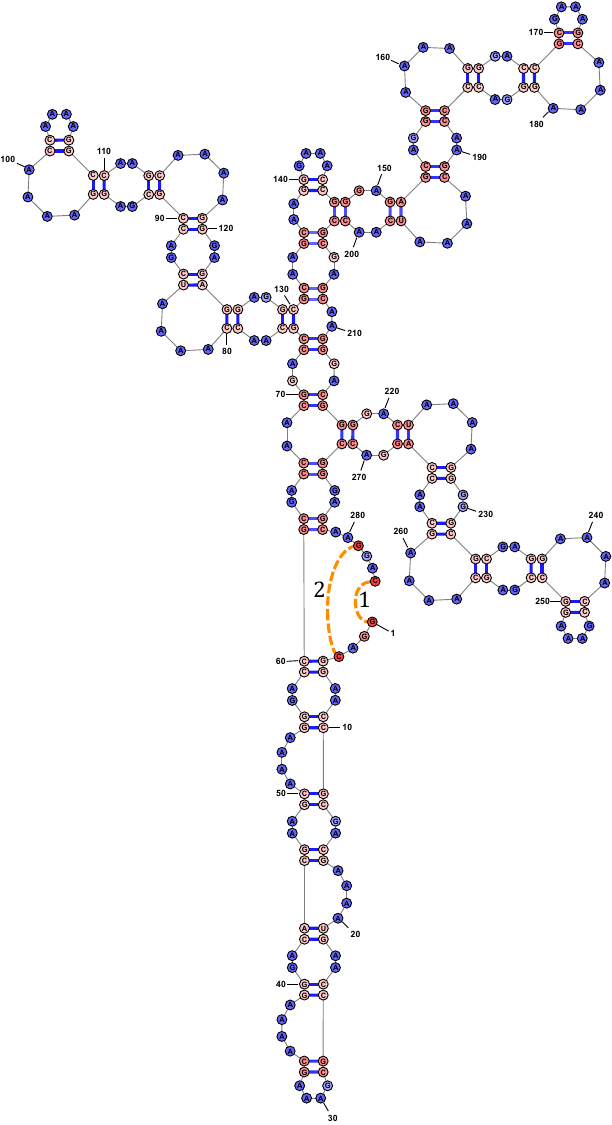}
      &
      \includegraphics[width=0.31\linewidth]{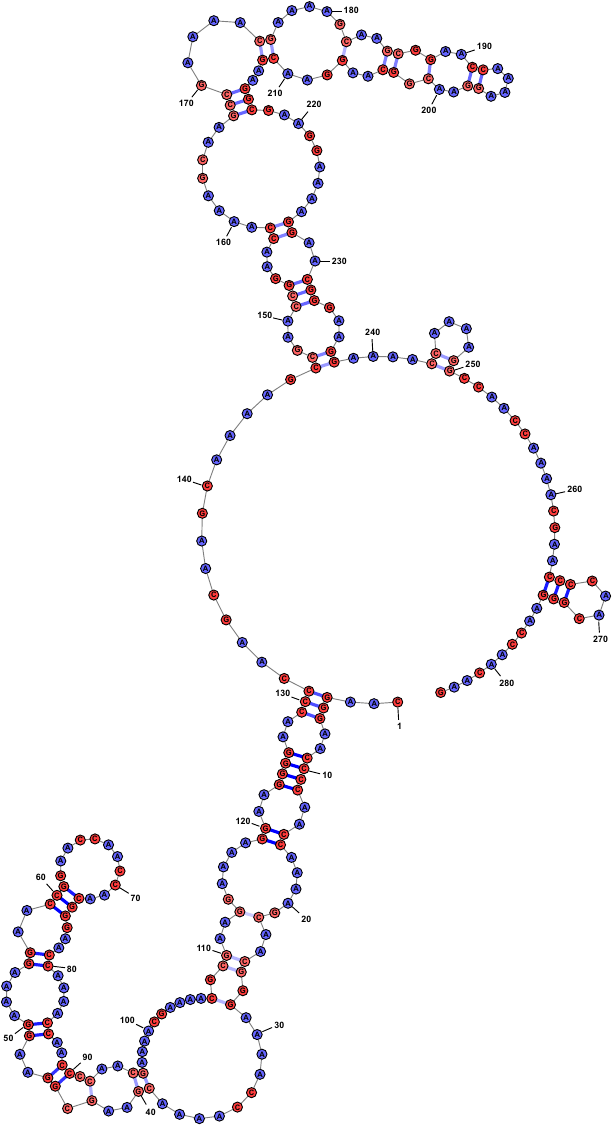}
    \end{tabular}\\[6pt]
    \begin{tabular}{cc}
        (e) Base-Pairing Probabilities (This Work) 
        &
        (f) Base-Pairing Probabilities (SAMFEO)
        \\
        \hspace{-.45cm}\raisebox{0.02cm}{
          \begin{tikzpicture}
              \node[anchor=south west,inner sep=0] (image) at (0,0) {\includegraphics[width=0.51\linewidth]{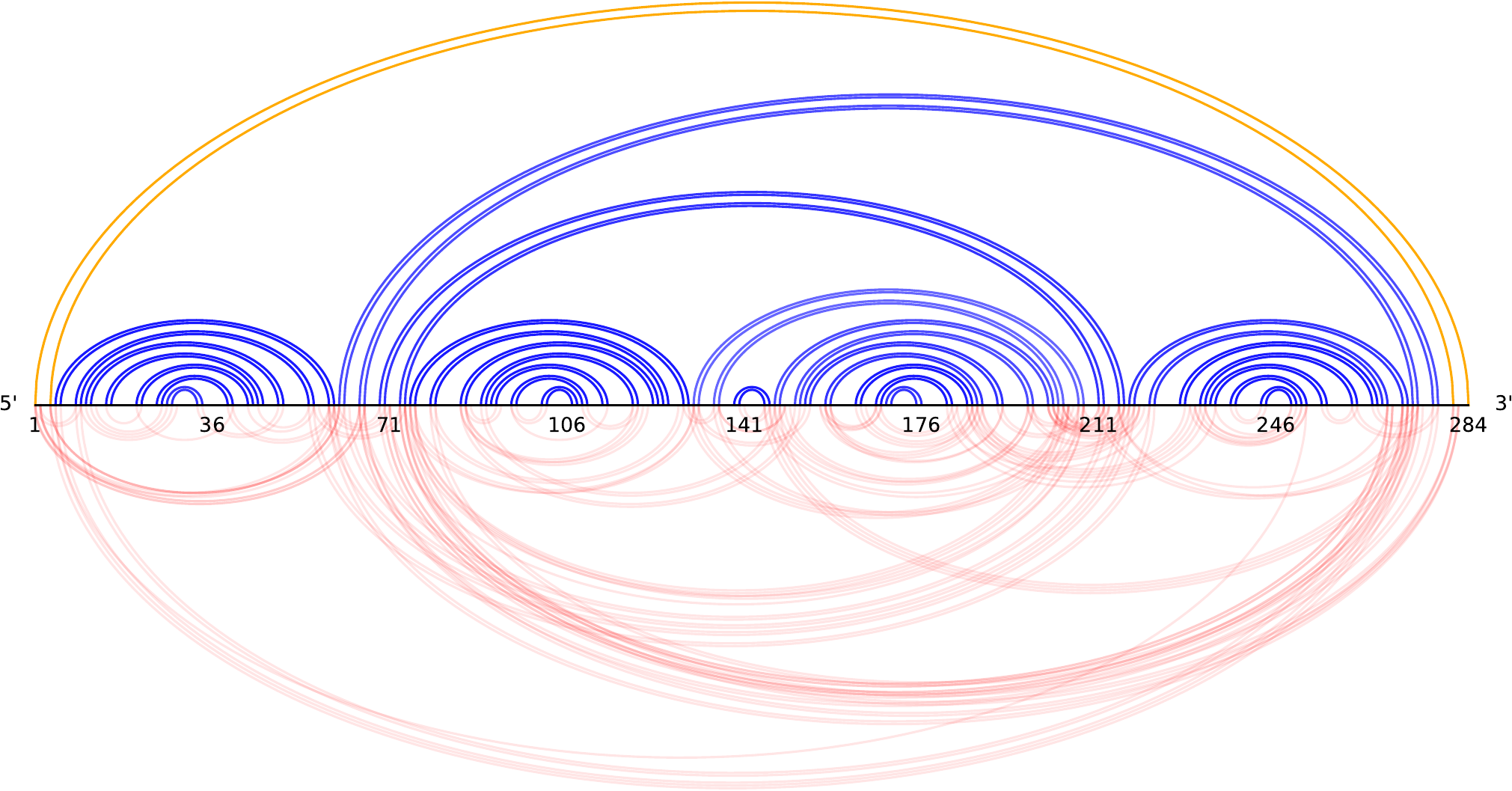}};
              \node at (4.0,4.87) {$1$};
              \node at (4.0,4.52) {$2$};
          \end{tikzpicture}
        }
        &
        \hspace{-.2cm}
        \includegraphics[width=0.51\linewidth]{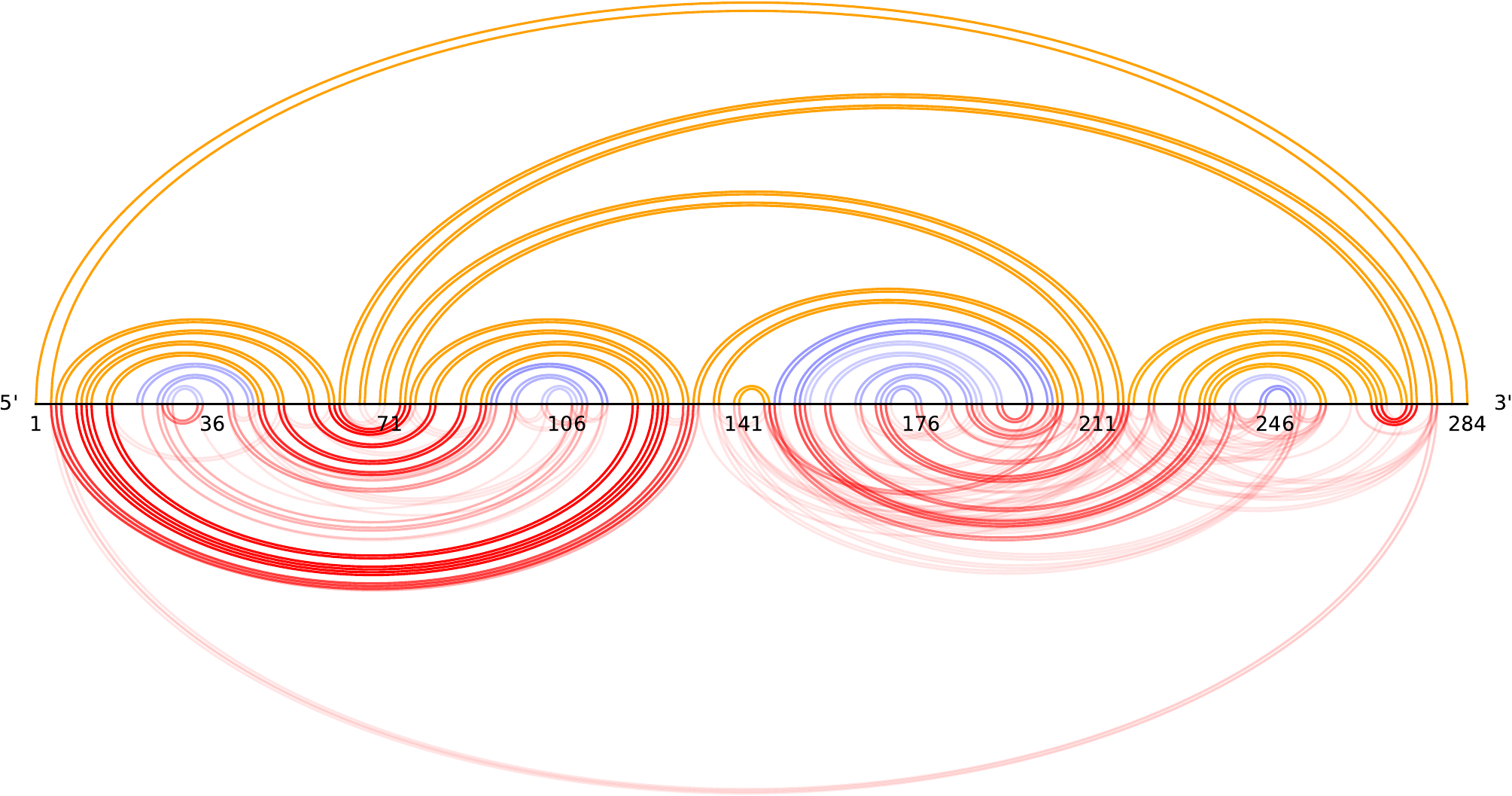}
    \end{tabular}
    \caption{Comparison of the best $p(\vecystar \mid \vecx)$ solution designed by this work vs.~SAMFEO for Puzzle 78 (``Mat - Lot 2-2 B'').
      (b) Target structure: pink-filled regions highlight loops that belong to an undesignable motif, while orange base pairs represent the missing pairs in Sampling's \MFE structure.
      (c) -- (d) \MFE structures of the best $p(\vecystar \mid \vecx)$ solutions from this work and SAMFEO.
      (e) -- (f) Base-pairing probabilities plots. Base-pairs are colored as follows: blue for correct pairs, red for incorrect pairs, with the intensity indicating pairing probability. Orange represents missing correct pairs (i.e.~correct pairs with a pairing probability below 0.1). Nucleotide colors range from blue to red, indicating positional defect. 
      $\vecytilde$ refers to the target structure with the (orange) base pairs from undesignable motifs removed (i.e.~pairs $1$ and $2$ are removed).}
    \label{fig:78}
\end{figure*}

%% file: conclusions.tex


We described a general framework for sampling-based continuous optimization with coupled-variable distributions that is applicable to optimizing any objective function for RNA design. Our method consistently outperformed other state-of-the-art methods across nearly all metrics, such as Boltzmann probability and ensemble defect, particularly on the long and hard-to-design puzzles in the Eterna100 benchmark. 

In the future, our work can be improved in ways such as:

\begin{enumerate}
    \item Our method evaluates the objective for each sample, which takes up the majority of runtime. Currently we cache full sequences (unique samples),
    but to speed up even further, we can cache subsequences shared among the samples. We can also use importance sampling \cite{kahn+harris:1951} to reduce the number of samples in each step.
    \item Our sequence distributions are products of independent distributions of coupled variables, which are both expressive and easy to sample from. But we can also consider more complex distributions using discriminative models \cite{lafferty+:2001}
 and neural models \cite{vaswani+:2017}, although sampling would be more difficult.
    \item We can extend this framework to protein design which should in principle work for arbitrary objective functions. The challenge is scalability due to much slower objective evaluation using protein folding engines \cite{alphafold2:2021,esmfold:2023}.
\end{enumerate}


%% file: appendix.tex
\newpage
\appendix
\onecolumn

  \begin{centering}
          \textbf{\Large Supporting Information}\\
    \vspace{0.5cm}
    \textbf{\Large Sampling-based Continuous Optimization with Coupled Variables for RNA Design}\\
    \vspace{0.5cm}
    \textbf{Wei Yu Tang, Ning Dai, Tianshuo Zhou, David H.~Mathews, and Liang Huang}
    
  \end{centering}

\setcounter{page}{1}

\setcounter{figure}{0}
\renewcommand{\thefigure}{S\arabic{figure}} 

\setcounter{table}{0}
\renewcommand{\thetable}{S\arabic{table}} 

\setcounter{section}{0}
\renewcommand{\thesection}{S\arabic{section}} 
\renewcommand{\thesubsection}{S\arabic{section}.\Alph{subsection}} 

\setcounter{footnote}{0}

\section{Derivation of Gradient}

\label{sec:derivation-gradient}

\subsection{Direct Parameterization}\label{sec:direct-gradient}
In the example of direct parameterization (Sec.~\ref{sec:opt_PGD}), our objective is to compute the gradient of the objective function with respect to the parameter associated with an unpaired position $i$ and nucleotide $\nucA$. As noted in Eq.~\ref{eq:grad-component}, this gradient can be approximated as follows:
\[
  \frac{\partial \obj(\vecTheta)}{\partial \thetau_{i,\nucA}} 
  \approx
  \frac{1}{|\SAMPLES|} \sum_{\vecx \in \SAMPLES} 
  \frac{\partial \log p_{\vecystar} (\vecx; \vecTheta)}
  {\partial \thetau_{i,\nucA}} f(\vecx, \vecystar)
\]
Here, $\log p_{\vecystar} (\vecx; \vecTheta)$ can be expressed as:
\begin{align*}
    \log p_{\vecystar} (\vecx; \vecTheta) 
    &= \log 
    \left( \prod_{i \in \unpaired(\vecy)} \thetau_{i,x_i} 
    \cdot
    \prod_{(i, j) \in \pairs(\vecy)} {\color{pair_color}{\thetap_{i,j,x_i x_j}}} 
    \cdot 
    \prod_{(i, j) \in \mismatches(\vecy)} \hspace{-.5cm} {\color{mismatch_color}{\thetam_{i,j,x_i x_j}}} \cdot \hspace{-.3cm}
    \prod_{(i, j, k) \in \trimismatches(\vecy)} \hspace{-.7cm} {\color{trimismatch_color}{\thetatm_{i,j,k,x_i x_j x_k} }}
    \right) \\
    &= \sum_{i \in \unpaired(\vecy)} \log(\thetau_{i,x_i}) 
    +
    \sum_{(i, j) \in \pairs(\vecy)} \log({\color{pair_color}{\thetap_{i,j,x_i x_j}}}) 
    +
    \sum_{(i, j) \in \mismatches(\vecy)} \hspace{-.5cm} \log({\color{mismatch_color}{\thetam_{i,j,x_i x_j}}}) 
    + \hspace{-.3cm}
    \sum_{(i, j, k) \in \trimismatches(\vecy)} \hspace{-.7cm} \log({\color{trimismatch_color}{\thetatm_{i,j,k,x_i x_j x_k} }})
\end{align*}
Since we are interested in the gradient with respect to a particular position $i$ and nucleotide \nucA, the other terms are constants. This simplifies the equation to:
\begin{align*}
    \frac{\partial \log p_{\vecystar} (\vecx; \vecTheta)}{\partial \thetau_{i,\nucA}} &= \frac{\partial}{\partial \thetau_{i,\nucA}} (\log \thetau_{i, x_i})
    = \frac{1}{\thetau_{i, x_i}} \mathbbm{1}[x_i = \nucA]
\end{align*}
Finally, the gradient approximation can be expressed as:
\begin{align*}
  \frac{\partial \obj(\vecTheta)}{\partial \thetau_{i,\nucA}}
  \approx
  \frac{1}{|\SAMPLES|} \sum_{\vecx \in \SAMPLES} 
  \mathbbm{1}[x_i = A]
  \frac{f(\vecx, \vecystar)}
  {\thetau_{i,\nucA}}
  = 
  \frac{1}{|\SAMPLES|} \sum_{\substack{\vecx \in \SAMPLES\\x_i = A}} 
  \frac{f(\vecx, \vecystar)}
  {\thetau_{i,\nucA}}
\end{align*}
We can also extend the gradient to other cases, including pairs, mismatches, and trimismatches. The gradients are listed below:
\begin{align*}
  \frac{\partial \obj(\vecTheta)}{\partial \thetap_{i,j,ab}}
  &\approx
  \frac{1}{|\SAMPLES|} \sum_{\substack{\vecx \in \SAMPLES\\x_i = a\\x_j=b}} 
  \frac{f(\vecx, \vecystar)}
  {\thetap_{i,j,ab}}
  , \qquad
  \frac{\partial \obj(\vecTheta)}{\partial \thetam_{i,j,ab}}
  \approx
  \frac{1}{|\SAMPLES|} \sum_{\substack{\vecx \in \SAMPLES\\x_i = a\\x_j=b}} 
  \frac{f(\vecx, \vecystar)}
  {\thetam_{i,j,ab}}
  , \qquad
  \frac{\partial \obj(\vecTheta)}{\partial \thetatm_{i,j,k,abc}}
  \approx
  \frac{1}{|\SAMPLES|} \sum_{\substack{\vecx \in \SAMPLES\\x_i = a\\x_j=b\\x_k=c}} 
  \frac{f(\vecx, \vecystar)}
  {\thetatm_{i,j,k,abc}}
\end{align*}

\subsection{Softmax Parameterization}

\label{sec:softmax-gradient}

In Section~\ref{sec:opt_softmax}, we introduce the softmax parameterization as an alternative approach to solving the optimization problem. The probability of a nucleotide $a\in\nucset$ at an unpaired position $i$ is defined in Eq.~\ref{eq:softmax-unpaired} as:
\[
  \pu_i(a; \vecthetau_i) \defeq  \frac{\exp(\thetau_{i,a})}{Z}, \quad \text{where } Z = \sum_{a' \in \nucset} \exp(\thetau_{i,a'})
\]
To compute the gradient of the objective function under the softmax parameterization (Eq.~\ref{eq:unconstraint-gradient}), we need to derive the derivative of the softmax function. Suppose that we are interested in the derivative of $\pu_i(a;\vecthetau_i)$ with respect to the parameter $\thetau_{i, \nucA}$. For the case where $a = \nucA$, applying the quotient rule yields:
\begin{align*}
\frac{\partial \pu_i(\nucA; \vecthetau_i)}{\partial \thetau_{i,\nucA}} &= \frac{\exp(\thetau_{i,\nucA})S - \exp(\thetau_{i,\nucA})\exp(\thetau_{i,\nucA})}{Z^2}
= \frac{\exp(\thetau_{i,\nucA})}{Z} \cdot \frac{Z - \exp(\thetau_{i,\nucA})}{Z}
= \pu_i(\nucA; \vecthetau_i) \cdot (1 - \pu_i(\nucA; \vecthetau_i))
\end{align*}
Similarly, for the case where $a \neq \nucA$, the derivative is given by:
\begin{align*}
  \frac{\partial \pu_i(a; \vecthetau_i)}{\partial \thetau_{i,\nucA}} &= \frac{0 - \exp(\thetau_{i,\nucA})\exp(\thetau_{i,a})}{Z^2}
  = -\frac{\exp(\thetau_{i,\nucA})}{Z} \cdot \frac{\exp(\thetau_{i,a})}{Z}
  = -\pu_i(\nucA; \vecthetau_i)\pu_i(a; \vecthetau_i)
  = \pu_i(a; \vecthetau_i)(0-\pu_i(\nucA; \vecthetau_i))
\end{align*}
Combining these results, we can express the derivative of the softmax function with respect to $\thetau_{i,\nucA}$ as:
\[
  \frac{\partial \pu_i(a; \vecthetau_i)}{\partial \thetau_{i,\nucA}} 
  = \pu_i(a; \vecthetau_i) 
  \cdot 
  (\mathbbm{1}[a = \nucA] - \pu_i(\nucA; \vecthetau_i))
\]
For other cases, such as pairs, mismatches, and trimismatches, the derivatives of the softmax function are provided below:
\begin{gather*}
  \frac{\partial \pp_{i,j}(ab; \vecthetap_{i,j})}{\partial \thetau_{i,j,a'b'}} 
  = \pp_{i,j}(ab; \vecthetap_{i,j}) 
  \cdot 
  (\mathbbm{1}[ab = a'b'] - \pp_{i,j}(a'b'; \vecthetap_{i,j}))
  , \qquad
  \frac{\partial \pm_{i,j}(ab; \vecthetam_{i,j})}{\partial \thetau_{i,j,a'b'}} 
  = \pm_{i,j}(ab; \vecthetam_{i,j}) 
  \cdot 
  (\mathbbm{1}[ab = a'b'] - \pm_{i,j}(a'b'; \vecthetam_{i,j}))
  ,\\
  \frac{\partial \ptm_{i,j,k}(abc; \vecthetatm_{i,j,k})}{\partial \thetau_{i,j,k,a'b'c'}} 
  = \ptm_{i,j,k}(abc; \vecthetatm_{i,j,k}) 
  \cdot 
  (\mathbbm{1}[abc = a'b'c'] - \ptm_{i,j,k}(a'b'c'; \vecthetatm_{i,j,k}))
\end{gather*}

\begin{table}[!h]
  \caption{Results of different RNA Design methods on the shortest structures in Eterna100 (up to 50 and 104 nucleotides, respectively).}
  \begin{tabular}{c|l|l|@{\hskip .2cm}cc c cccc}
    & Methods                        & Objective & \multicolumn{2}{c}{${p}(\vecystar \mid \vecx)(\uparrow)$}  
    & $\NED(\vecx, \vecystar)$ & $d(\MFE(\vecx), \vecystar)$ & $\DDG(\vecx, \vecystar)$ & \# of & \# of \\
    &  & & mean & geom.$\dagger$  & $(\downarrow)$  & $(\downarrow)$ & $(\downarrow)$ &  MFE $(\uparrow)$ & uMFE $(\uparrow)$ \\ 
    
    \hline
    \hline

    \multirow{4}{*}{\rotatebox[origin=c]{90}{Length $\leq 50$}} & Matthies et al.~\cite{matthies+:2023} & $-\! \log\!\left[ \frac{\E_{\vecx}[e^{-\freeenergy(x, y) / RT}]}{\E_{\vecx}[Q(x)]}\! \right]$\!\! 
    &  0.545 & 0.088 & {0.120} & {3.72} & {1.12} & {11} & {11} \\[2pt] 
    
    \cline{2-10}
  
    & SAMFEO                         & $1 - p(\vecystar \mid \vecx)$ & 0.712 & \underline{0.314} & \textbf{0.046} & \textbf{0.56} & \textbf{0.49} & \textbf{16} & \textbf{16} \\[2pt] 
    
    \cline{2-10}
    
    & This Work ({\em projection}) & $\E_{\vecx} [-\log p(\vecystar \mid \vecx)]$ & \underline{0.715} & \textbf{0.317} & 0.049 & \underline{0.78} & \underline{0.50} & \underline{15} & 14 \\[2pt] 
    
    \cline{2-10}
  
    & This Work ({\em softmax})    & $\E_{\vecx} [-\log p(\vecystar \mid \vecx)]$ & \textbf{0.716} & \textbf{0.317} & \underline{0.048} & 0.83 & 0.52 & \underline{15} & \underline{15} \\[2pt] 

  \hline 
  \hline
  
    \multirow{4}{*}{\rotatebox[origin=c]{90}{Length $\leq 104$}} & Matthies et al.~\cite{matthies+:2023} & $-\! \log\!\left[ \frac{\E_{\vecx}[e^{-\freeenergy(x, y) / RT}]}{\E_{\vecx}[Q(x)]}\! \right]$\!\! 
    &  0.447 & 0.135 & \textit{0.110} & \textit{6.08} & \textit{1.38} & \textit{27} & \textit{27} \\[2pt] 
    
    \cline{2-10}
  
    & SAMFEO                         & $1 - p(\vecystar \mid \vecx)$ & 0.675 & 0.699 & \textbf{0.038} & \textbf{1.08} & 0.37 & \textbf{42} & \textbf{40} \\[2pt] 
    
    \cline{2-10}
    
    & This Work ({\em projection}) & $\E_{\vecx} [-\log p(\vecystar \mid \vecx)]$ & \textbf{0.681} & \textbf{0.722} & \underline{0.040} & \underline{1.35} & \textbf{0.33} & \textbf{42} & \underline{39} \\[2pt] 
    
    \cline{2-10}
  
    & This Work ({\em softmax})    & $\E_{\vecx} [-\log p(\vecystar \mid \vecx)]$ & \underline{0.680} & \underline{0.717} & 0.041 & 1.43 & \underline{0.34} & \underline{41} & \textbf{40} \\[2pt] 
    
  \hline
  \end{tabular}
  \label{tab:104nucs}
\end{table}

    
  
    
  
    
    
    
  
    

\begin{figure*}[!htb]
  \centering
  \begin{tabular}{c}
      (a) $p(\vecystar \mid \vecx)(\uparrow)$ of puzzles up to 104 nucleotides.
      \\
      \includegraphics[width=0.7\linewidth]{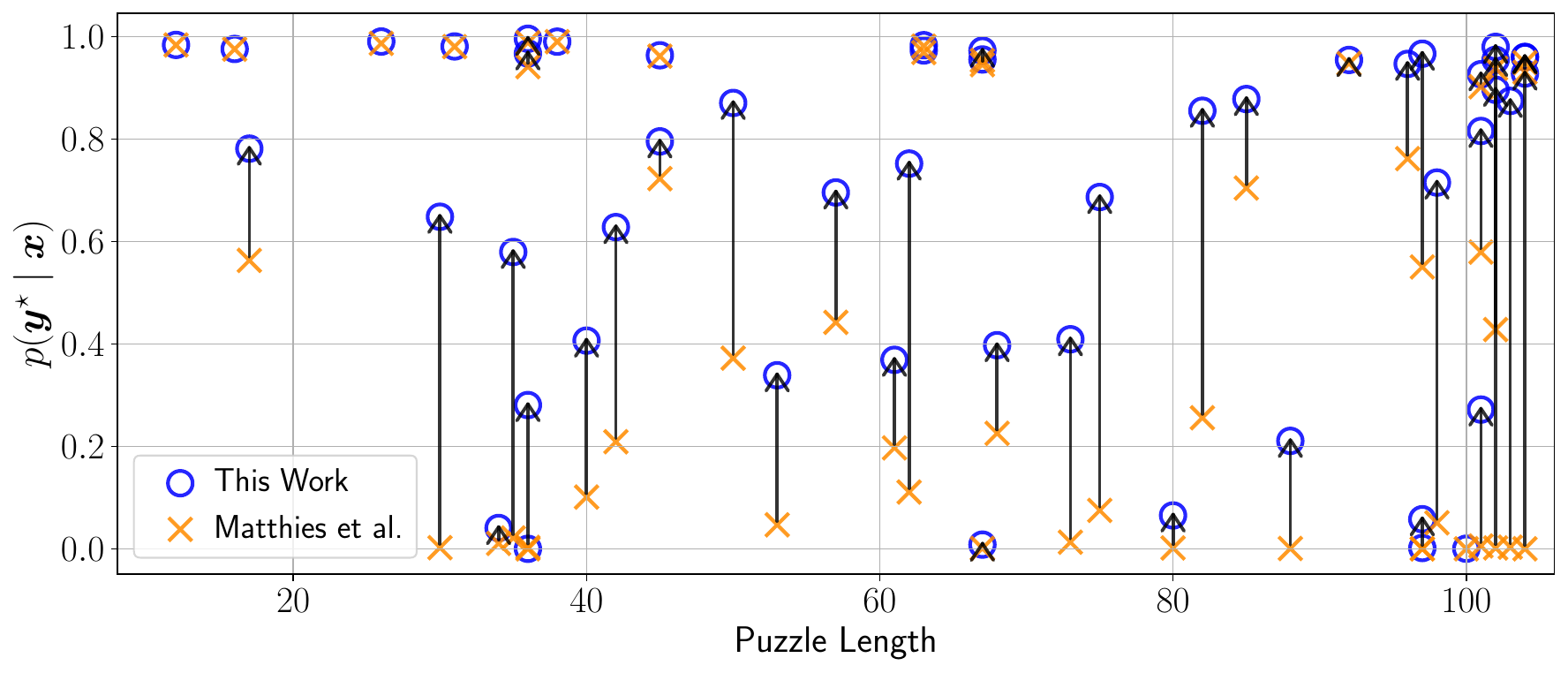}
  \end{tabular}
  \caption{$p(\vecystar \mid \vecx)$ of solutions designed by this work vs.~Matthies et al.~\cite{matthies+:2023} on the 51 shortest structures in Eterna100 (up to 104 nucleotides).}
  \label{fig:max}
\end{figure*}


\input{fig-si-mainresults}
\input{fig-si-nemo}

\begin{table*}[!h]
  \centering
  \caption{Comparison of solutions designed by this work vs.~SAMFEO on some long and hard-to-design Eterna100 puzzles.}
  \begin{tabular}{l|l|cccccc}
    Puzzle & Method & $p(\vecystar \mid \vecx)$ & $\NED(\vecx, \vecystar)$ & $d(\MFE(\vecx), \vecystar)$ & $\DDG(\vecx, \vecystar)$ & is MFE & is uMFE \\ 
    & & $(\uparrow)$ & $(\downarrow)$ & $(\downarrow)$ & $(\downarrow)$ & & \\
    \hline
    \hline

    \texttt{\#73} (370 \nts)& This Work & \textbf{0.005} & \textbf{0.055} & \;\,\, \textbf{0} & \, \textbf{0.0 kcal/mol} & \textbf{Yes} & \textbf{Yes} \\
    Figure \ref{fig:73} & SAMFEO & $3 \times 10^{-27}$ & 0.417 & 138 & 33.2 kcal/mol & No & No \\

    \hline

    \texttt{\#76} (393 \nts) & This Work & \textbf{0.035} & \textbf{0.054} & \; \textbf{0} & \; \textbf{0.0 kcal/mol} & \textbf{Yes} & \textbf{Yes} \\
    Figure \ref{fig:76} & SAMFEO & $7 \times 10^{-8}$ & 0.239 & 26 & \; 4.2 kcal/mol & No & No \\

    \hline

    \texttt{\#78} (284 \nts) & This Work & \textbf{0.001} & \textbf{0.123} & \;\,\, \textbf{4} & \; \textbf{2.6 kcal/mol} & No & No \\
    Figure \ref{fig:78} & SAMFEO & $9 \times 10^{-25}$ & 0.452 & 140 & 29.2 kcal/mol & No & No \\

    \hline

    \texttt{\#91} (392 \nts) & This Work & \textbf{0.0001} &\textbf{0.034} & \; \textbf{8} & \; \textbf{3.4 kcal/mol} & No & No \\
    Figure \ref{fig:91} & SAMFEO & $2 \times 10^{-20}$ & 0.128 & 52 & 25.2 kcal/mol & No & No \\

    \hline

    \texttt{\#99} (364 \nts) & This Work & $\mathbf{8 \times 10^{-11}}$ & \textbf{0.111} & \textbf{20} & \; \textbf{9.8 kcal/mol} & No & No\\
    Figure \ref{fig:99} & SAMFEO & $3 \times 10^{-28}$ & 0.197 & 80 & 34.4 kcal/mol & No & No\\

    \hline
    \hline

  \end{tabular}
  \label{tab:comparison}
\end{table*}

\input{fig-si-76}
\input{fig-si-91}
\input{fig-si-99}

\input{fig-si-steps}

\input{fig-si-learning}

\begin{table*}[!h]
  \centering
  \caption{Ablation studies. Results of this work (\textit{softmax}) optimizing for $p(\vecystar \mid \vecx)$ without mismatch and trimismatch. Targeted initialization only. $\dagger$: geometric mean without 18 undesignable puzzles.}
  \label{tab:ncrna_results}
  \resizebox{.99\textwidth}{!}{
  \begin{tabular}{l|l|ccccccc}
    Methods & Distribution & \multicolumn{2}{c}{${p}(\vecystar \mid \vecx)(\uparrow)$} & $\NED(\vecx, \vecystar)$ & $d(\MFE(\vecx), \vecystar)$ & $\DDG(\vecx, \vecystar)$ & \# of & \# of \\
    & & mean & geom.$\dagger$ & $(\downarrow)$ & $(\downarrow)$ & $(\downarrow)$ & \MFE$(\uparrow)$ & \UMFE$(\uparrow)$ \\ 
      
    \hline 
    \hline

    This Work  & v3: default & \textbf{0.589} & \textbf{0.502} & \textbf{0.035} & \textbf{2.96} & \textbf{0.81} & \textbf{78} & \textbf{75} \\ [2pt] 
                        & v2: base-pair \& mismatch & 0.578 & 0.467 & 0.039 & 3.19 & 0.94 & 75 & 72 \\ [2pt]
                        & v1: only base-pair & 0.566 & 0.422 & 0.042 & 3.93 & 0.97 & 75  & 70  \\ [2pt] 
                                    
    \hline
    \hline
  \end{tabular}
  }
\label{tab:ablation}
\end{table*}

%% file: fig-si-mainresults.tex
\begin{figure*}[!htb]
    \centering
    \setlength{\tabcolsep}{2pt}
    \begin{tabular}{cc}
        (a) Arithmetic mean of $p(\vecystar \mid \vecx)(\uparrow)$, grouped by length
        &
        (b) Average $\DDG(\vecx, \vecystar)(\downarrow)$, grouped by length
        \\
        \includegraphics[width=0.45\linewidth]{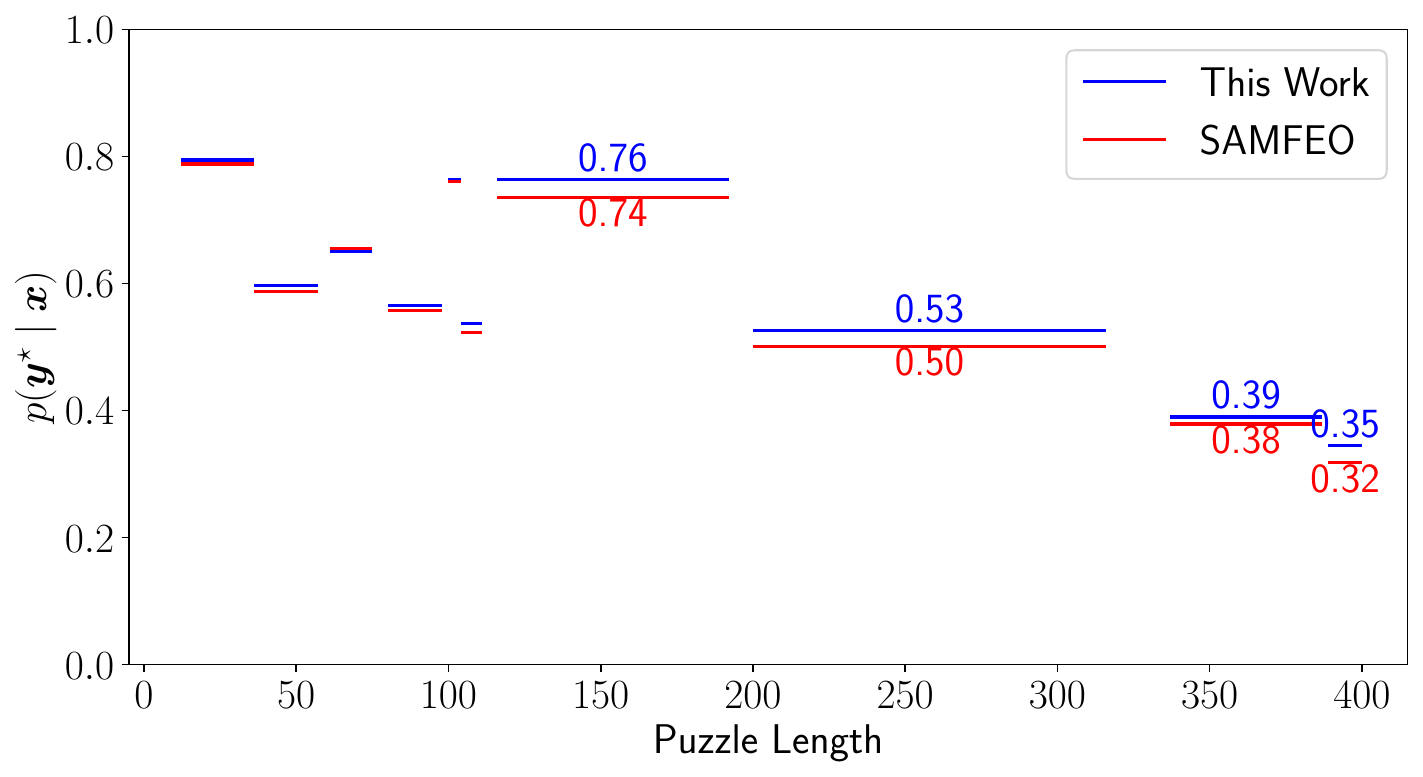}
        &
        \includegraphics[width=0.45\linewidth]{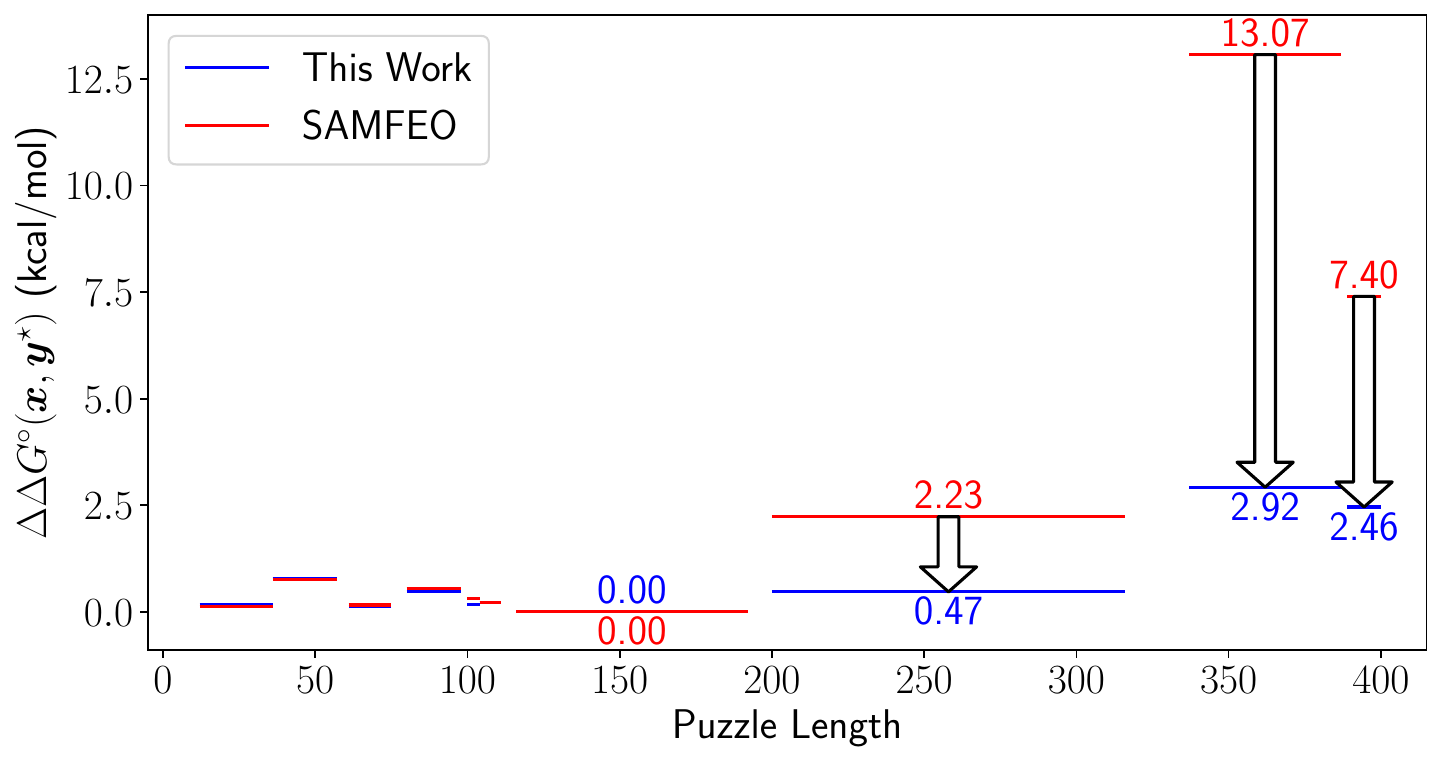}
        \\[8pt]
        (c) Average $d(\MFE(\vecx), \vecystar)(\downarrow)$, grouped by length
        &
        (d) $\NED(\vecx, \vecystar)(\downarrow)$ of individual puzzles
        \\
        \hspace{-.2cm}
        \begin{tikzpicture}
            \node [] {\includegraphics[width=0.45\linewidth]{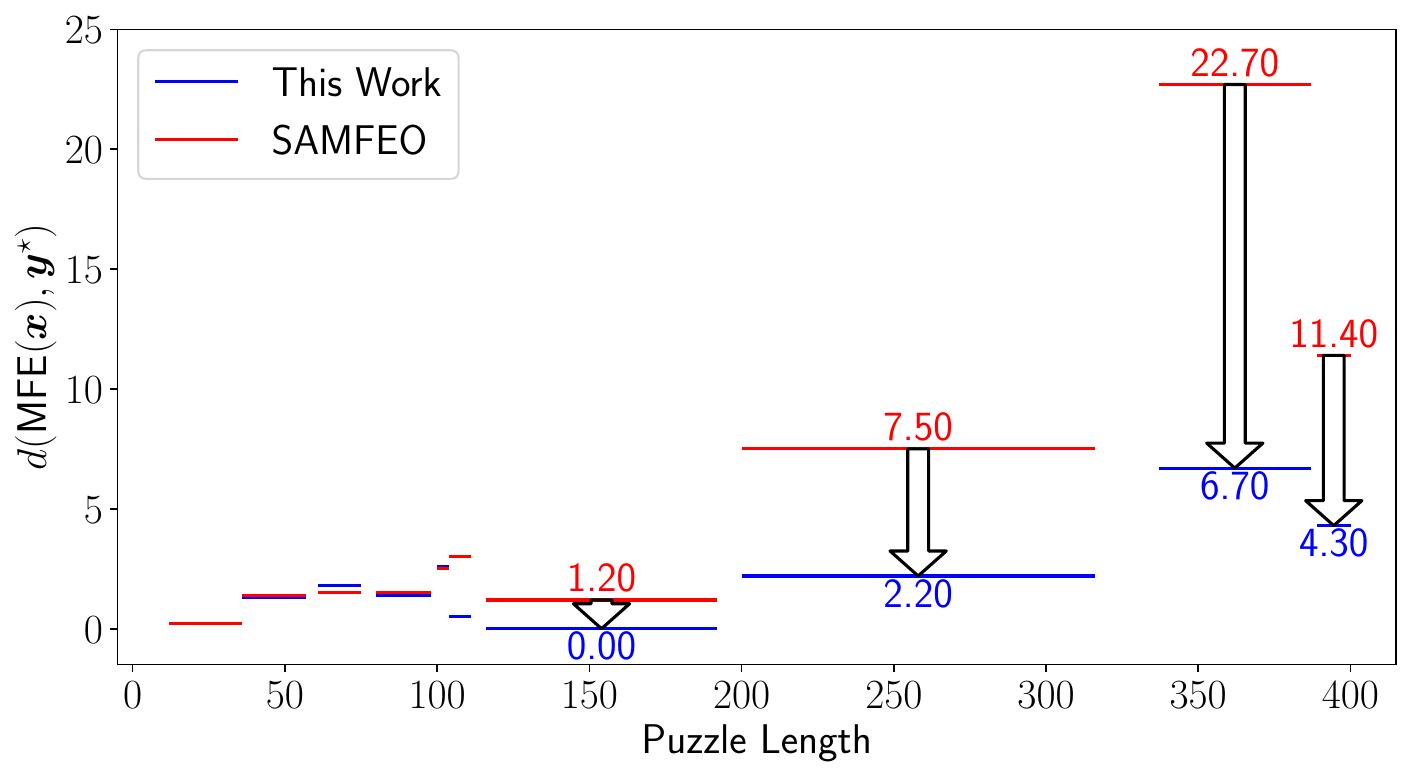}};
        \end{tikzpicture}
        &
        \begin{tikzpicture}
            \node [] {\includegraphics[width=0.45\linewidth]{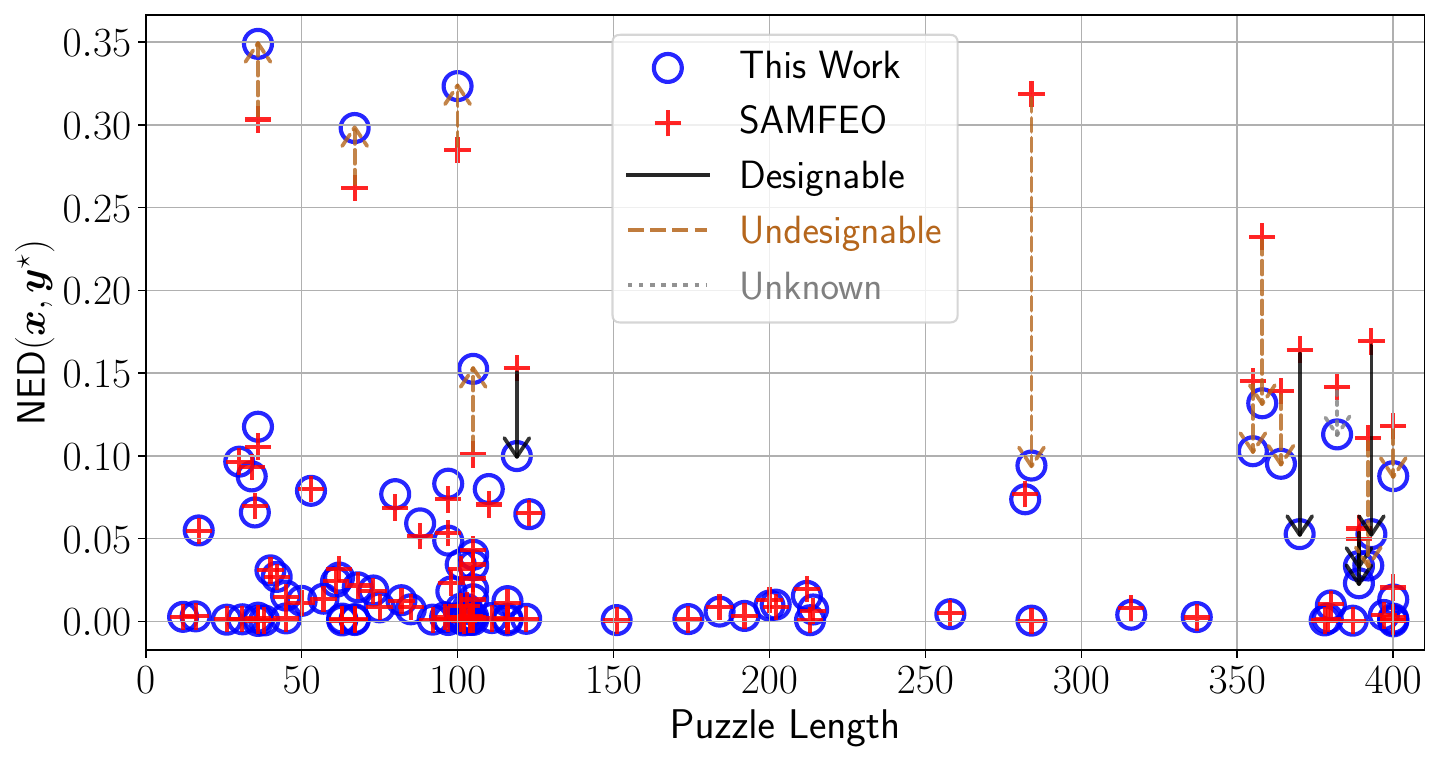}};
            \node at (1.4,1.7) {\small \hyperref[fig:78]{\color{undesignable}{\texttt{{\#78$^\star$}}}}};
            \node at (2.8,1.0) {\small \color{undesignable}{\texttt{\#96}}};
            \node at (3.2,0.35) {\small \texttt{\hyperref[fig:73]{\color{designable}{{\#73$^\star$}}}}};
            \node at (3.85,0.4) {\small \texttt{\hyperref[fig:76]{\color{designable}{{\#76$^\star$}}}}};
        \end{tikzpicture}
    \end{tabular}
    \caption{(a) -- (c) Average of metrics when puzzles are grouped by length, with each group consisting of 10 puzzles. (d) $\NED(\vecx, \vecystar)$ of solutions designed by this work vs.~SAMFEO. Figure \ref{fig:main_result} displays similar grouped-by-length plots and scatterplots for other metrics.}
    \label{fig:si-main-results}
\end{figure*}

%% file: fig-si-nemo.tex
\begin{figure*}[!h]
    \centering   
    (a)  \begin{tabular}{c|c|c|c|c|c|c}
      \texttt{\#71} (88 \nts) & $p(\vecystar \mid \vecx)$ & NED$(\vecx, \vecystar)$ & $d(\MFE(\vecx), \vecystar)$ & $\DDG(\vecx, \vecystar)$ & is MFE & is uMFE\\
      & $(\uparrow)$ & $(\downarrow)$ & $(\downarrow)$ & $(\downarrow)$ & & \\
      \hline
      This Work & \textbf{0.211} & \textbf{0.059} & 4 & 0.4 kcal/mol & No & No\\
      NEMO & 0.051 & 0.180 & \textbf{0} & \textbf{0.0 kcal/mol} & \textbf{Yes} & \textbf{Yes}\\
    \end{tabular}
    
    \includegraphics[width=.7\linewidth]{./figs/color_legend-crop}\\[6pt]
      \begin{tabular}{cc}
          (b) Target Structure \vecystar and \MFE Structure (NEMO)
          &
          (c) MFE Structure $\vecyhat_1$ (This Work)
          \\
          \begin{tikzpicture}
            \node[anchor=south west,inner sep=0] (image) at (0,0) {\includegraphics[width=0.52\linewidth]{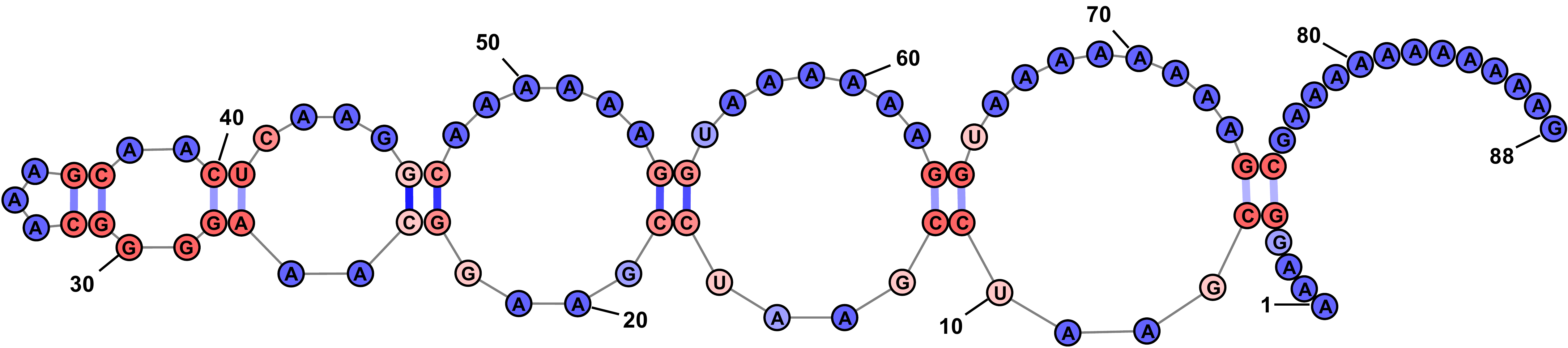}};
            \node at (7.1, 0.95) {$1$};
            \node at (7.75, 0.95) {$2$};
            \draw[dashed] (1.49,1.3) rectangle (0.0,0.5) node[] {};
          \end{tikzpicture}
          &
          \raisebox{-0.0cm}{
            \begin{tikzpicture}
              \node[anchor=south west,inner sep=0] (image) at (0,0) {\includegraphics[width=0.46\linewidth]{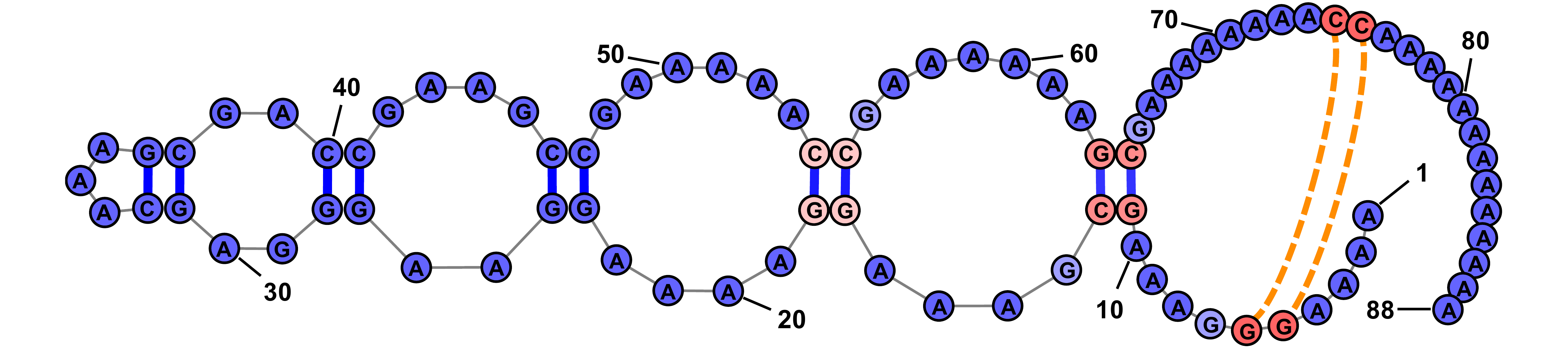}};
              \node at (6.7, 1.2) {$1$};
              \node at (7.2, 1.1) {$2$};
              \draw[dashed] (1.95,1.33) rectangle (0.3,0.42) node[] {};
            \end{tikzpicture}
          }
          \\
          (d) Base Pairing Probabilities (NEMO)
          &
          (e) Base Pairing Probabilities (This Work)
          \\
          \begin{tikzpicture}
            \node[anchor=south west,inner sep=0] (image) at (0,0) {\includegraphics[width=0.45\linewidth]{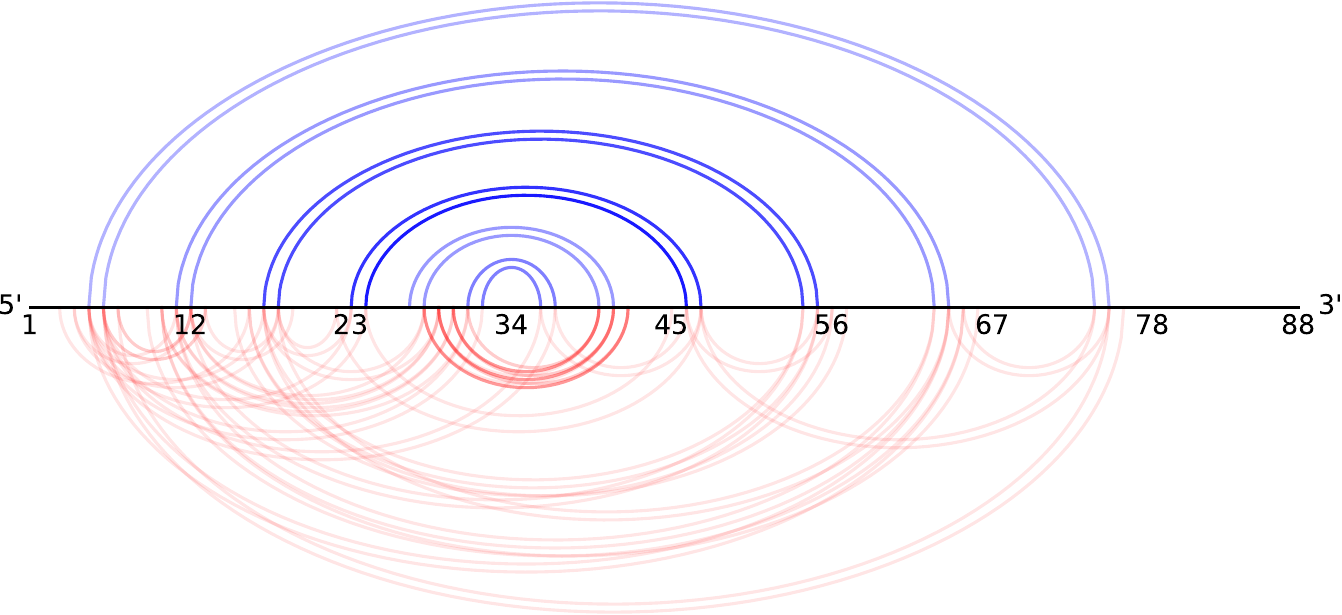}};
            \node at (5.8, 2.8) {$1$};
            \node at (6.1, 3.1) {$2$};
            \draw[dashed] (2.4,2.35) rectangle (3.8,1.3) node[] {};
          \end{tikzpicture}
          &
          \raisebox{0.02cm} {
            \begin{tikzpicture}
              \node[anchor=south west,inner sep=0] (image) at (0,0) {\includegraphics[width=0.45\linewidth]{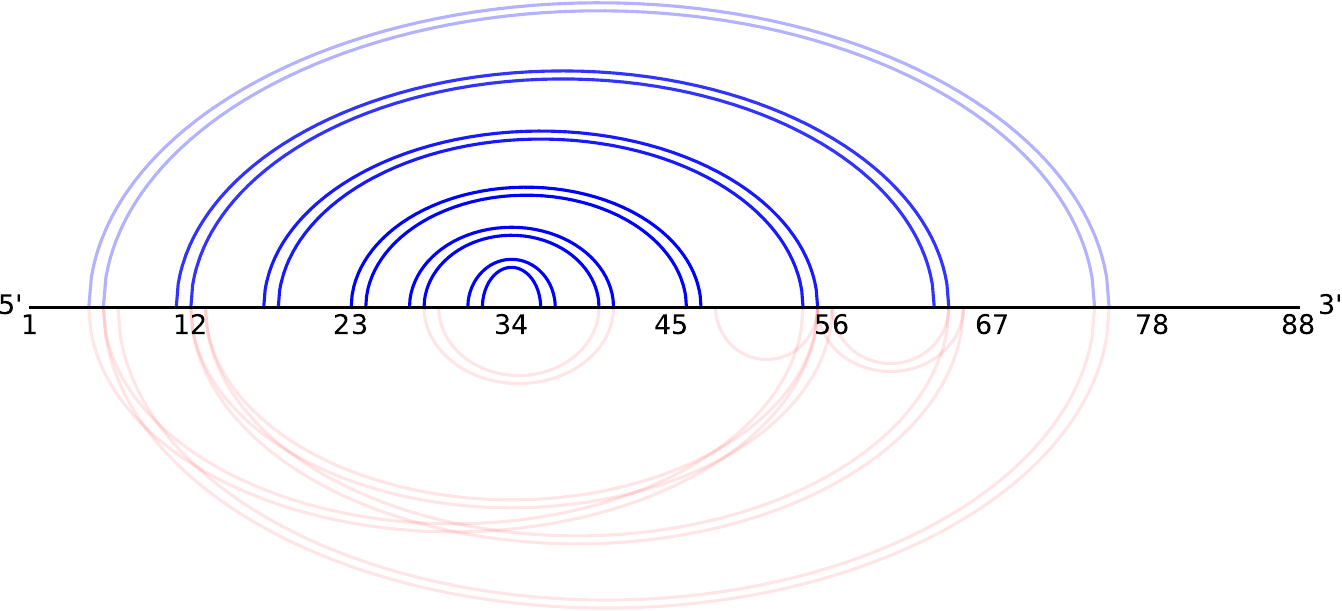}};
              \node at (5.8, 2.8) {$1$};
              \node at (6.1, 3.1) {$2$};
              \draw[dashed] (2.4,2.35) rectangle (3.8,1.3) node[] {};
            \end{tikzpicture}
          }        
      \end{tabular}
      \\[6pt]

      \hrulefill
      \\[6pt]

      (f)  \begin{tabular}{c|c|c|c|c|c|c}
        \texttt{\#79} (101 \nts) & $p(\vecystar \mid \vecx)$ & NED$(\vecx, \vecystar)$ & $d(\MFE(\vecx), \vecystar)$ & $\DDG(\vecx, \vecystar)$ & is MFE & is uMFE\\
        & $(\uparrow)$ & $(\downarrow)$ & $(\downarrow)$ & $(\downarrow)$ & & \\
        \hline
        This Work & \textbf{0.272} & \textbf{0.041} & 4 & \textbf{0.0 kcal/mol} & \textbf{Yes} & No \\
        NEMO & 0.092 & 0.113 & \textbf{0} & \textbf{0.0 kcal/mol} & \textbf{Yes} & \textbf{Yes} \\
    \end{tabular}\\[6pt]

      \begin{tabular}{cc}
          (g) Target Structure \vecystar and \MFE Structure (NEMO)
          &
          (h) MFE Structure $\vecyhat_2$ (This Work)
          \\
          \begin{tikzpicture}
            \node[anchor=south west,inner sep=0] (image) at (0,0) {\includegraphics[width=0.45\linewidth]{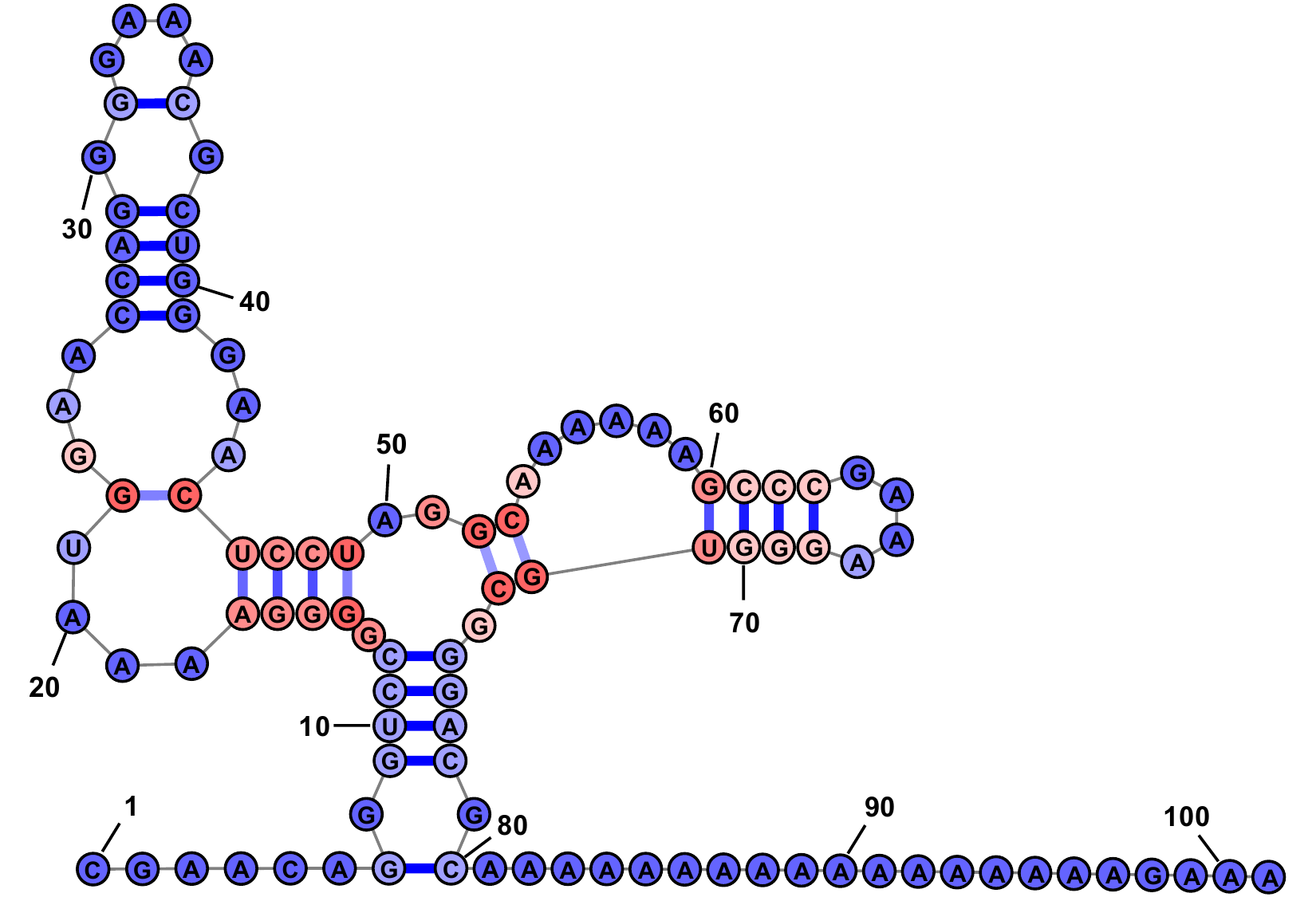}};
            \node at (2.7, 2.0) {$1$};
            \node at (3.3, 2.2) {$2$};
            \draw[dashed] (1.38,2.25) rectangle (2.2,1.58) node[] {};
          \end{tikzpicture}
          &
          \!\!\!
          \begin{tikzpicture}
            \node[anchor=south west,inner sep=0] (image) at (0,0) {\includegraphics[width=0.47\linewidth]{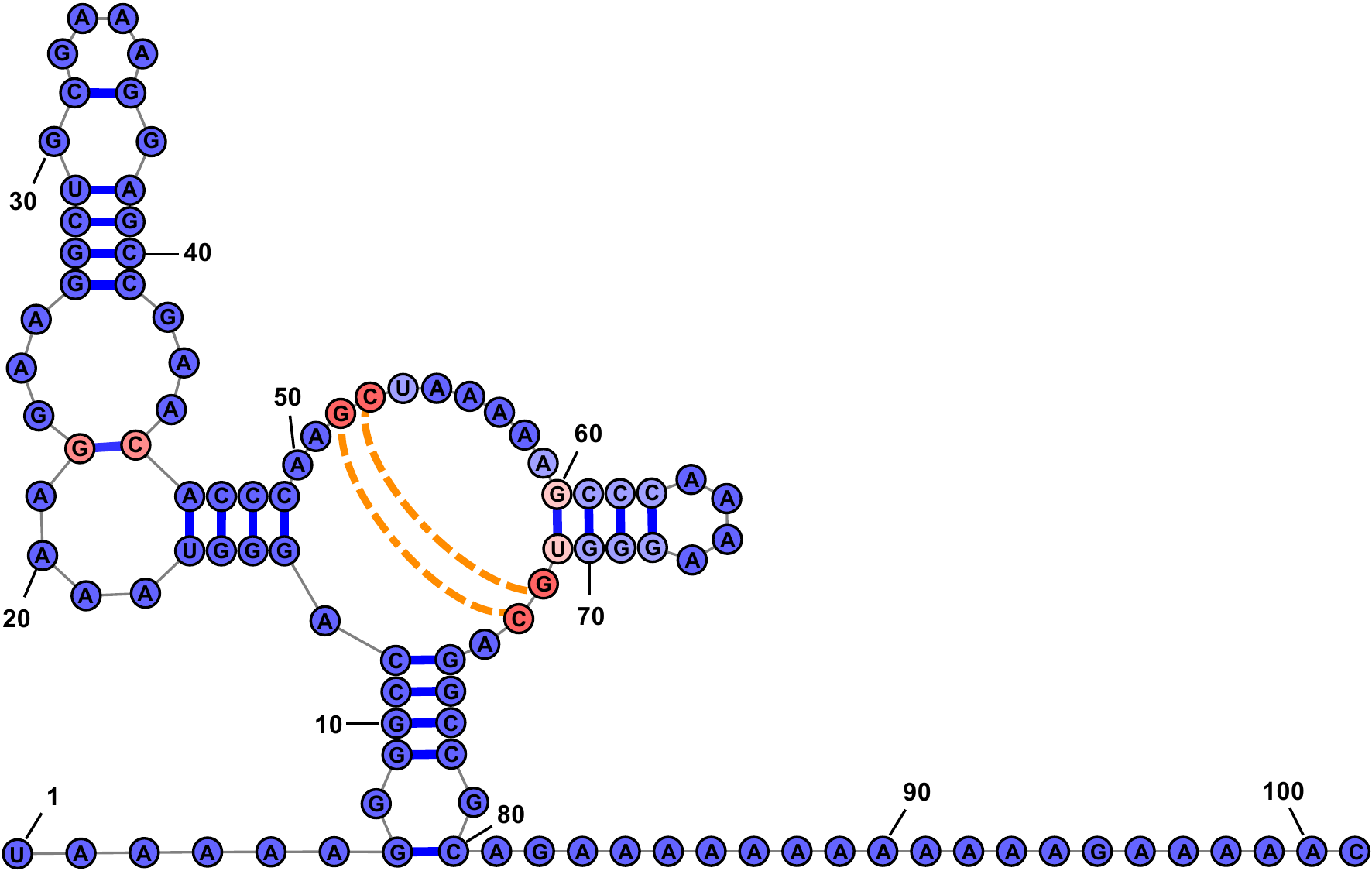}};
            \node at (2.4, 1.75) {$1$};
            \node at (2.9, 2.2) {$2$};
            \draw[dashed] (1.0,2.4) rectangle (1.9,1.8) node[] {};
          \end{tikzpicture}
          \\
          (i) Base Pairing Probabilities (NEMO)
          &
          (j) Base Pairing Probabilities (This Work)
          \\
          \raisebox{0.15cm} {
            \begin{tikzpicture}
              \node[anchor=south west,inner sep=0] (image) at (0,0) {\includegraphics[width=0.49\linewidth]{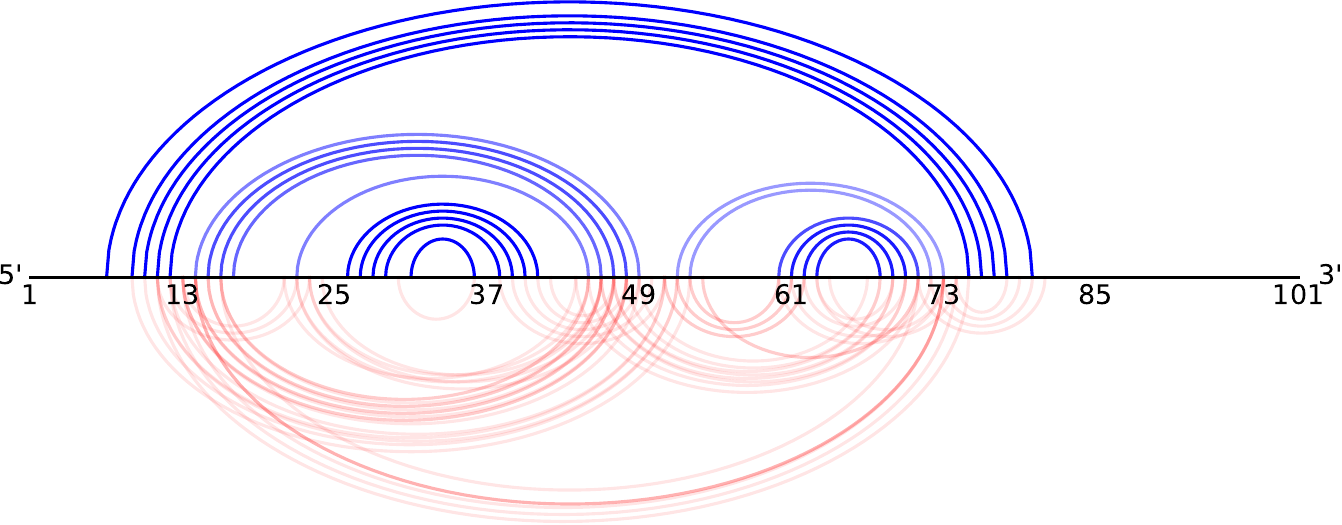}};
              \node at (4.85, 2.35) {$1$};
              \node at (5.0, 2) {$2$};
              \draw[dashed] (1.2,2.55) rectangle (4.2,0.65) node[] {};
            \end{tikzpicture}
          }
          &
          \!\!\!
          \begin{tikzpicture}
            \node[anchor=south west,inner sep=0] (image) at (0,0) {\includegraphics[width=0.49\linewidth]{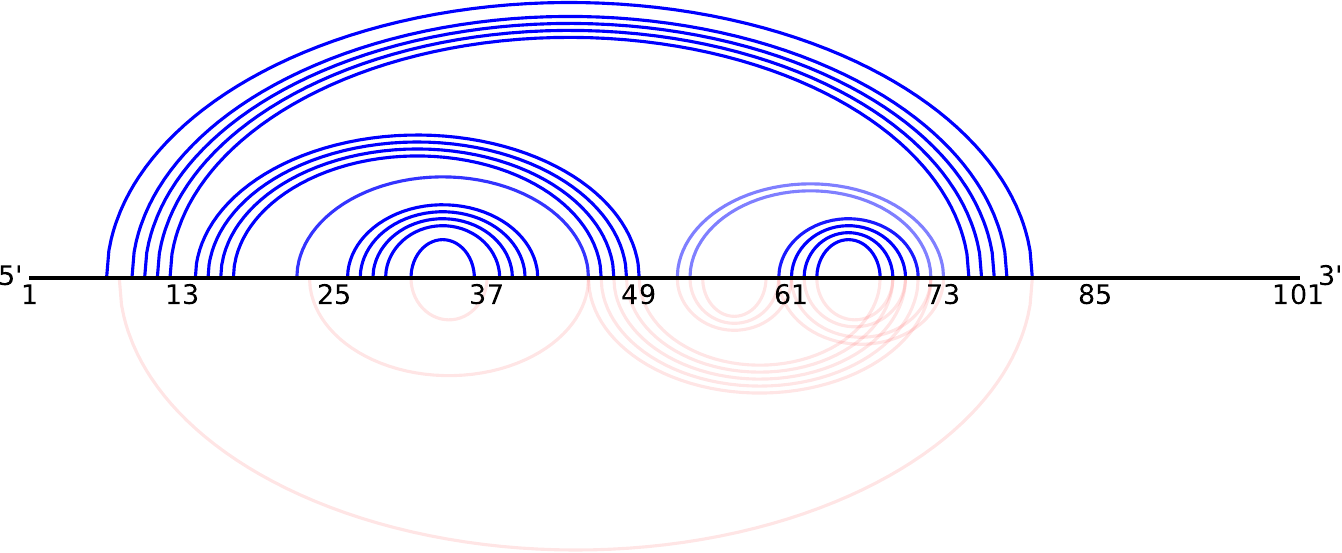}};
            \node at (4.85, 2.5) {$1$};
            \node at (5.0, 2.2) {$2$};
            \draw[dashed] (1.2,2.7) rectangle (4.2,0.8) node[] {};
          \end{tikzpicture}
          \\
      \end{tabular}
      \caption{Puzzles \texttt{\#71}, \texttt{\#79} are solved by NEMO (but not by this work) under the UMFE criterion. MFE Structures: Base pairs are colored blue for correct pairs and red for incorrect pairs, with color intensity indicating pairing probability. Nucleotides are colored using a blue-to-red gradient to represent positional defects. Orange dashed lines in our \MFE structure indicate the missing pairs from target structure. The dashed box highlights regions where NEMO's solution shows poor positional defects and many base-pairing competitions due to alternative structures. For puzzle \texttt{\#79}, although our design is not a \UMFE solution, it is still an \MFE solution, meaning that $\vecyhat_2 \in \MFE_s(\vecx)$ and $\freeenergy(\vecx, \vecystar) = \freeenergy(\vecx, \vecyhat_2)$.}
      \label{fig:nemo_examples}
\end{figure*}

%% file: fig-si-76.tex
\begin{figure*}[!h]
    \centering   
      (a)  \begin{tabular}{c|c|c|c|c|c|c}
        \texttt{\#76} (393 \nts) & $p(\vecystar \mid \vecx)$ & NED$(\vecx, \vecystar)$ & $d(\MFE(\vecx), \vecystar)$ & $\DDG(\vecx, \vecystar)$ & is MFE & is uMFE\\ 
        & $(\uparrow)$ & $(\downarrow)$ & $(\downarrow)$ & $(\downarrow)$ & & \\
        \hline
  
        This Work & \textbf{0.035} & \textbf{0.054} & \; \textbf{0} & \;\textbf{0.0 kcal/mol} & \textbf{Yes} & \textbf{Yes}\\
        SAMFEO & $7 \times 10^{-8}$ & 0.239 & 26 & 4.2 kcal/mol & No & No\\
      \end{tabular}
    
      \includegraphics[width=.8\linewidth]{./figs/color_legend-crop}
      \begin{tabular}{cc}
          (b) Target Structure \vecystar and MFE Structure (This Work)
          &
          (c) MFE Structure (SAMFEO)
          \\
          \includegraphics[width=0.45\linewidth]{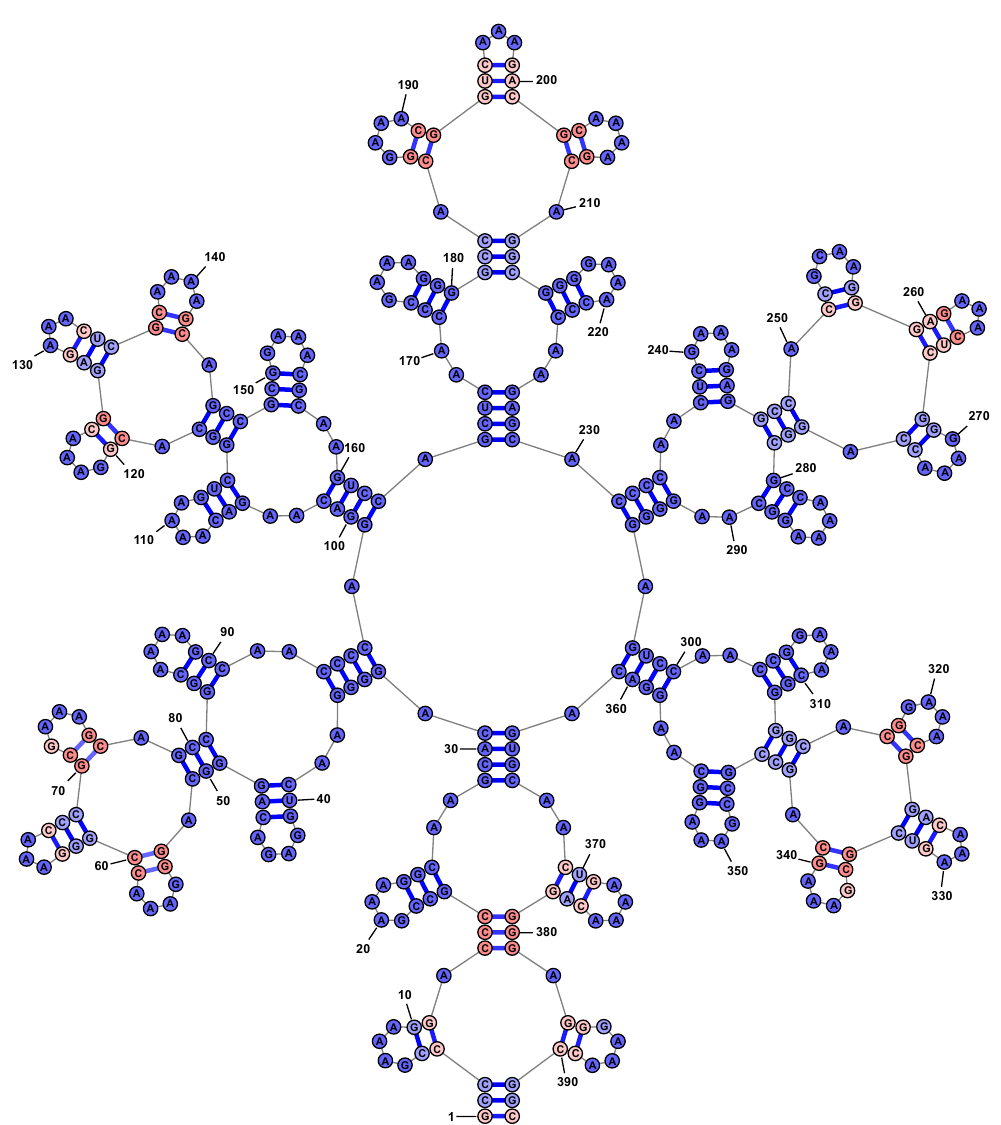}
          &
          \includegraphics[width=0.45\linewidth]{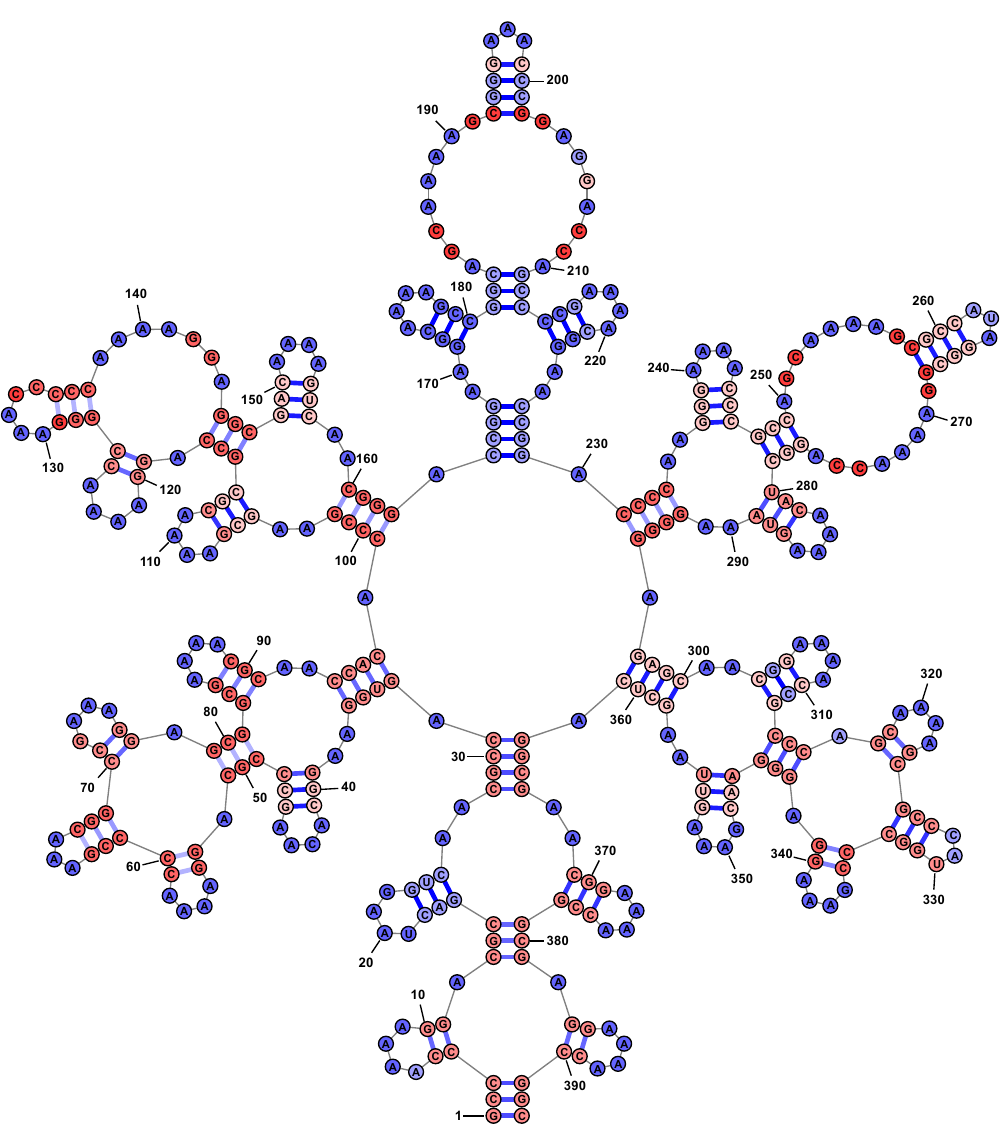}
          \\[6pt]
          (d) Base-Pairing Probabilities (This Work)
          &
          (e) Base-Pairing Probabilities (SAMFEO)
          \\
          \hspace{-0.5cm}
          \includegraphics[width=0.51\linewidth]{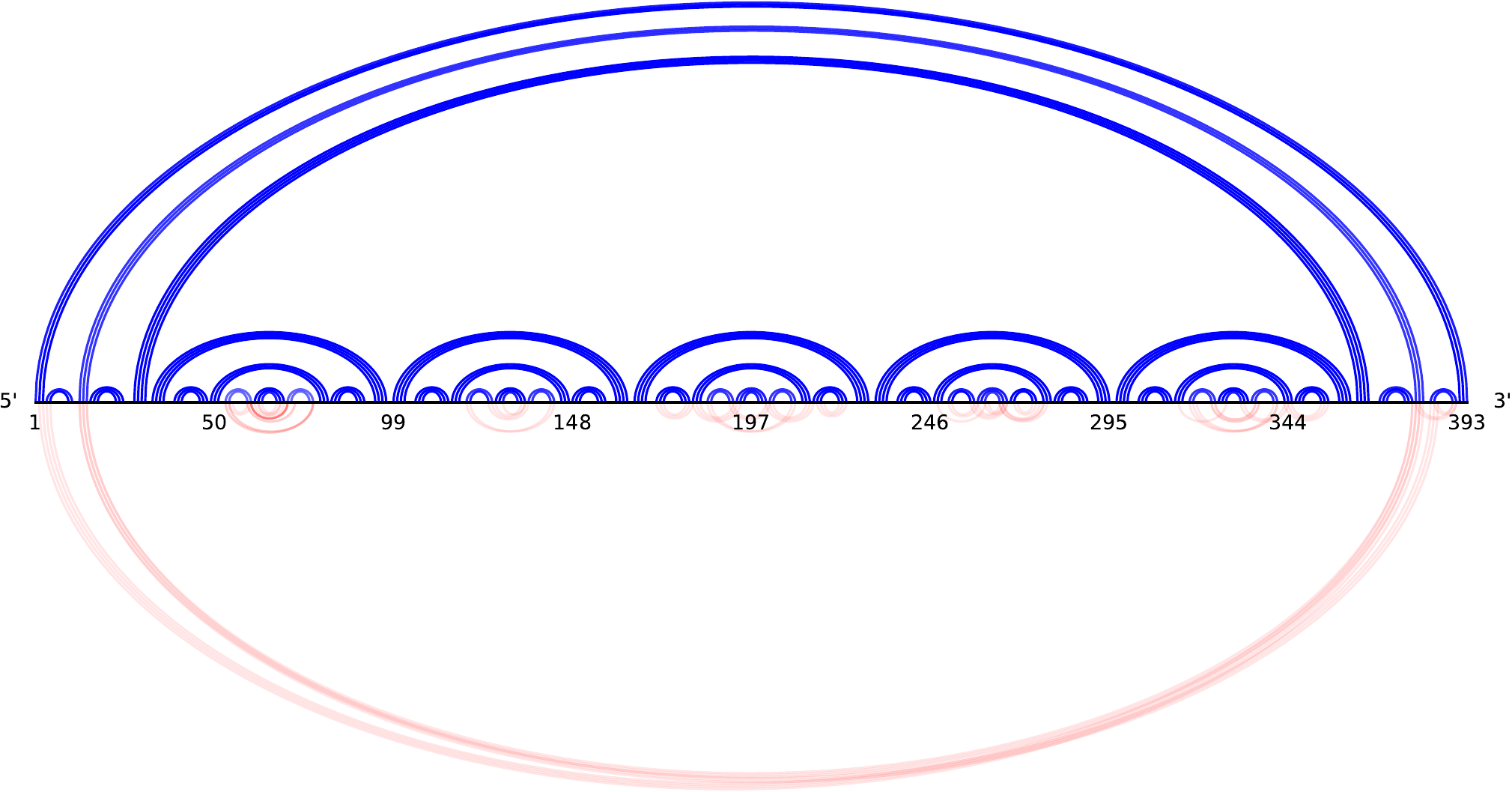} 
          \!\!\!\!\!
          &
          \!\!\!\!\!
          \includegraphics[width=0.51\linewidth]{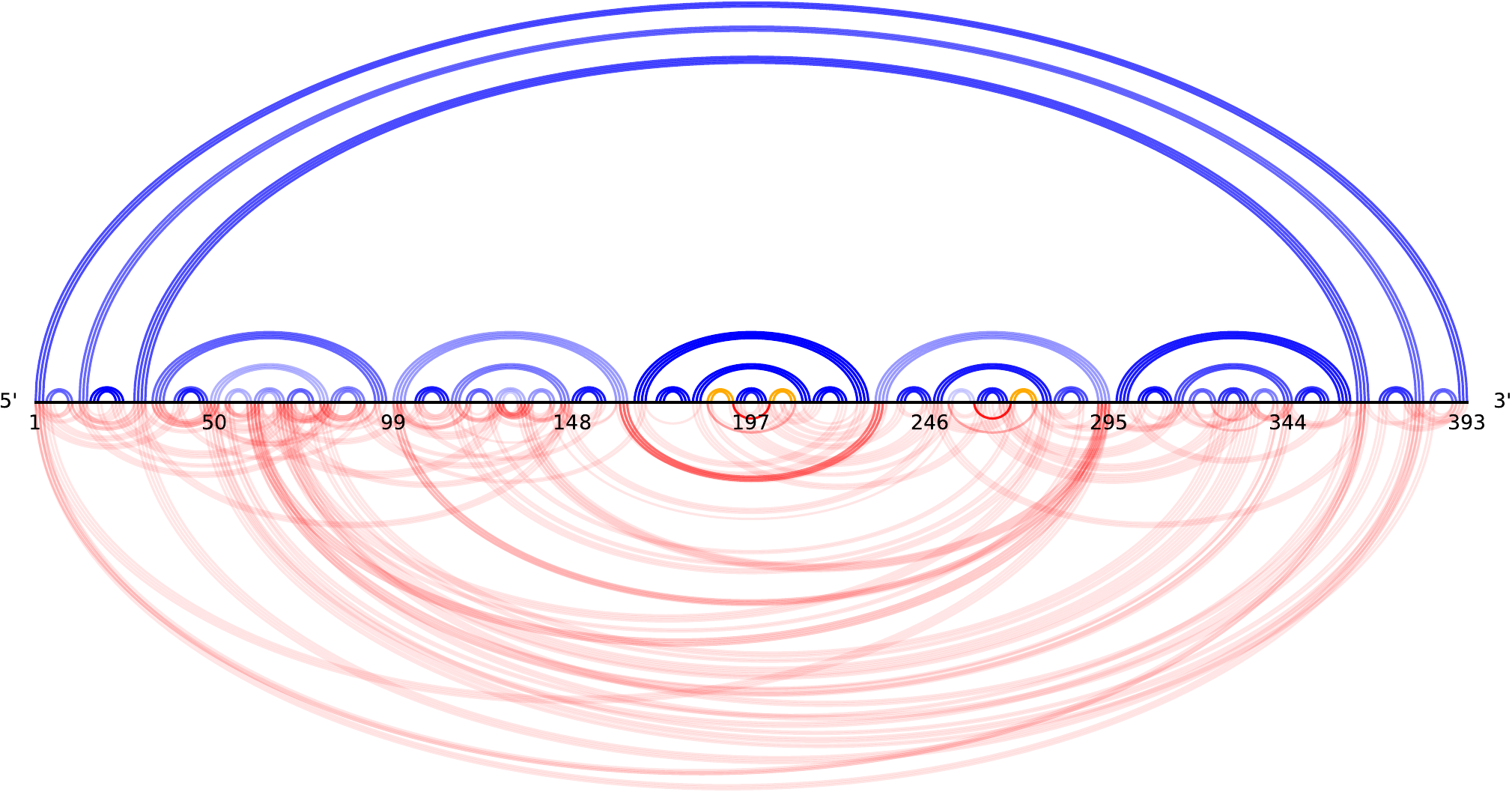}
      \end{tabular}
      \caption{Comparison of the best $p(\vecystar \mid \vecx)$ solutions designed by this work vs.~SAMFEO for Puzzle 76 (``Snowflake 3''). (b) -- (c) \MFE structures of the solutions from this work and SAMFEO. Base-pairs are colored as follows: blue for correct pairs, red for incorrect pairs, with the intensity indicating pairing probability. Nucleotide colors range from blue to red, indicating positional defect. (d) -- (e) Base-pairing probabilities of this work and SAMFEO. Orange represents missing correct pairs (i.e.~correct pairs with a pairing probability below 0.1).}
      \label{fig:76}
    \end{figure*}

%% file: fig-si-91.tex
\begin{figure*}[!h]
    \centering   
      (a)  \begin{tabular}{c|c|c|c|c|c|c||c}
        \texttt{\#91} (392 \nts)& $p(\vecystar \mid \vecx)$ & NED$(\vecx, \vecystar)$ & $d(\MFE(\vecx), \vecystar)$ & $\DDG(\vecx, \vecystar)$ & is MFE & is uMFE & $p(\vecytilde \mid \vecx)$\\
        & $(\uparrow)$ & $(\downarrow)$ & $(\downarrow)$ & $(\downarrow)$ & & & $(\uparrow)$ \\
        \hline
        This Work & \textbf{0.0001} & \textbf{0.034} & \; \textbf{8} & \; \textbf{3.4 kcal/mol} & No & No & \textbf{0.037}\\
        SAMFEO & $2 \times 10^{-20}$ & 0.128 & 52 & 25.2 kcal/mol & No & No & $8 \times 10^{-13}$\\
    \end{tabular}

      \includegraphics[width=.8\linewidth]{./figs/color_legend-crop}
      \begin{tabular}{ccc}
        (b) Target Structure \vecystar
        &
        (c) MFE Structure (This Work)
        &
        (d) MFE Structure (SAMFEO)
        \\
        \includegraphics[width=0.32\linewidth]{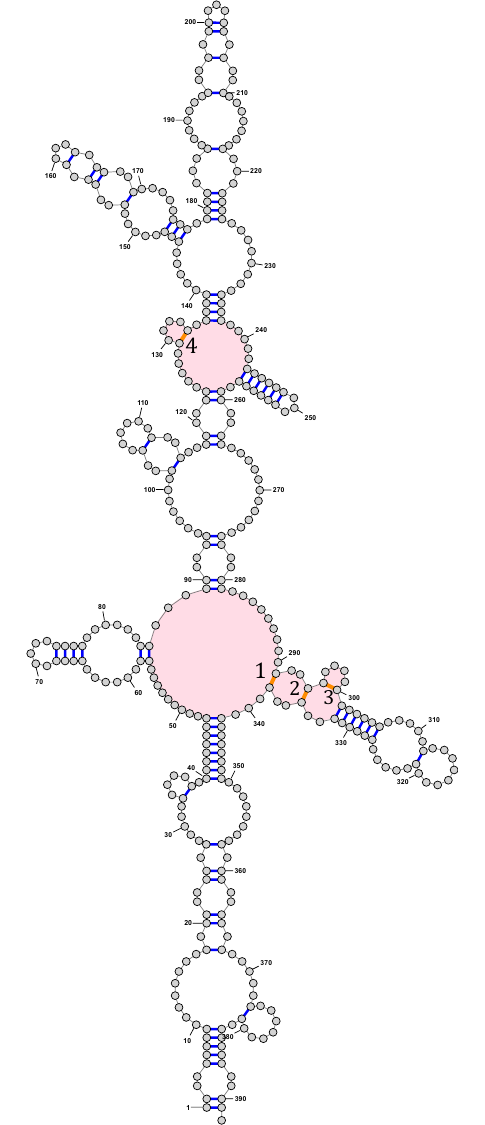}
        &
        \includegraphics[width=0.32\linewidth]{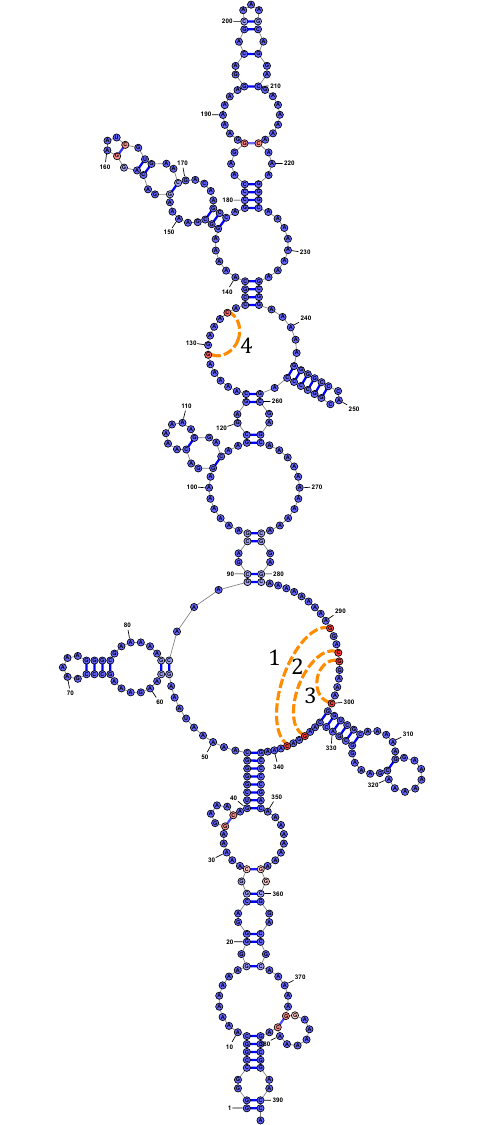}
        &
        \includegraphics[width=0.32\linewidth]{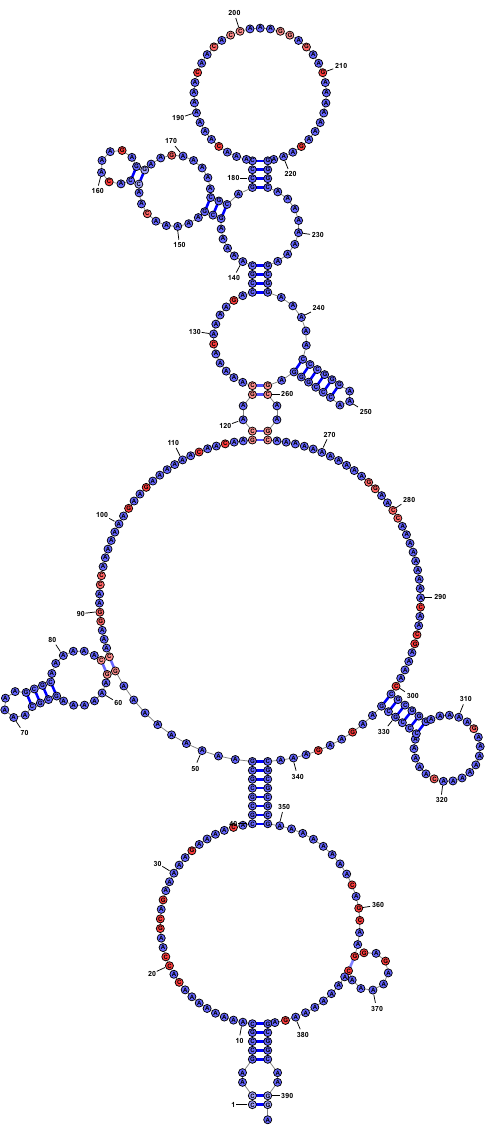}
      \end{tabular}\\[6pt]
      \begin{tabular}{cc}
        (d) Base-Pairing Probabilities (This Work)
        &
        (e) Base-Pairing Probabilities (SAMFEO)
        \\
          \hspace{-0.5cm}
          \begin{tikzpicture}
            \node[anchor=south west,inner sep=0] (image) at (0,0) {\includegraphics[width=0.5\linewidth]{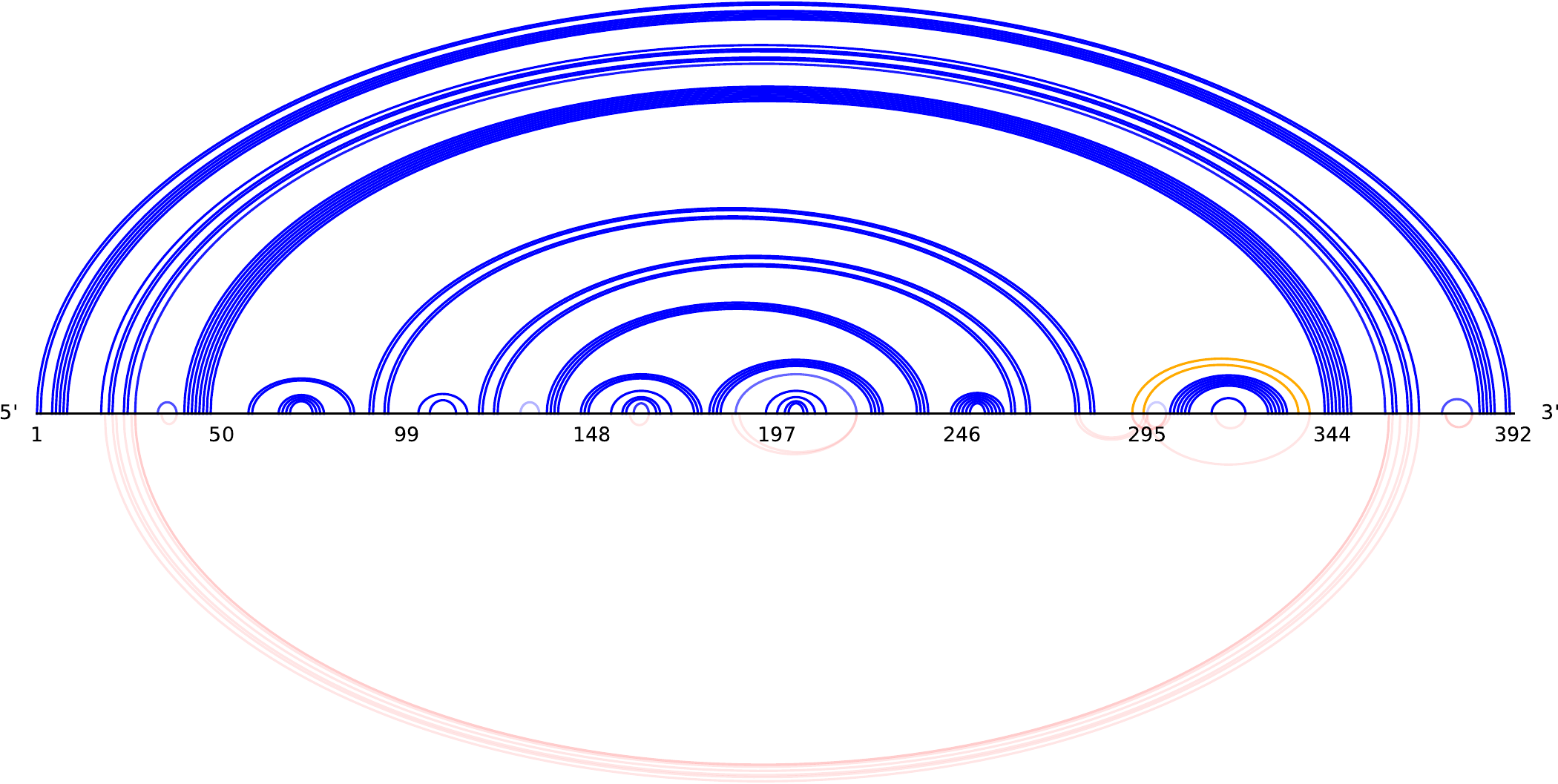}};
            \node at (6.68, 2.45) {$\scriptstyle 1$};
            \node at (7.34, 2.33) {$\scriptscriptstyle 2$};
            \node at (6.6, 2.22) {$\scriptscriptstyle 3$};
            \node at (3.0, 2.27) {$\scriptstyle 4$};
          \end{tikzpicture}
          \!\!\!
          &
          \!\!\!\!\!
          \raisebox{.7cm}{\includegraphics[width=0.52\linewidth]{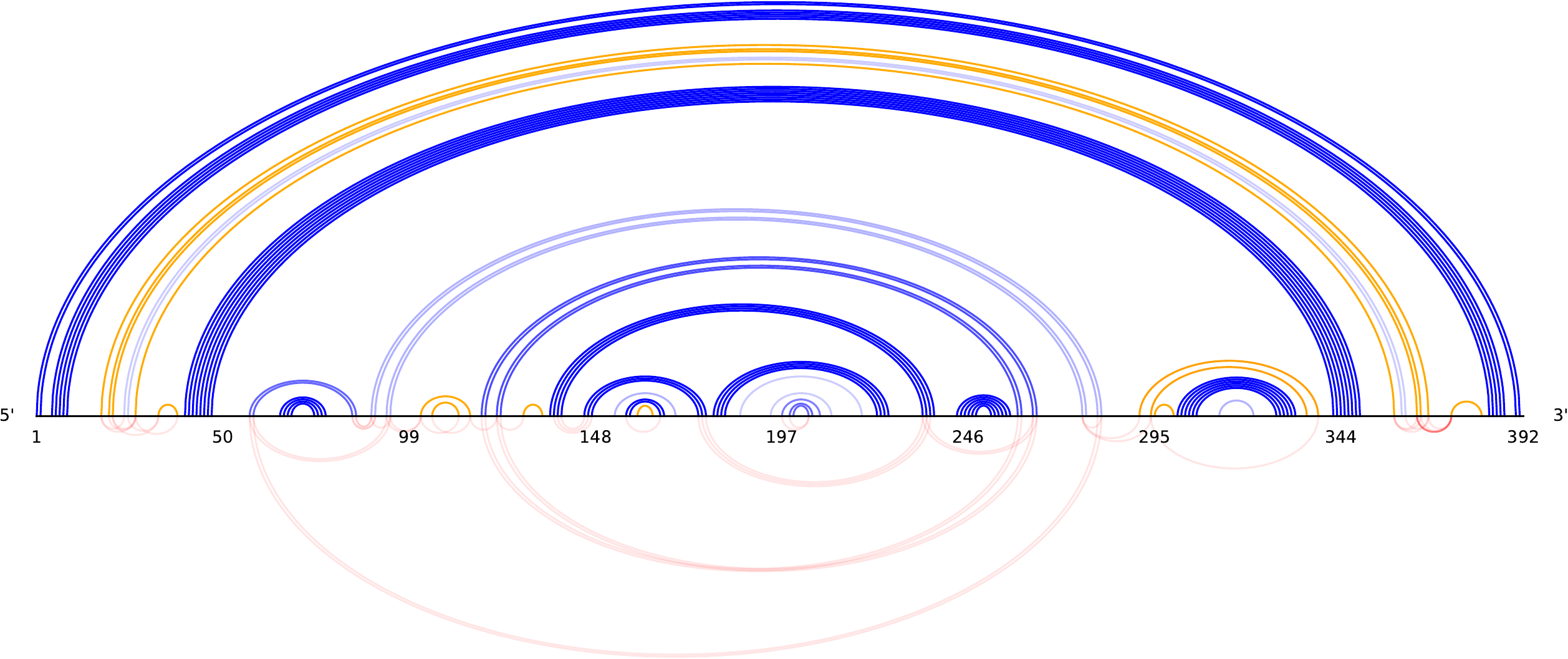}}
      \end{tabular}
      \caption{Comparison of the best $p(\vecystar \mid \vecx)$ solution designed by this work vs.~SAMFEO for Puzzle 91 (``Thunderbolt'').
      (b) Target structure: pink-filled regions highlight loops that belong to an undesignable motif, while orange base pairs represent the missing pairs in Sampling's \MFE structure.
      (c) -- (d) \MFE structures of the best $p(\vecystar \mid \vecx)$ solutions from this work and SAMFEO. (e) -- (f) Base-pairing probabilities plots.
      Base-pairs are colored as follows: blue for correct pairs, red for incorrect pairs, with the intensity indicating pairing probability. Orange represents missing correct pairs (i.e.~correct pairs with a pairing probability below 0.1). 
      Nucleotide colors range from blue to red, indicating positional defect. 
      $\vecytilde$ refers to the target structure with the (orange) base pairs from undesignable motifs removed (i.e.~pairs $1$, $2$, $3$ and $4$ are removed).}
      \label{fig:91}
\end{figure*}

%% file: fig-si-99.tex
\begin{figure*}[!h]
    \centering   
      (a)  \begin{tabular}{c|c|c|c|c|c|c||c}
        \texttt{\#99} (364 \nts) & $p(\vecystar \mid \vecx)$ & NED$(\vecx, \vecystar)$ & $d(\MFE(\vecx), \vecystar)$ & $\DDG(\vecx, \vecystar)$ & is MFE & is uMFE & $p(\vecytilde \mid \vecx)$\\
        & $(\uparrow)$ & $(\downarrow)$ & $(\downarrow)$ & $(\downarrow)$ & & & $(\uparrow)$ \\
        \hline
        This Work & $\mathbf{8 \times 10^{-11}}$ & \textbf{0.111} & \textbf{20} & \; \textbf{9.8 kcal/mol} & No & No & $\mathbf{2 \times 10^{-10}}$\\
        SAMFEO & $3 \times 10^{-28}$ & 0.197 & 80 & 34.4 kcal/mol & No & No & $4 \times 10^{-26}$\\
    \end{tabular}

      \includegraphics[width=.8\linewidth]{./figs/color_legend-crop}
      \\[6pt]
      \begin{tabular}{ccc}
          (b) Target Structure \vecystar
          &
          (c) MFE Structure (This Work)
          &
          (d) MFE Structure (SAMFEO)
          \\
          \includegraphics[width=0.31\linewidth]{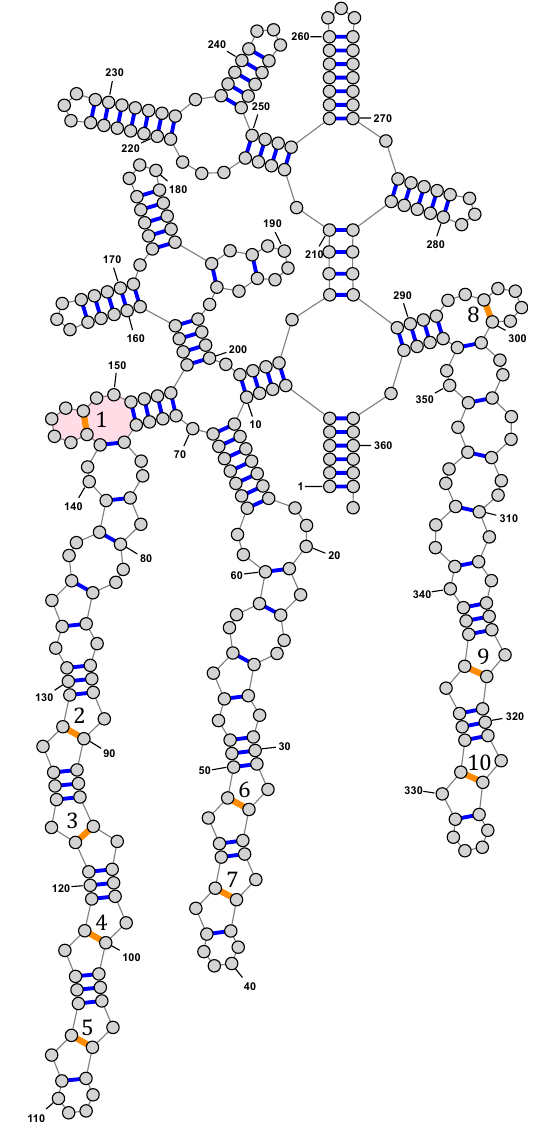}
          &
          \includegraphics[width=0.31\linewidth]{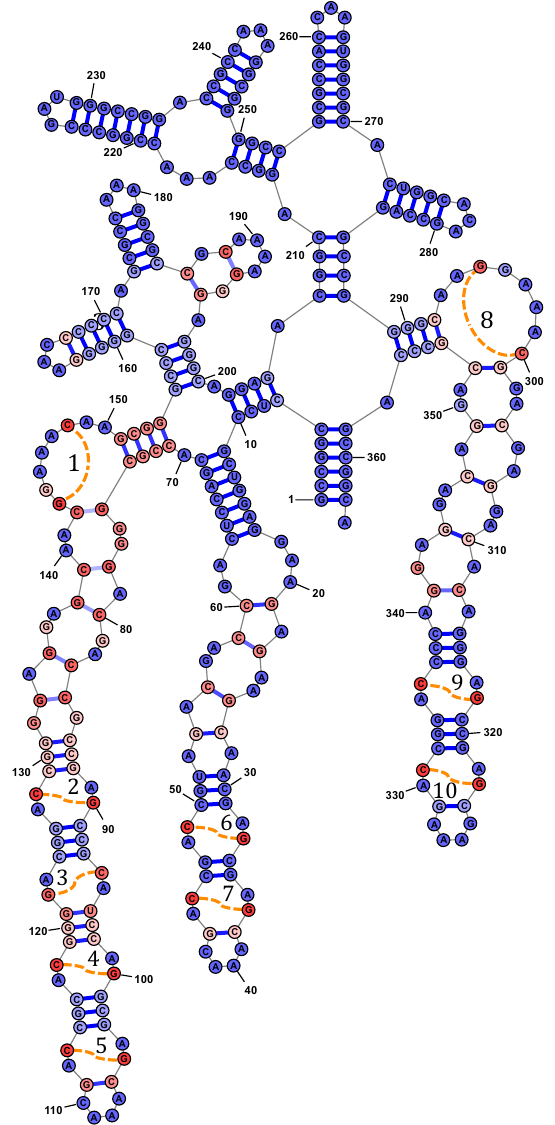}
          &
          \includegraphics[width=0.33\linewidth]{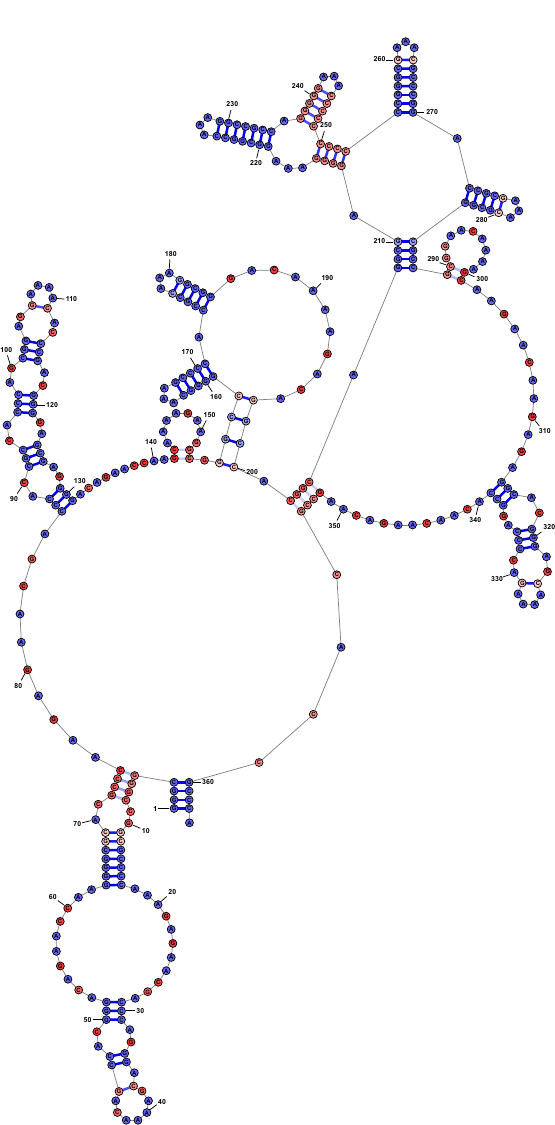}
      \end{tabular}
      \\[6pt]
      \begin{tabular}{cc}
        (d) Base-Pairing Probabilities (This Work)
        &
        (e) Base-Pairing Probabilities (SAMFEO)
        \\
            \hspace{-0.53cm}
            \raisebox{1.29cm}{
              \begin{tikzpicture}
                \node[anchor=south west,inner sep=0] (image) at (0,0) {\includegraphics[width=0.53\linewidth]{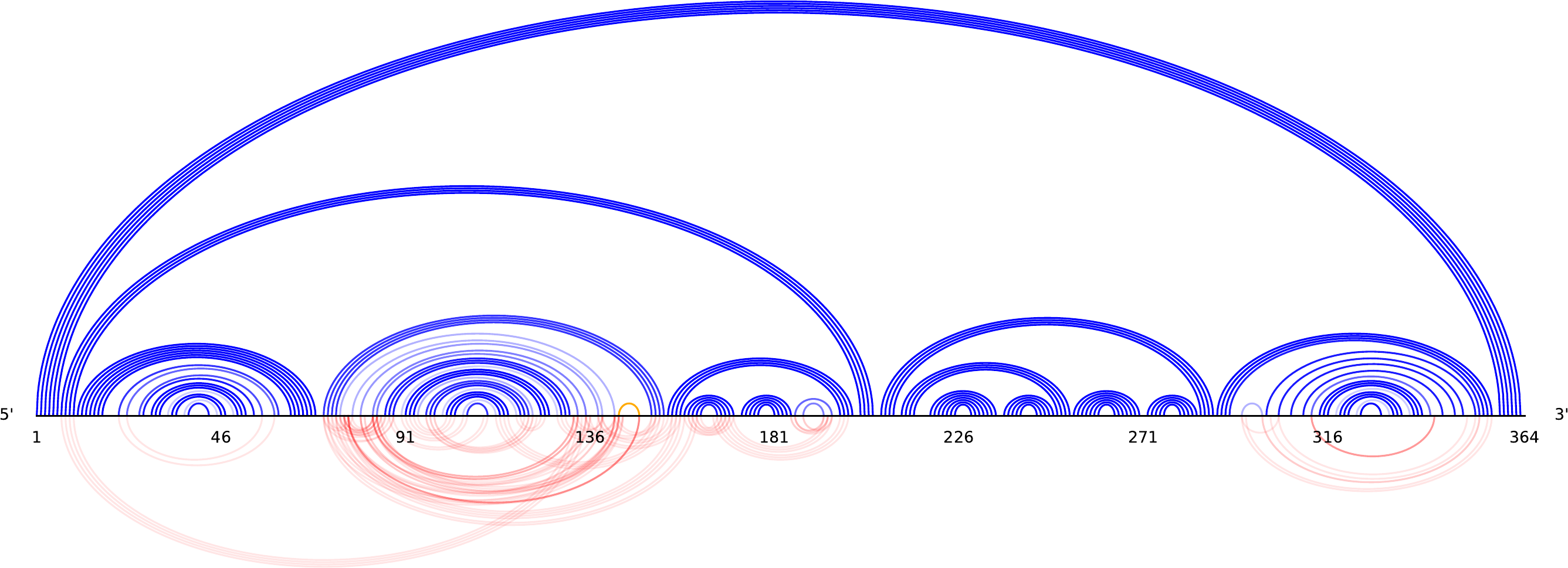}};
                \node at (3.78, 1.1) {$\scriptstyle 1$};
                \node at (7.5, 1.1) {$\scriptstyle 8$};
              \end{tikzpicture}
            }
            \!\!\!
          &
            \!\!\!
            \includegraphics[width=0.48\linewidth]{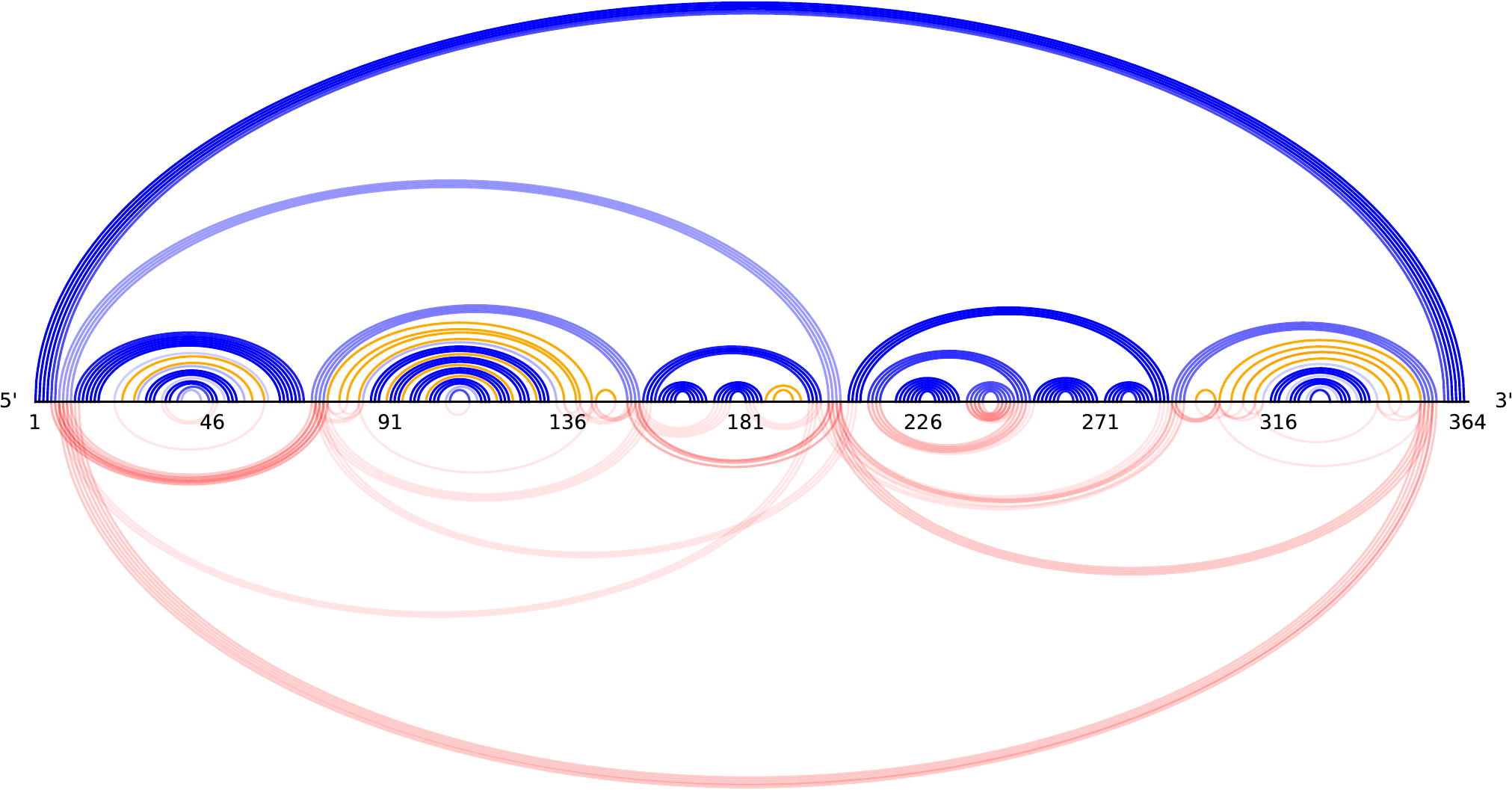}
      \end{tabular}
      \caption{Comparison of the best $p(\vecystar \mid \vecx)$ solution designed by this work vs.~SAMFEO for Puzzle 99 (``Shooting Star'').
      (b) Target structure: pink-filled regions highlight loops that belong to an undesignable motif, while orange base pairs represent the missing pairs in Sampling's \MFE structure.
      (c) -- (d) \MFE structures of the best $p(\vecystar \mid \vecx)$ solutions from this work and SAMFEO. (e) -- (f) Base-pairing probabilities plots. 
      Base-pairs are colored as follows: blue for correct pairs, red for incorrect pairs, with the intensity indicating pairing probability. Orange represents missing correct pairs (i.e.~correct pairs with a pairing probability below 0.1). 
      Nucleotide colors range from blue to red, indicating positional defect. 
      $\vecytilde$ refers to the target structure with the (orange) base pairs from undesignable motifs removed (i.e.~pair $1$ is removed).}
      \label{fig:99}
\end{figure*}

%% file: fig-si-steps.tex
\begin{figure*}[!htb]
    \centering
    \begin{tabular}{cc}
        (a) Results at different cutoff step
        &
        (b) Number of steps for each puzzle (sorted by length $\rightarrow$)
        \\
        \hspace{-.5cm}
        \multirow{2}{*}[3.5cm]{
            \includegraphics[width=0.47\linewidth]{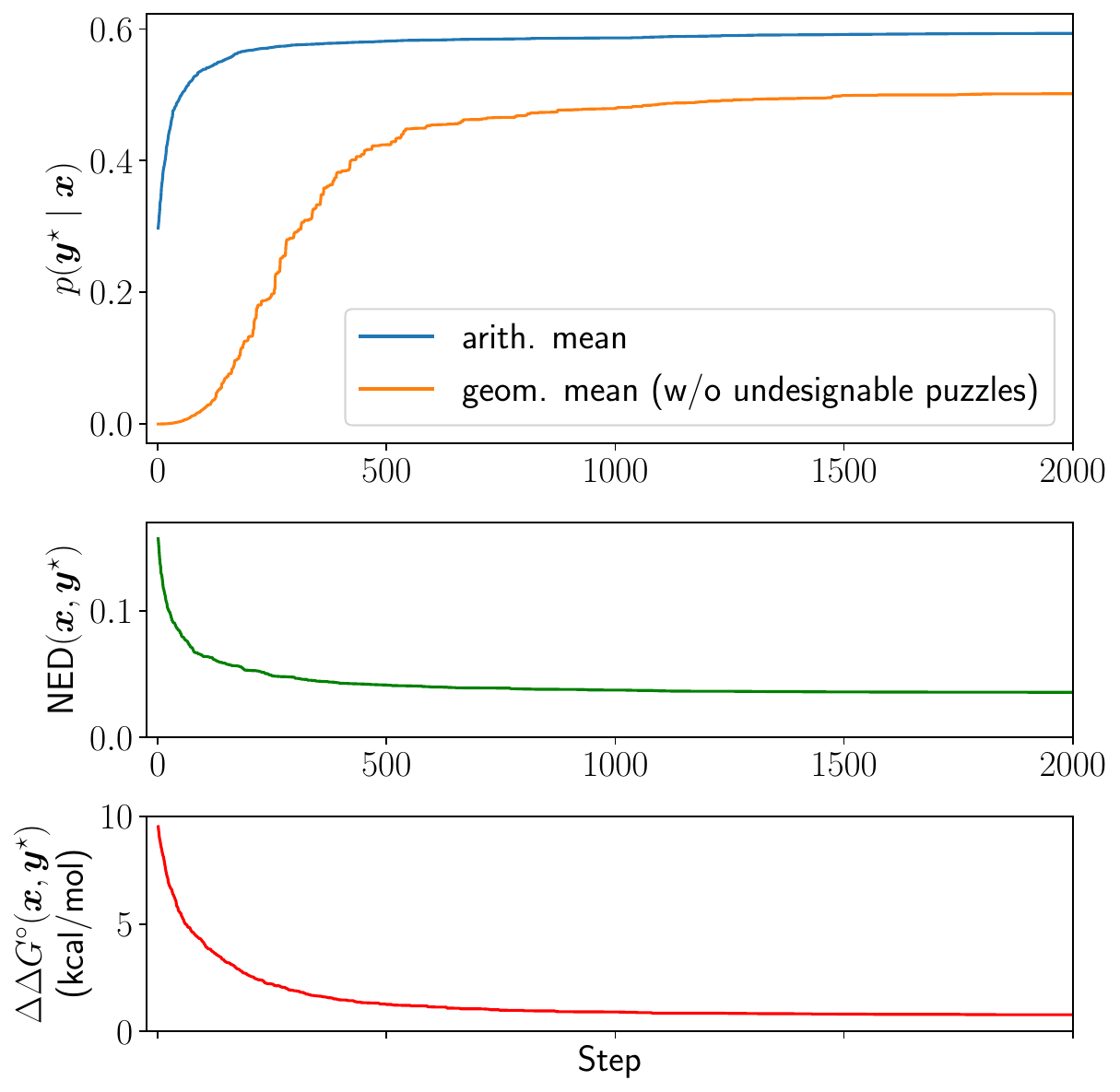}
        }
        &
        \includegraphics[width=0.5\linewidth]{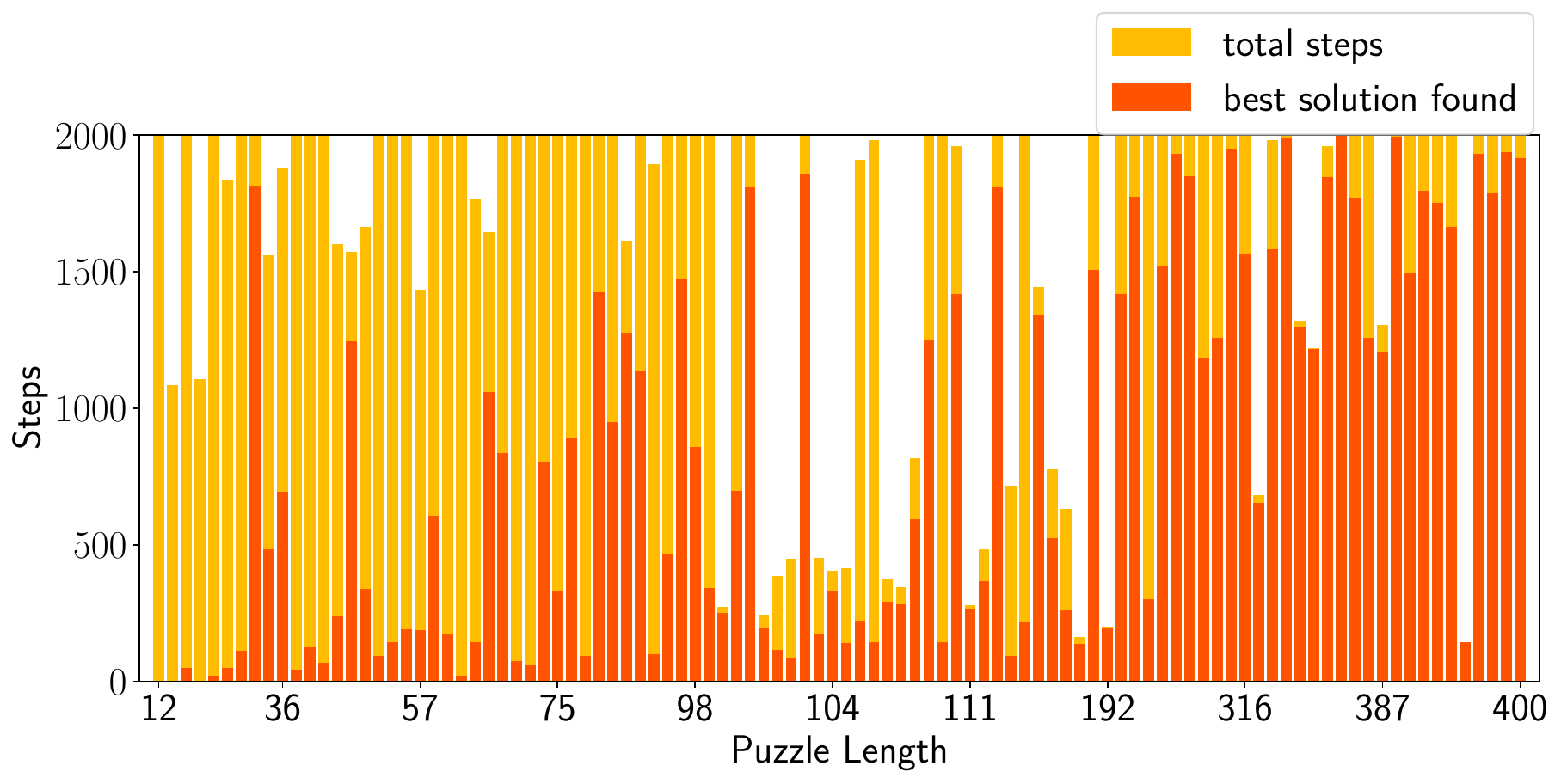}
        \\
        &
        (c) Total time for each puzzle
        \\
        &
        \begin{tikzpicture}
            \node [] at (0, 0) {\includegraphics[width=0.5\linewidth]{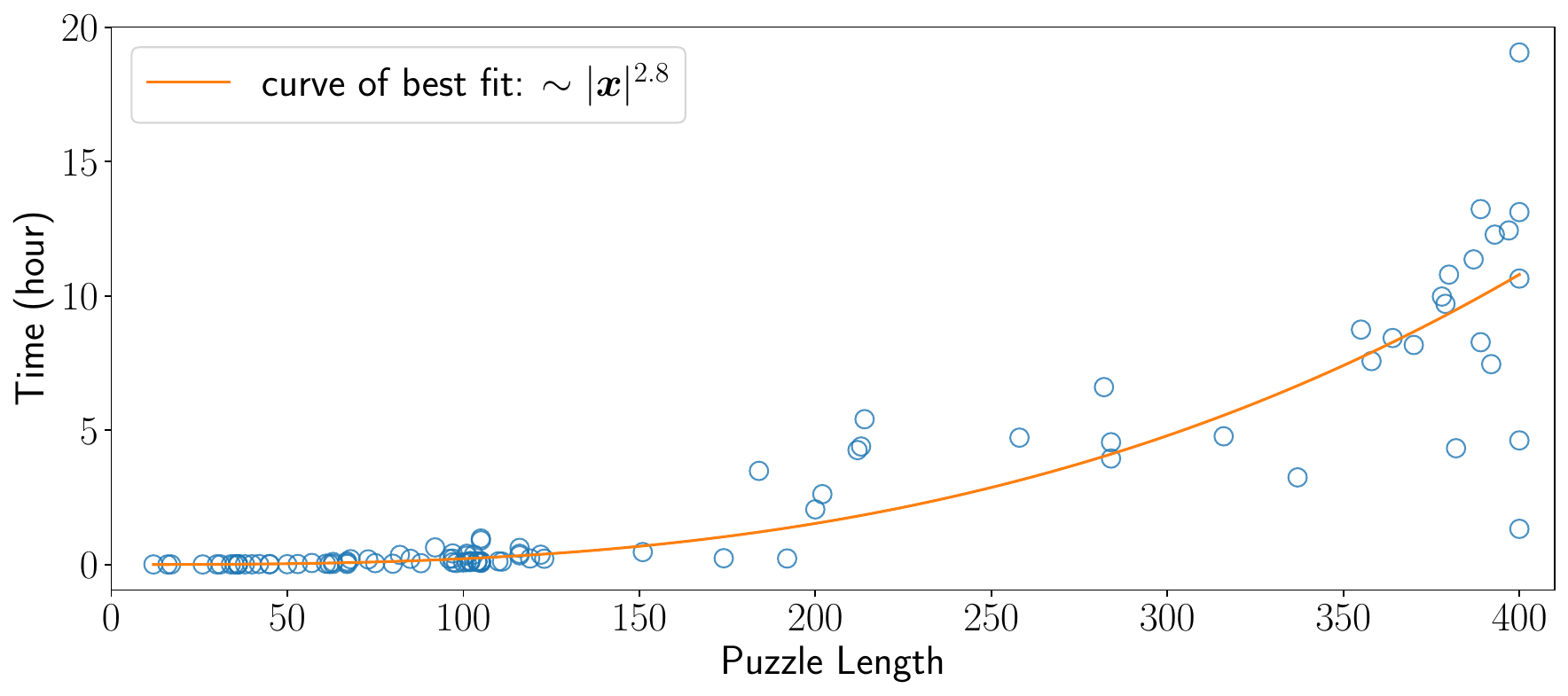}};
        \end{tikzpicture}
    \end{tabular}
    \caption{(a) Metrics of the puzzles over different step cutoffs (up to 2000 steps). (b) Number of steps taken to solve each puzzle with the stopping criteria: 50 steps in which the objective function does not improved and the total number of steps is limited to 2000. (c) Total time taken to solve each puzzle, ran on a server with 28 physical cores.}
    \label{fig:steps}
\end{figure*}


%% file: fig-si-learning.tex
\begin{figure*}[!hb]
    \centering
      \begin{tabular}{cc}
        (a) Puzzle \texttt{\#97} (400 \nts)
        &
        (b) Puzzle \texttt{\#98} (202 \nts)
        \\
        \includegraphics[width=0.48\linewidth]{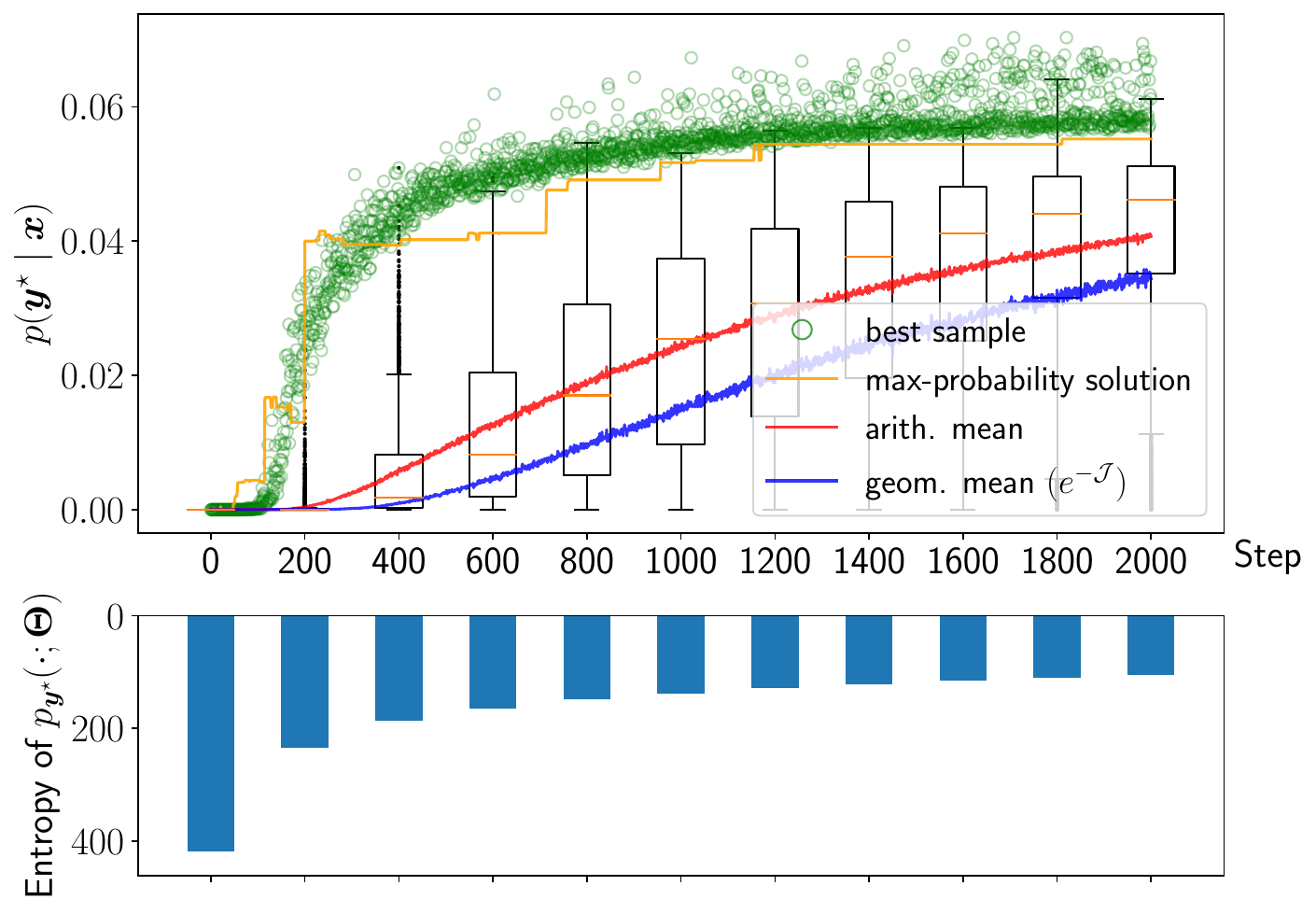}
        &
        \includegraphics[width=0.48\linewidth]{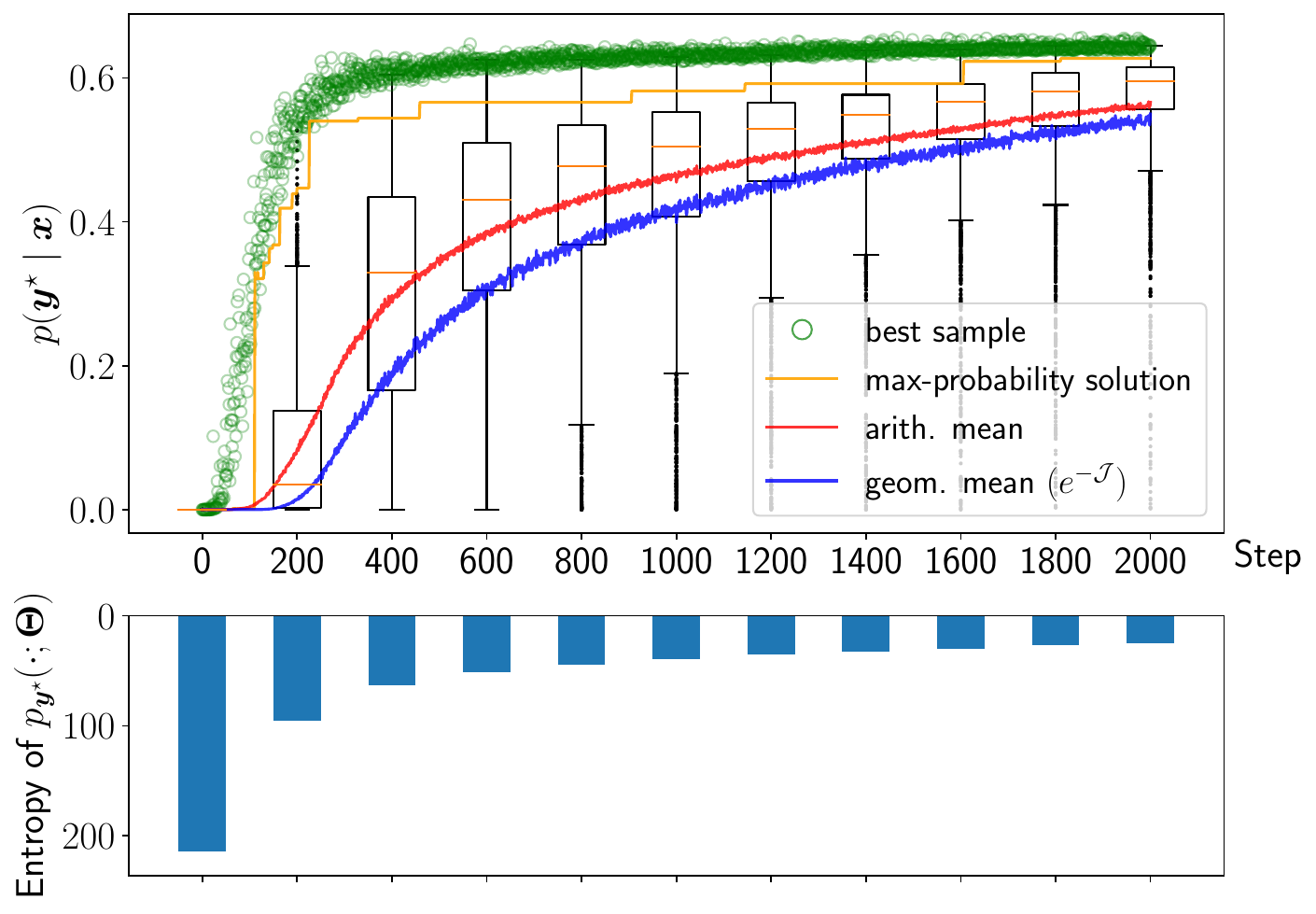}
      \end{tabular}
      \includegraphics[width=0.8\linewidth]{./figs/color_legend-crop}
      \\[6pt]
      \begin{tabular}{cc}
        \hspace{-.5cm}
        (c) \MFE Structure $\vecyhat_1$ (This Work) for Puzzle \texttt{\#97}
        &
        (d) \MFE Structure $\vecyhat_2$ (This Work) for Puzzle \texttt{\#98}
        \\
        \hspace{-.5cm}
        \raisebox{-1.5cm}{
          \includegraphics[width=0.54\linewidth]{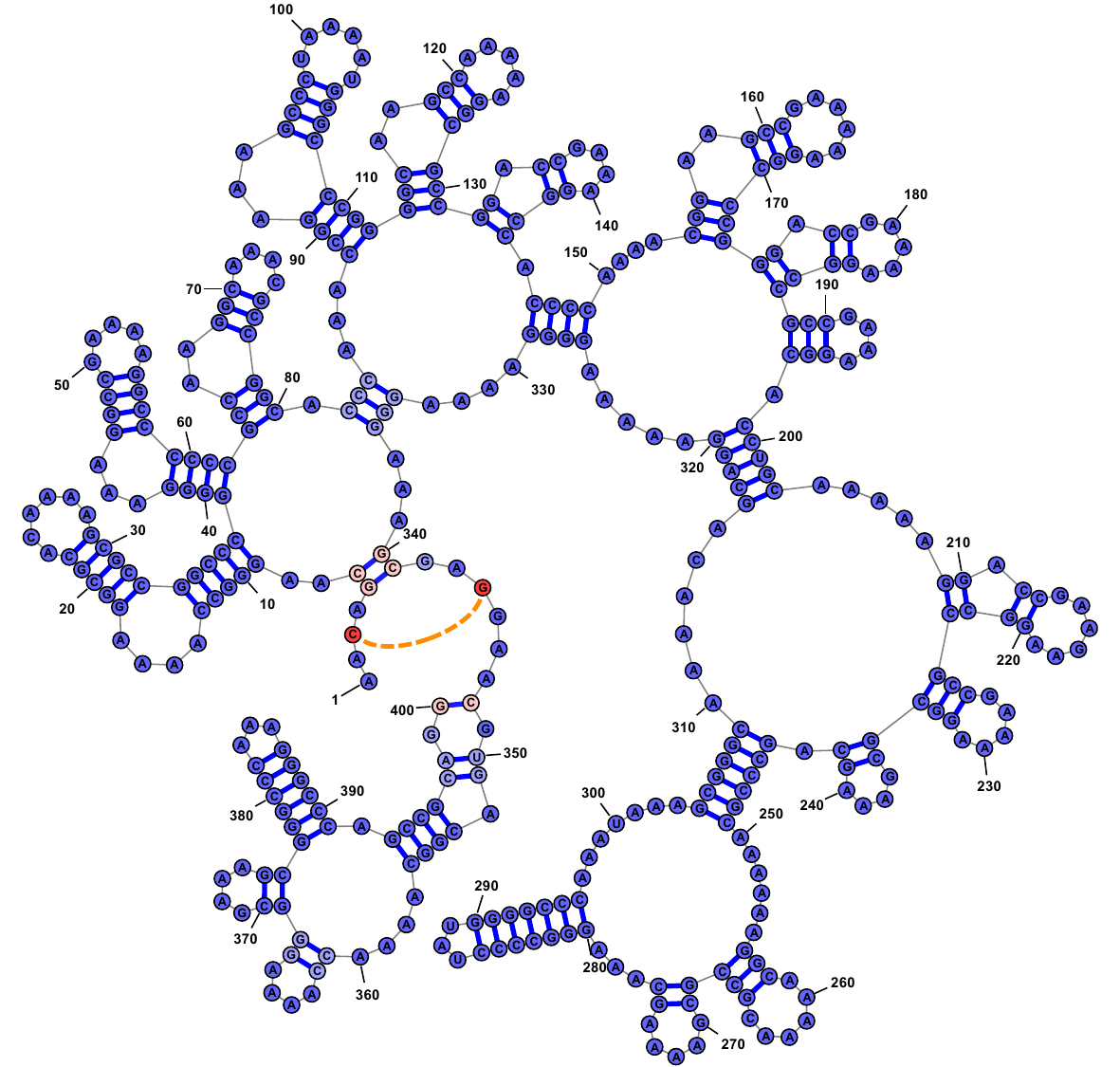}
        }
        &
        \hspace{-.5cm}\includegraphics[width=0.38\linewidth]{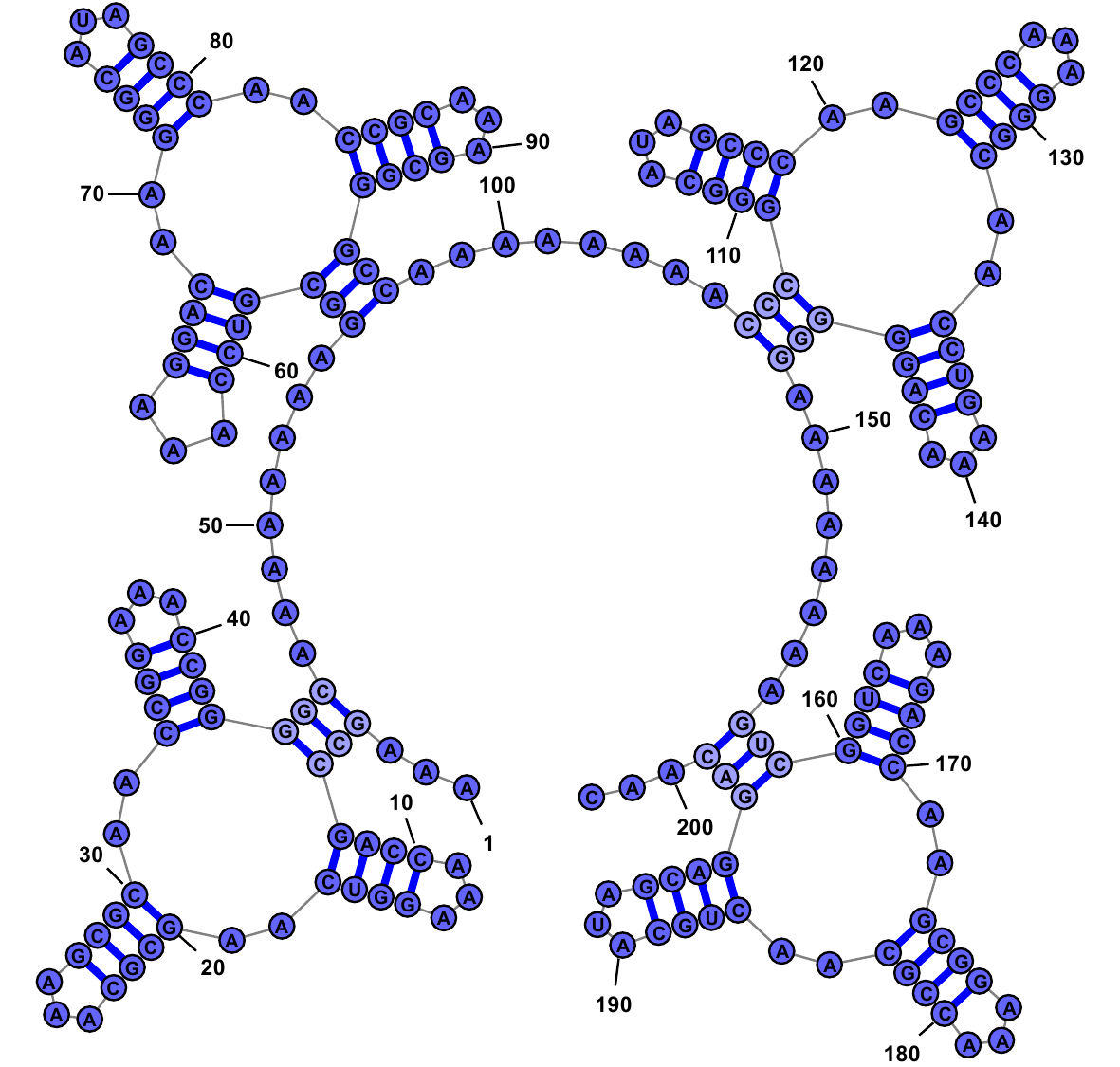}
      \end{tabular}
      \caption{(a) -- (b) Learning curves of puzzles \texttt{\#97} and \texttt{\#98}. Each step illustrates the $p(\vecystar \mid \vecx)$ for both the best sample and the integral solution with the arithmetic and geometric means of $p(\vecystar \mid \vecx)$ across all samples. The boxplots depict the distribution of $p(\vecystar \mid \vecx)$ across all samples every 200 step, showing the interquartile range and the median of the samples. The barplots correspond to the entropy of the distribution every 200 step. (c) -- (d) The \MFE structures of the best $p(\vecystar \mid \vecx)$ solution from our method. Puzzle \texttt{\#97} is ``unknown'' in the sense that it does not have a known \UMFE solution and has not yet been proven to be undesignable. The \MFE structure of puzzle \texttt{\#97}, $\vecyhat_1$, is missing a base pair from the target structure, denoted by the orange dashed line. For this puzzle, $p(\vecystar \mid \vecx) = 0.078$ and $p(\vecyhat_1 \mid \vecx) = 0.337$.}
      \label{fig:learning-curves}
    \end{figure*}